\pdfoutput=1
\documentclass[nofootinbib,aps]{revtex4}
\usepackage{amssymb,amsmath,graphicx,color,microtype,bm}
\usepackage{enumerate}
\usepackage{slashed}
\usepackage{xcolor}
\usepackage{multirow}
\usepackage{booktabs}
\usepackage{placeins}
\usepackage{empheq}
\usepackage[linktoc=page]{hyperref}
\hypersetup{
	colorlinks=true,
	linkcolor=blue,
	pdftitle={Chemical-potential-assisted particle production in FRW spacetimes},
	bookmarks=true,
	citecolor=[rgb]{0.54,0,0}
}

\newcommand*\widefbox[1]{\fbox{\hspace{2em}#1\hspace{2em}}}
\def\Re{\mathrm{Re}\,}
\def\Im{\mathrm{Im}\,}
\def\Res{\mathrm{Res}\,}
\usepackage[paperwidth=210mm,paperheight=297mm,centering,hmargin=2cm,vmargin=2.5cm]{geometry}
\graphicspath{{fig/}}

\begin{document}
\title{Chemical-Potential-Assisted Particle Production in FRW Spacetimes}
\author{Chon Man Sou$^{1,2}$}
\email{cmsou@connect.ust.hk}
\author{Xi Tong$^{1,2}$}
\email{xtongac@connect.ust.hk}
\author{Yi Wang$^{1,2}$}
\email{phyw@ust.hk}
\affiliation{${}^1$Department of Physics, The Hong Kong University of Science and Technology, \\
	Clear Water Bay, Kowloon, Hong Kong, P.R.China}
\affiliation{${}^2$The HKUST Jockey Club Institute for Advanced Study, The Hong Kong University of Science and Technology, \\
	Clear Water Bay, Kowloon, Hong Kong, P.R.China}

\begin{abstract}
	We analyze gravitational particle production assisted by chemical potential. By utilizing the uniformly smoothed Stokes-line method and Borel summation, we gain insight into the fine-grained history of enhanced particle production. Analytic/semi-analytic formulae describing the production amount, time and width are obtained for both spin-1 and spin-1/2 particles in various FRW spacetimes. Our work also serves as a concrete demonstration of the uniformly smoothed Stokes-line method applied to cosmology.
\end{abstract}

\maketitle

\tableofcontents
\section{Introduction}
The study of particle production in cosmology is, in a sense, the study of the origin of matter itself. Arguably, the most natural choice of the initial state of the universe is the one in which strict spatial homogeneity is maintained, if space is to retain its physical meaning at all. Such a perfectly homogeneous state must also be devoid of any ordinary matter, which we view as a collection of fundamental particles. 
In addition, on the classical level, this strict homogeneity will be kept throughout the cosmic evolution if the Lagrangian respects translational symmetry. This would naively suggest a dull universe with no matter but zero modes only. However, this is not the whole story. 
The dynamics of the background spacetime can bring the vacuum fluctuations to reality, breaking the spatial homogeneity on a quantum level. This spontaneous breaking of spatial translational invariance is accompanied with the appearance of perturbative modes in spacetime, which upon quantization, become real particles. 

Such is the scenario for the paradigm of inflation \cite{Guth:1980zm,Linde:1981mu,Albrecht:1982wi,Starobinsky:1980te}. The initial condition of the inflationary universe is chosen as a Bunch-Davis (BD) vacuum, $i.e.$, a coherent state of the inflaton field that is annihilated by the positive-frequency part of all fields with non-zero momenta. However, the expansion of the spacetime stretches the wavelength of different modes $\phi_{\mathbf{k}}$ and produces a squeezed state with occupation number $|\beta(k)|^2\neq 0$. For $\phi$ being the inflaton or the graviton, this occupation number is exponentially large so that these particles decohere and become essentially classical waves, which source the primordial inhomogeneity for the later cosmic evolutions. For $\phi$ being heavier degrees of freedom, the occupation number is finite and typically suppressed by an exponentially small factor $|\beta(k)|^2\propto e^{-2\pi m/H}$, where $H\lesssim 10^{13}$GeV is the Hubble parameter during inflation. These gravitationally produced particles are the very first matter emergent from the BD vacuum in the inflationary universe and their interactions leave characteristic non-Gaussian imprints on the Cosmic Microwave Background (CMB) as well as the Large Scale Structure (LSS). This recently thriving field known as cosmological collider physics \cite{Chen:2009we,Chen:2009zp,Baumann:2011nk,Noumi:2012vr,Arkani-Hamed:2015bza,Lee:2016vti} thus has an intimate relation with the phenomenon of particle production from the vacuum. For example, the signal strength is directly proportional to the square root of the production amount, $S\propto |\beta(k)|$. Thus heavy particles with $m\gg H$ are extremely difficult to probe if they are produced in the purely gravitational way. 

This problem motivates several proposals where the exponential suppression can be alleviated. One possibility is to consider special inflation models. For instance, in axion-monodromy inflation \cite{Silverstein:2008sg,McAllister:2008hb}, time-dependent mass terms violate adiabaticity and lead to a dramatic amplification of particle number density and thus the size of non-Gaussianity \cite{Flauger:2016idt}. In warm inflation \cite{Berera:1995wh,Berera:1995ie}, thermally produced heavy particles are Boltzmann-suppressed by an alternative temperature much higher than the Hubble scale, also giving rise to enhanced cosmological collider signals \cite{Tong:2018tqf}. 

Alternatively, one can consider the interesting possibility of introducing a chemical potential. This can be naturally achieved via a rolling scalar field $\phi$ coupled to massive fields through operators of the form $\partial_\mu\phi J^\mu$, where $J^\mu$ is a certain current made from massive fields. For massive spin-1 particles, choosing the Chern-Simons current $J_{CS}$ results in an enhanced production of vector boson \cite{Turner:1987bw,Garretson:1992vt,Barnaby:2010vf,Liu:2019fag,Wang:2020ioa}. For massive spin-1/2 fermions, choosing the chiral current $J_5$ leads to a natural amplification of fermion production rate \cite{Adshead:2015kza,Chen:2018xck,Adshead:2018oaa,Hook:2019zxa}. Similar amplification is also found for charged scalar particles if we generalize the chemical potential operator to $\kappa_\mu J^\mu$, and reinterpret the chemical potential $\kappa_\mu$ as a background gauge field creating Schwinger pairs \cite{Kobayashi:2014zza,Geng:2017zad,Chua:2018dqh}. These mechanisms of chemical-potential-assisted particle production typically break parity or rotational invariance, hence leaving sizable and distinctive signatures in cosmological observables. 

In addition to physics during inflation, chemical potential can also play a role in the late universe. The aforementioned chemical potential introduced by a rolling Axion-Like Particle (ALP) generates a tachyonic instability in the gauge boson sector. This can, for example, efficiently convert the ALP to Dark Photon Dark Matter (DPDM) \cite{Co:2018lka}, and produce chiral gravitational waves \cite{Machado:2018nqk,Machado:2019xuc,Salehian:2020dsf}. In the fermion sector, chemical potential also sources the helicity asymmetry of fermion numbers, which can be important for baryogenesis \cite{Adshead:2015kza,Adshead:2015jza}.

Thus, chemical-potential-assisted particle production is a generic phenomenon that appears in many different contexts and setups. A general and systematic investigation of particle production in the presence of chemical potential is necessary. 

In this work, we re-derive Berry's uniformly smoothed Stokes-line method \cite{Berry:1989zz,Berry:1990,Berry:1990histories,Berry_1993} and apply it to analyze the \textit{fine-grained} production history of particles of mass $m$ with chemical potential $\kappa$. By fine-grained production history, we mean the production amount, time and width (duration) exact to the leading order in the super-adiabatic expansion. We found that for a constant chemical potential $\kappa=\text{const}$, spin-1/2 fermions and spin-1 vector bosons share a similar production history, with a simple yet subtle replacement rule $m^2\leftrightarrow m^2+\kappa^2$. We also derive analytic/semi-analytic formulae for the production history in five common FRW spacetimes. 

We note that there are many past studies in the literature that utilize the Stokes phenomenon to study particle production. For example, it is applied to the Sauter-Schwinger effect \cite{Dumlu:2010ua,Dumlu:2010vv,Dabrowski:2016tsx,KIM:2018rvw,CAI:2019vow,Taya:2020dco}, to Hawking radiation \cite{Dumlu:2020wvd}, to the adiabatic particle number in global de Sitter (dS) spacetime \cite{Kim:2010xm,Kim:2013cka,Dabrowski:2014ica}, to dark matter production at the end of inflation \cite{Li:2019ves}, to preheating \cite{Enomoto:2020xlf,Enomoto:2021hfv}, and to particle production triggered by vacuum decay \cite{Hashiba:2020rsi}. An excellent review is recently given in \cite{Hashiba:2021npn}. However, we point out that most of them (with the exception of \cite{Dabrowski:2014ica,Dabrowski:2016tsx,Hashiba:2020rsi}) focus on the asymptotic production amount far away from the Stokes line. To our knowledge, the analytic calculation of production time and width presented in this work is a new ingredient in this area, with or without chemical potential. Understanding these fine production details is useful, for example, in the loop-level estimation of cosmological collider signals, or in estimation of backreaction time scales. We hope this work also serves to demonstrate the application of the uniformly smoothed Stokes-line method applied to cosmology.

This paper is organized as follows. In Sect.~\ref{ChemPotentialGeneralities}, we first discuss the generalities of chemical potential and define the model we work with. Then in Sect.~\ref{StokesLineMethodReview}, we re-derive the uniformly smoothed Stokes phenomenon for both bosonic and fermionic systems and justify its validity for the case with significant particle production. In Sect.~\ref{TheLongTechinicalSection}, we move on to work out the production details for spin-1,1/2 particles in five common FRW spacetimes. At last, we summarize and give outlooks in Sect.~\ref{SummaryOutlook}. For readers who wish to skip the detailed analysis to directly look up the results, we assemble our formulae into a checklist in Appendix~\ref{ChecklistAppendix}.

\section{Chemical potential: Generalities and dS solutions}\label{ChemPotentialGeneralities}
In this section, loosely following \cite{Wang:2019gbi}, we give some general discussions on chemical potential in cosmology. In thermodynamics, chemical potential is originally introduced by Gibbs to describe the change of internal energy of a system with respect to the change of particle numbers when the entropy and volume are held fixed. More formally in statistical mechanics, it is identified with the Lagrange multiplier $\kappa$ of total particle number when the systems in a grand canonical ensemble are allowed to exchange particles with each other. Starting from the partition function for a grand canonical ensemble, $Z=e^{-(H-\kappa N)/T}$, we can straightforwardly generalize it to the field theory context as a path integral in the phase space of a field $\phi$,
\begin{equation}
	Z=\int\mathcal{D}\phi\mathcal{D}\pi e^{-i\int dt(H[\phi,\pi]-\kappa N[\phi,\pi])}~,
\end{equation}
where $H$ is the Hamiltonian and $N$ is the particle number operator associated with a certain symmetry (which may or may not be exact). Throughout this work, we will assume a spatially flat FRW spacetime background with $ds^2=-dt^2+a(t)^2d\mathbf{x}^2$. Going into the field configuration space and write the chemical potential term as the integral over a local density, we have
\begin{equation}
	Z=\int\mathcal{D}\phi e^{i\int dt d^3x\sqrt{|g|}(\mathcal{L}(\phi,\partial\phi)+\kappa J^0(\phi,\partial\phi))}~,
\end{equation}
where $N\equiv\int d^3x a^3 J^0$. Now we can turn on the spacetime dependence of $\kappa$ and interpret it as the zeroth component of a local vector field $\kappa_\mu(t,\mathbf x)$. Thus the general form of a chemical potential as a background field coupled to the matter field is
\begin{equation}
	\Delta\mathcal{L}_{\text{chem}}\equiv \kappa_\mu(x)J^\mu(x)~.
\end{equation}

Now if one inspects the effect of introducing such a chemical potential term into the matter Lagrangian, the result will depend on two aspects. First, in the absence of $\kappa_\mu$, if the matter current is derived from an exact symmetry, it is conserved as an operator identity, $i.e.$, $\langle \nabla_\mu J^\mu\rangle=0$, where $\nabla$ is a covariant derivative with respect to the metric $g$. Then one can always consistently gauge this symmetry by minimally coupling the conserved current to a vector potential. We are free to choose $\kappa_\mu(x)$ to be such a background gauge field. If the chemical potential $\kappa=\kappa_\mu dx^\mu$ is closed in the 1-form sense, $d_D \kappa=d\kappa+\kappa\wedge\kappa=0$, where $D$ is compatible with the gauge field connection $\kappa$, the background gauge field then has no field strength and is gauge-equivalent to vacuum. Thus we can perform a gauge transformation to eliminate $\kappa$ locally. Such is the case if $\kappa=\kappa(t)$ is Abelian and spatially homogeneous. However, we point out that the elimination of the chemical potential term is not completely trivial in the scalar case, as we will see below, since it shifts the scalar mass in a quadratic way. The second possibility is that if the matter current is not built from an exact symmetry and is hence not conserved. This suggests that there is no consistent way of coupling it to a background gauge field and thus interpreting it as the chemical potential. In summary, we give the following necessary condition of a chemical potential term that has physical effects other than quadratically shifting the mass of the matter particle,
\begin{equation}
	\text{either  }d_D \kappa=d\kappa+\kappa\wedge\kappa\neq 0,~~~\text{or  }\nabla\cdot J\neq 0~.\label{necessaryCondi}
\end{equation}
In the following discussion, we provide several examples that satisfy the above criterion and have interesting particle production features. We will start the general discussions in flat FRW spacetime and retreat to exact dS spacetime when solving the Equations of Motion (EoMs). The general FRW EoMs will be discussed in later sections.

\subsection{Spin-0}
Consider a complex scalar field $\sigma$, the only non-trivial chemical potential term we can find at quadratic level is with $J_\mu=i(\sigma^*\partial_\mu{\sigma}-\sigma\partial_\mu{\sigma}^*)$. If the original Lagrangian is $U(1)$-symmetric, this current is conserved. Therefore, according to (\ref{necessaryCondi}), we have to go for the first possibility\footnote{In fact, even if the $U(1)$ symmetry is explicitly broken with $\nabla\cdot J\neq 0$, it can be shown that there is only a quadratic mass-shift and no enhancement particle production is present \cite{Wang:2019gbi}. This is why (\ref{necessaryCondi}) is only a necessary condition, rather than being sufficient. However, we note that a different opinion is given in \cite{Bodas:2020yho}, where it is argued that a scalar chemical potential can also bring isotropic enhancement to particle production.}. We can write the following action in the FRW background,
\begin{equation}
	S_0=\int d^4 x\sqrt{-g}\left[-g^{\mu\nu}\partial_\mu\sigma^*\partial_\nu\sigma-m^2\sigma^*\sigma+\kappa_\mu J^\mu\right]~,
\end{equation}
where $g_{\mu\nu}\equiv a(\tau)^2\eta_{\mu\nu}$ using comoving coordinates. Absorbing this chemical potential term into the derivative term and shifting the mass term accordingly, we have
\begin{equation}
	S_0=\int d^4 x\sqrt{-g}\left[-g^{\mu\nu}(\partial_\mu-i\kappa_\mu)\sigma^*(\partial_\nu+i\kappa_\nu)\sigma-(m^2-\kappa_\mu \kappa^\mu)\sigma^*\sigma\right]~.
\end{equation}
As mentioned above, if $\kappa_\mu$=$\kappa_0(\tau)\delta_\mu^0$, the whole system can be considered as a charged scalar moving in the vacuum with a new mass $M^2(\tau)\equiv m^2-\frac{\kappa_0(\tau)^2}{a(\tau)^2}$. There is no enhancement of particle production unless $M^2(\tau)$ becomes negative or its time dependence violates adiabatic condition.

If the first criterion is satisfied, then there is a non-zero field strength $F_{\mu\nu}=\partial_\mu \kappa_\nu-\partial_\nu \kappa_\mu$. For instance, with $\kappa_0=0,\kappa_i=\kappa_i(\tau)$, there is a uniform (time-dependent) electric field $F_{i0}=-\kappa_i'(\tau)$, where prime denotes a derivative with respect to the conformal time $\tau$. The enhancement of particle production is exactly the Schwinger effect of this electric field \cite{Chua:2018dqh}. The EoM of $\sigma$ in momentum space reads
\begin{equation}
	(a\sigma_{\mathbf{k}})''+\left[k^2-2\mathbf{k}\cdot\bm{\kappa}a+m^2 a^2-\frac{a''}{a}\right](a \sigma_{\mathbf{k}})=0~,\label{spin0EoM}
\end{equation}
where $\mathbf{\kappa}\equiv \frac{\kappa_i}{a}\hat{e}_i$. Clearly the second term in the square bracket breaks rotational symmetry and stands for the effect of the background electric field. It introduces an angular dependence in the effective mass of different modes. Those with lighter effective mass tend to get produced more easily, especially for the mode traveling at the same direction as $\mathbf{\kappa}$. We shall see that this term linear in momentum is typical in the presence of chemical potentials. They represent the \textit{bias} on the effective mass of different modes introduced by the chemical potential. For large enough $|\bm\kappa|$, there is even a tachyonic instability and the pair creation rate becomes exponentially large. To be more quantitative, we set $|\bm{\kappa}|=\text{const}$ and limit ourselves to dS by taking $a=-\frac{1}{H\tau}$. Then the solution to the EoM is given by
\begin{equation}
	\sigma_{\mathbf{k}}=-\frac{e^{\pi\mathbf{\hat{k}}\cdot\bm{\tilde{\kappa}}/2}}{\sqrt{2k}H\tau}W_{-i\mathbf{\hat{k}}\cdot\bm{\tilde{\kappa}},i\mu}(2ik\tau)~,~~~\mu\equiv\sqrt{\tilde{m}^2-\frac{9}{4}}~,
\end{equation}
where $\mathbf{\hat {k}}=\frac{\mathbf{k}}{|\mathbf{k}|}$, $\bm{\tilde\kappa}\equiv\frac{\bm\kappa}{H}$, $\tilde{m}\equiv\frac{m}{H}$ and $W$ is the Whittaker function that matches the BD initial condition at $\tau\to-\infty$.
The late-time expansion reveals an angular-dependent production amount
\begin{equation}
	|\beta(\mathbf{k})|^2
	=\frac{e^{2 \pi  \left(\mu+ \mathbf{\hat{k}}\cdot\bm{\tilde{\kappa}}\right)}+1}{e^{4 \pi \mu}-1}~.
\end{equation}
To gain an intuitive understanding of the enhancement, we can go to the large mass limit $m\gg H$. Then the leading order particle number is 
\begin{equation}\label{dSScalarApproxProdAmount}
	|\beta(\mathbf{k})|^2\xrightarrow{\mu\gg 1}e^{-2\pi(\mu-\mathbf{\hat{k}}\cdot\bm{\tilde{\kappa}})}~.
\end{equation}
Therefore, the direct consequence of a chemical potential is a \textit{linear} and biased (rather than quadratic and un-biased) shift of the effective mass of the particle modes, making them easier or harder to produce\footnote{See an alternative understanding of chemical potential as a non-trivial modification of dispersion relation in \cite{Wang:2019gbi}.}.

\subsection{Spin-$\frac{1}{2}$}\label{fermionGeneralities}
In the massive spin-$\frac{1}{2}$ case, the second possibility of (\ref{necessaryCondi}) can be satisfied by choosing the axial current $J^\mu_5=e_{~a}^\mu\bar\Psi\gamma^a\gamma_5\Psi$. Hence a time-like chemical potential $\kappa_\mu(\tau)\propto\delta_\mu^0$ has no interpretation as a trivial background pure-gauge and can become physically relevant. To illustrate how it assists gravitational particle production, we and choose a Majorana fermion model \cite{Chen:2018xck} written in a Weyl basis $\Psi=\left(\begin{smallmatrix}
	\psi\\\psi^\dagger
\end{smallmatrix}\right)$. The Dirac fermion case can be obtained by combing two Majorana fermions with analogous behaviors. The action of a Majorana fermion with chemical potential reads
\begin{equation}
	S_{1/2}=\int d^4x\sqrt{-g}\left[i\psi^\dagger\bar{\sigma}^a e^\mu_{~a}\mathcal{D}_\mu\psi-\frac{1}{2}m(\psi\psi+\psi^\dagger\psi^\dagger)+\kappa_\mu e^\mu_{~a} \psi^\dagger\bar{\sigma}^a\psi\right]~,
\end{equation}
where $\mathcal{D}_\mu\equiv\partial_\mu-\frac{i}{4}\omega_{\mu ab}\sigma^{ab},\omega_{\mu ab}=e^\nu_{~b}\nabla_\mu e_{\nu}^{~c}\eta_{ac}$. This chemical potential term can arise from, for instance, a dimension-5 coupling to a rolling scalar,
\begin{equation}
	\Delta \mathcal{L}_{\text{chem}}=-\frac{\partial_\mu \phi J^\mu_5}{2\Lambda} =\kappa_\mu e^\mu_{~a}\psi^\dagger\bar{\sigma}^a\psi,~~\kappa_\mu\equiv\frac{\partial_\mu\phi}{\Lambda}~.
\end{equation}
After choosing a tetrad $e_\mu^{~a}=a(\tau)\delta_\mu^a, e^\mu_{~a}=a(\tau)^{-1}\delta^\mu_a$ and $\kappa_\mu=a(\tau)\kappa\delta_\mu^0$ with $\kappa=\frac{\dot\phi}{\Lambda}=\text{const}$, the action simplifies to
\begin{equation}
	S_{1/2}=\int d\tau d^3x\left[i\psi^\dagger\bar{\sigma}^a \delta^\mu_{a}\partial_\mu\psi-\frac{1}{2}am(\psi\psi+\psi^\dagger\psi^\dagger)+a\kappa\psi^\dagger\bar{\sigma}^0\psi\right]~,\label{fermionLagrangian}
\end{equation}
where for simplicity, we have rescaled $\psi\to a^{-3/2}\psi$. In momentum space, we can perform a standard decomposition into helicity eigenmodes,
\begin{equation}
	\psi(\tau,\mathbf{x})=\int\frac{d^3k}{(2\pi)^3}\sum_{s=\pm}\left[h^s(\mathbf{\hat{k}})u_s(\tau,k)e^{i\mathbf{k}\cdot\mathbf{x}}b_{\mathbf{k}}^s+i\sigma^2 h^{s*}(\mathbf{\hat{k}}) v_s(\tau,k)^* e^{-i\mathbf{k}\cdot\mathbf{x}}b_{\mathbf{k}}^{s\dagger}\right]~,
\end{equation}
where $\bm{\sigma}\cdot\mathbf{\hat{k}}h^s(\mathbf{\hat{k}})=s h^s(\mathbf{\hat{k}})$. The EoM reads
\begin{eqnarray}
	\nonumber i u_\pm'&=&(\mp k-a\kappa)u_\pm+a m v_\pm\\
	i v_\pm'&=&a m u_\pm+(\pm k+a\kappa)v_\pm~.
\end{eqnarray}
This set of equations is exactly solvable in dS. With a BD initial condition, the mode functions take the form
\begin{eqnarray}
	\nonumber&&u_+(\tau,k)=\frac{\tilde{m}e^{-\pi\tilde{\kappa}/2}}{\sqrt{-2k\tau}}W_{-\frac{1}{2}+i\tilde{\kappa},~i\sqrt{\tilde{m}^2+\tilde{\kappa}^2}}(2ik\tau),~~~~~~u_-(\tau,k)=\frac{e^{\pi \tilde{\kappa}/2}}{\sqrt{-2k\tau}}W_{\frac{1}{2}-i\tilde{\kappa},~i\sqrt{\tilde{m}^2+\tilde{\kappa}^2}}(2ik\tau)\\
	&&v_+(\tau,k)=\frac{e^{-\pi\tilde{\kappa}/2}}{\sqrt{-2k\tau}}W_{\frac{1}{2}+i\tilde{\kappa},~i\sqrt{\tilde{m}^2+\tilde{\kappa}^2}}(2ik\tau),~~~~~~~~~~v_-(\tau,k)=\frac{\tilde{m}e^{\pi \tilde{\kappa}/2}}{\sqrt{-2k\tau}}W_{-\frac{1}{2}-i\tilde{\kappa},~i\sqrt{\tilde{m}^2+\tilde{\kappa}^2}}(2ik\tau)~,
\end{eqnarray}
where $\tilde{m}\equiv\frac{m}{H}$ and $\tilde{\kappa}=\frac{\kappa}{H}$. Similar to the scalar case, performing an IR expansion at $\tau=0$ and matching the Bogoliubov coefficients, one can arrive at the production amount formula \cite{Adshead:2015kza}
\begin{equation}
	|\beta_\pm(k)|^2=
	\frac{e^{2 \pi  \left(\sqrt{\tilde{m}^2+\tilde{\kappa }^2}\mp \tilde{\kappa }\right)}-1}{e^{4 \pi  \sqrt{\tilde{m}^2+\tilde{\kappa}^2}}-1}~.
\end{equation}
When the mass is large and chemical potential is small, the leading order particle number again takes the form of a Boltzmann factor with linearly biased effective mass, 
\begin{equation}
	|\beta_\pm(k)|^2\xrightarrow{\tilde{m}\gg |\tilde{\kappa}|,~\tilde{m}\gg 1}e^{-2\pi(\tilde{m}\pm\tilde{\kappa})}~.
\end{equation}
For a positive chemical potential, the negative-helicity mode gets amplified whereas the positive-helicity mode is suppressed. However, when the chemical potential is larger than the mass scale, the enhancement in the negative-helicity particle production begins to saturate, 
\begin{equation}
	|\beta_-(k)|^2\xrightarrow{\tilde{\kappa}\gg\tilde{m}\gg 1} e^{-2\pi(\sqrt{\tilde{m}^2+\tilde{\kappa}^2}-\tilde{\kappa})}\approx e^{-\pi \tilde{m}^2/\tilde{\kappa}}=e^{-\frac{\pi m^2}{\kappa H}}~.\label{dSFermionSaturation}
\end{equation}
This is essentially the Pauli blocking phenomenon generic to all fermionic systems. The exclusion principle forbids any mode being occupied more than once.

Actually, fermion production with constant chemical potential in general FRW spacetimes can be understood in an elegant way. In terms of the physical time $t$, the EoM is essentially a two-state system evolving according to a Schrödinger equation
\begin{equation}
	i\frac{\partial}{\partial t}
	\left(\begin{array}{ccc}
		u_\pm\\
		v_\pm
	\end{array}\right)=\left(\begin{array}{ccc}
		\mp\frac{k}{a}-\kappa & m\\
		m & \pm\frac{k}{a}+\kappa
	\end{array}\right)\left(\begin{array}{ccc}
		u_\pm\\
		v_\pm
	\end{array}\right)~.\label{fermionEoM}
\end{equation}
Without loss of generality, let us consider the negative helicity mode with $s=-$ (the $s=+$ helicity mode is obtained by $k\to-k$). The unitary evolution governed by a Schrödinger equation (\ref{fermionEoM}) preserves the normalization $|u_-|^2+|v_-|^2=1$. The BD initial condition selects the positive frequency mode $u_-$ in the early time limit $t\rightarrow -\infty$. If $\kappa>0$, there can be a time when the physical wavelength $\frac{k}{a}$ of the mode is comparable to the chemical potential scale $\kappa$. In the language of quantum physics, we have an avoided crossing. Namely when the diagonal elements of the Hamiltonian vanish, its instantaneous eigenvalues approach each other, but a complete degeneracy is avoided due to the off-diagonal terms. If this process occurs adiabatically, according to the adiabatic theorem, the state smoothly maintains its positive-frequency trajectory and there is not much particle production. However, if $\kappa$ is large, the avoided crossing becomes a non-adiabatic one, with almost all positive-frequency crossed into negative-frequency part and thus nearly maximal particle production. Viewed in this way, Pauli exclusion principle is a built-in feature of fermions such that the evolution of $(u,v)$ is \textit{unitary}, as opposed to the \textit{symplectic} evolution of bosons $(\sigma,\dot{\sigma})$, which can be unbounded from above and exponentiating ($e.g.$, if $\kappa\gg m$ in (\ref{dSScalarApproxProdAmount})).

Interestingly, the large-chemical potential limit in the spin-1/2 fermion case enjoys a universal behavior in general FRW spacetimes. To see this more explicitly, we can assume $\kappa\gg m$ and expand around the crossing time $t_*$, where $\kappa=k/a(t_*)$,
\begin{equation}
	i\frac{\partial}{\partial t}
	\left(\begin{array}{ccc}
		u_-\\
		v_-
	\end{array}\right)\approx\left(\begin{array}{ccc}
		\kappa H(t_*) (t-t_*) & m\\
		m & -\kappa H(t_*) (t-t_*)
	\end{array}\right)\left(\begin{array}{ccc}
		u_-\\
		v_-
	\end{array}\right)~. \label{eq:LZ_approximation}
\end{equation}

\begin{figure}[h!]
	\centering
	\includegraphics[width=9cm]{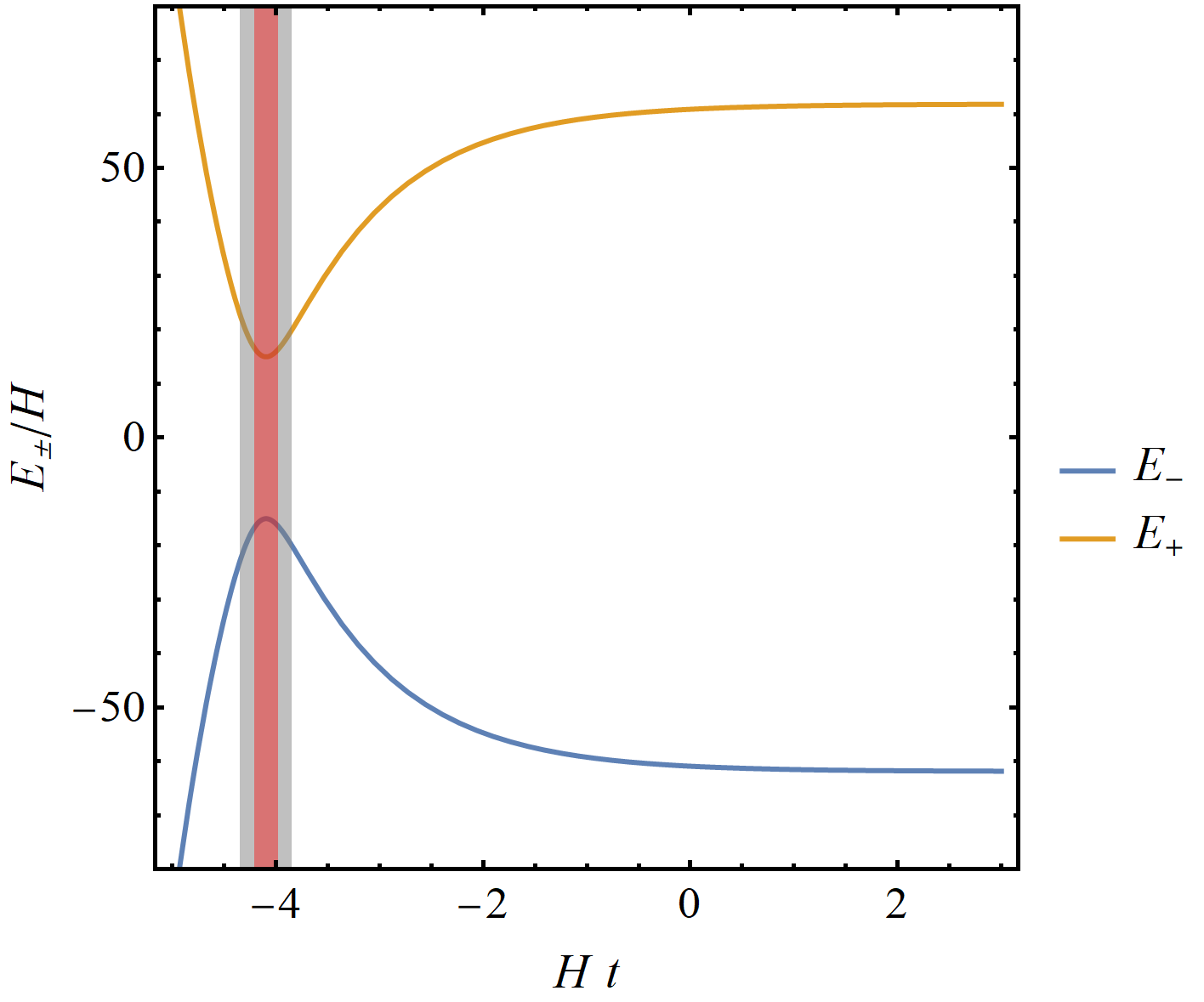}
	\caption{The instantaneous eigenvalues of the Hamiltonian in dS. Here we have chosen $m=15H$, $\kappa=60H$, or, after translating to LZ model parameters, $\eta=15H$, $\gamma=120H^2$. The gray band corresponds to the naive time scale $\frac{4\eta}{\gamma}=0.5H^{-1}$ while the red band corresponds to the actual production width $\Delta t_*=\sqrt{\frac{2\pi}{\gamma}}=0.23H^{-1}$. As long as $\kappa\gg m$, the light pink band is narrow enough so that the expansion of spacetime becomes irrelevant, and the instantaneous energy eigenvalues approach to that of a LZ model.}\label{largeChemLZ}
\end{figure}

This is none other than a Landau-Zener (LZ) model with $\eta=m$ and $\gamma=2\kappa H(t_*)$ \cite{10011873546,Zener:1932ws}. The corresponding LZ parameter describing adiabaticity is $z=\frac{\eta^2}{\gamma}=\frac{m^2}{2\kappa H(t_*)}$, where $z\gg 1$ corresponds to adiabatic transitions and $0<z\ll 1$ corresponds to diabatic transitions. See Fig.~\ref{largeChemLZ} for justification of approximation using an LZ transition in dS. Thus the crossing probability is given by the exponential factor 
\begin{equation}
	|\beta_{-}(k)|^2=e^{-2\pi z}=e^{-\frac{\pi m^2}{\kappa H(t_*(k))}}~,\label{fermionLZLargeChem}
\end{equation}
in agreement with the exact dS result (\ref{dSFermionSaturation}). Furthermore, we obtain the production time $t_*$ as the solution to the equation $\kappa=\frac{k}{a(t_*)}$. Naively, we would expect the production width to be the time scale at which the LZ resonance happen, $i.e.$, $\frac{4\eta}{\gamma}=\frac{2m}{\kappa H(t_*)}$. However, we shall see in Sect.~\ref{StokesLineMethodReviewFermionic} that this is not the case. The actual particle production width is $\Delta t_*=\sqrt{\frac{2\pi}{\gamma}}=\frac{1}{\sqrt{\kappa H(t_*)/\pi}}$, which is typically shorter than the LZ time scale \cite{Berry:1990histories}.

\subsection{Spin-1}
One can also easily find a non-conserved current in the Abelian massive vector boson case, namely the Chern-Simons current $J_{CS}^\mu=\mathcal{E}^{\mu\nu\rho\sigma}A_\nu F_{\rho\sigma}=\frac{1}{\sqrt{-g}}\epsilon^{\mu\nu\rho\sigma}A_\nu F_{\rho\sigma}$. The chemical potential term looks like
\begin{equation}
	\Delta\mathcal{L}_{\text{chem}}=\frac{1}{2}\int d^4x \sqrt{-g}\kappa_\mu J_{CS}^\mu=\frac{1}{2}\int d\tau d^3x \kappa_\mu\epsilon^{\mu\nu\rho\sigma}A_\nu F_{\rho\sigma}~.
\end{equation}
For a choice of $\kappa_\mu=\theta'(\tau)\delta_\mu^0\equiv a(\tau) \kappa \delta_\mu^0$, or $\kappa\equiv\dot{\theta}=\text{const}$, an integration-by-part gives a time-dependent $\theta$-term (a rolling axion),
\begin{equation}
	\Delta\mathcal{L}_{\text{chem}}=\frac{1}{2}\int d^4x \sqrt{-g}\kappa_\mu J_{CS}^\mu=-\frac{1}{4}\int d\tau d^3x\theta(\tau)\epsilon^{\mu\nu\rho\sigma}F_{\mu\nu} F_{\rho\sigma}~.
\end{equation}
Thus the system is described by an axion electrodynamics Lagrangian \cite{Wilczek:1987mv}
\begin{equation}
	S_1=\int d^4x\sqrt{-g}\left[-\frac{1}{4}F_{\mu\nu}F^{\mu\nu}+\frac{1}{2}m^2A_\mu A^\mu-\frac{1}{4}\theta(\tau)\mathcal{E}^{\mu\nu\rho\sigma}F_{\mu\nu} F_{\rho\sigma}\right]~.
\end{equation}
A massive vector boson has three degrees of freedom, two transverse modes and a longitudinal mode, from which the time-like component is solved using the constraint $\nabla_\mu A^\mu=0$. Decomposing the spatial components into helicity eigenstates, we have
\begin{equation}
	A_i(\tau,\mathbf{x})=\int\frac{d^3k}{(2\pi)^3}e^{i\mathbf{k}\cdot\mathbf{x}}\sum_{s=0,\pm}\epsilon_i^s(\mathbf{\hat{k}})\left[f_s(\tau,k)a_{\mathbf{k}}^s+f_s(\tau,k)^* a_{-\mathbf{k}}^{s\dagger}\right]~,
\end{equation}
The EoM then reads
\begin{equation}
	f_{\pm}''+\left(k^2\pm 2k \kappa a +m^2 a^2\right)f_{\pm}=0~,~~~f_{0}''+\left(k^2 +m^2 a^2\right)f_{0}=0\label{spin1EoM}
\end{equation}
As a result, the transverse modes are affected by the chemical potential while the longitudinal mode is not. Focusing on the transverse modes, we can solve the EoMs analogously as (\ref{spin0EoM}),
\begin{equation}
	f_\pm(\tau,k)=\frac{e^{\mp\pi\tilde{\kappa}/2}}{\sqrt{2k}}W_{-i\tilde{\kappa},i\mu}(2ik\tau)~,~~~\mu\equiv\sqrt{\tilde{m}^2-\frac{1}{4}}~.
\end{equation}
The transverse particle production amount is then
\begin{equation}
	|\beta_\pm(k)|^2=\frac{e^{2 \pi  \left(\mu\mp \tilde{\kappa }\right)}+1}{e^{4 \pi \mu}-1}~.\label{spin0ProdAmountExact}
\end{equation}
In the large mass limit, the effect of chemical potential again simplifies to a linear bias over the effective mass,
\begin{equation}
	|\beta_{\pm}(k)|^2\xrightarrow{\mu\gg 1}e^{-2\pi(\mu\pm\tilde{\kappa})}~.\label{spin0ProdAmountExactLargeMass}
\end{equation}
Here, without the protection of the exclusion principle, the production amount starts to become exponentially large if $|\tilde\kappa|>\mu$. This is commonly recognized as a tachyonic instability in the study of axions. In this parameter regime, the backreaction to the rolling $\theta(\tau)$ must be taken into account.

\section{The uniformly smoothed Stokes-line method}\label{StokesLineMethodReview}
In this section, we derive the uniformly smoothed Stokes-line method for both spin-1 vector bosons and spin-1/2 Majorana fermions, providing a framework to analyze the histories of particle production with chemical potentials. Mathematically, the evolution of spin-1 vector bosons (\ref{spin1EoM}) and spin-1/2 Majorana fermions (\ref{fermionEoM}) behave as second order differential equations and the Schr{\"o}dinger equation with two quantum states respectively, and such systems can experience the emergence of negative-frequency part starting from an initial positive-frequency solution, known as the Stokes phenomenon:
\begin{align}
	\Psi^+(\tau)\to \alpha(\tau)\Psi^+(\tau)+\beta(\tau)\Psi^-(\tau) \ , \label{eq:illustrate_solution}
\end{align} 
where $\Psi^{\pm} (\tau)$ are the positive/negative-frequency parts of either bosons or fermions, and the two time-dependent functions $\alpha(\tau)$ and $\beta(\tau)$ can be regarded as the Bogoliubov coefficients, associated with the particle production. The Stokes phenomena in second order differential equations \cite{Berry:1990,Dabrowski:2014ica,Dabrowski:2016tsx,Li:2019ves,Winitzki:2005rw,dingle:1973asymptotic} and transitions between two quantum states \cite{Berry:1990histories,Berry_1993} have been studied in many works with the assumption that the magnitude of the emergent part $|\beta(\tau)|$ is exponentially small, and all of these works point out that the singulant 
\begin{align}
	F(\tau)=-2i\int^\tau_{\tau_c} \omega(\tau_1)d\tau_1 \ ,
\end{align} defined as the difference between the positive and negative phases accumulated from the complex turning point $\tau_c$ satisfying $\omega(\tau_c)=0$, are important to describe the details of the Stokes phenomena. To be specific, when the systems evolve near the Stokes line, defined as the line linking $\tau_c$ and $\tau_c^*$ with ${\rm Im}F(\tau)=0$, the negative-frequency part starts to produce, and the production histories including the amounts and widths can be calculated from the singulant $F(\tau)$, proved with the technique of optimally truncating the asymptotic series solution of (\ref{eq:illustrate_solution}).

In the following subsections, we first follow previous studies to apply the optimally truncated asymptotic series solution to calculate the particle production for vector bosons and fermions respectively, and this framework works properly when the particle production is exponentially small. We then analyze the situation when the exponential particle production is significantly enhanced by the chemical potential, so that the optimal truncation technique is not applicable, and the Borel summation technique should be applied to obtain the particle production.

\subsection{Bosonic case}\label{StokesLineMethodReviewBosonic}
Consider the mode expansion for the transverse component of a massive vector boson,
\begin{equation}
	A_i^\bot=\int\frac{d^3k}{(2\pi)^3}e^{i\mathbf{k}\cdot\mathbf{x}}\sum_{s=\pm}\epsilon_i^s(\mathbf{\hat{k}})\left[f_s(\tau,k)a_{\mathbf{k}}^s+f_s(\tau,k)^* a_{-\mathbf{k}}^{s\dagger}\right]~,
\end{equation}
The mode function satisfies a second order EoM of the general form
\begin{align}
	\frac{d^2}{dz^2}f(z)+\lambda^2 w^2(z)f(z)=0\ , ~~w^2=1+\frac{2s\kappa a}{k}+\frac{m^2a^2}{k^2}~, \label{eq:EoM_Boson}
\end{align}
where we use the dimensionless variable\footnote{In dS, it is more convenient to choose $z=-k\tau$. Then the form of the singulant integral will differ by a sign and the $z_c$ will be on the upper half complex plane. These convention differences do not change the physical results and one can choose the convention that best-suit the problem. We follow the guideline that $\Im F$ decreases as conformal time increases, and that the singulant integral always starts with the turning point on the lower half conformal time plane.} $z=k\tau$, and we denote $'$ as the derivative with respect to $z$ starting from here. The helicity label $s$ is also omitted for simplicity. The asymptotic parameter $\lambda$, defined for later analysis of asymptotic series, is supposed to be large, and we can choose $\lambda=m$ if the mass is the largest parameter in the problem. (\ref{eq:EoM_Boson}) has the same structure as the one-dimensional time-independent Schr{\"o}dinger equation with a barrier, and therefore we adopt the method of analyzing waves near Stokes lines \cite{Berry:1990} to study the particle production. We can express the solution of (\ref{eq:EoM_Boson}) with the WKB form
\begin{align}
	f(z)=\frac{C}{\sqrt{2W(z)}}e^{-i\lambda\int^z_{z_c}W(z_1)dz_1} \ , \label{eq:WKB_form}
\end{align}
where $C$ is a constant fixed by the initial and normalization conditions, $z_c$ is the complex turning point defined by $w(z_c)=0$ and located at the lower-half complex plane, and the function $W(z)$ satisfies
\begin{align}
	W^2(z)=w^2(z)-\frac{1}{\lambda^2}\left[\frac{W''(z)}{2W(z)}-\frac{3}{4}\left(\frac{W'(z)}{W(z)}\right)^2\right] \ . \label{eq:exact_W}
\end{align}
In general, we cannot obtain the exact solution of $W(z)$, but we can apply the iterative adiabatic expansion to approximate it  \cite{Dabrowski:2014ica,Dabrowski:2016tsx}
\begin{align}
	W^{(n+1)}(z)&=\sqrt{w^2(z)-\frac{1}{\lambda^2}\left[\frac{{W''}^{(n)}(z)}{2W^{(n)}(z)}-\frac{3}{4}\left(\frac{{W'}^{(n)}(z)}{W^{(n)}(z)}\right)^2\right]} \nonumber  \\
	&=\sqrt{w^2(z)-\frac{\sqrt{W^{(n)}(z)}}{\lambda^2}\frac{d^2}{dz^2}\left(\frac{1}{\sqrt{W^{(n)}(z)}}\right)} \ , \label{eq:iterative_W}
\end{align}
with $W^{(0)}(z)=w(z)$. Such an iterative relation can be used to derive the asymptotic series solution of $W(z)$
\begin{align}
	W(z)=w(z)\sum_{n=0}^\infty \frac{\varphi_{2n}(z)}{\lambda^{2n}} \ , \label{eq:asymptotic_W}
\end{align}
and the series truncation at the $\mathcal{O}(\lambda^{-2n})$ is $W^{(n)}$ defined in (\ref{eq:iterative_W}). Applying (\ref{eq:asymptotic_W}) can express the oscillating phase integral in (\ref{eq:WKB_form}) with the asymptotic series, and we will see later in Sect.~\ref{dSAnalysis} that this is important for obtaining the $1/4$ correction to the vector boson's dS effective mass. On the other hand, it is well-known that the terms in the asymptotic series (\ref{eq:asymptotic_W}) keep increasing when $n$ is sufficiently large \cite{Berry:1990,Dabrowski:2014ica,Dabrowski:2016tsx,Winitzki:2005rw}
\begin{align}
	\frac{\varphi_{2n}}{\lambda^{2n}}\approx-\frac{(2 n-1)!}{\pi F^{2n}} \ ,
\end{align}
where
\begin{align}
	F(z)=-2i\lambda\int^z_{z_c}w(z_1)dz_1 \ , \label{eq:singulant_F}
\end{align}
is Dingle's singulant variable \cite{dingle:1973asymptotic}. So we can truncate the series sum at a suitable order $n$ to approximate the solution (\ref{eq:WKB_form}).

The asymptotic series solution with an optimal truncation order $n$, $f^{(n)}(z)=e^{-i\lambda\int^z_{z_c}W^{(n)}(z_1)z_1}/\sqrt{2W^{(n)}(z)}$, cannot fully represent the WKB solution (\ref{eq:WKB_form}), but we can choose to expand the exact solution with the super-adiabatic basis formed by $f^{(n)}$ and ${f^*}^{(n)}$,
\begin{align}
	f(z)&=\alpha(z)\frac{e^{-i\lambda\int_{z_i}^{z_c}W^{(n)}(z_1)dz_1}}{\sqrt{\lambda}}f^{(n)}(z)+\beta(z)\frac{e^{i\lambda\int_{z_i}^{z_c}W^{(n)}(z_1)dz_1}}{\sqrt{\lambda}}{f^*}^{(n)}(z) \nonumber \\
	\nonumber&=\frac{\alpha(z)e^{-i\lambda\int^{z}_{z_i}W^{(n)}(z_1)dz_1}+\beta(z)e^{i\lambda\int^{z}_{z_i}W^{(n)}(z_1)dz_1}}{\sqrt{2\lambda W^{(n)}(z)}}\\
	&\equiv\alpha(z)g(z)+\beta(z)g^*(z) \ , \label{eq:superadiabatic_ansatz}
\end{align}
where $z_i$ in the value of $z$ at initial time and $g(z)$ can be viewed as the instantaneous positive-frequency solution. Now the vector field can be expanded in an alternative form using $g(z)$:
\begin{equation}
	A_i^\bot=\int\frac{d^3k}{(2\pi)^3}e^{i\mathbf{k}\cdot\mathbf{x}}\sum_{s=\pm}\epsilon_i^s(\mathbf{\hat{k}})\left[g_s(\tau,k)b_{\mathbf{k}}^s(\tau)+g_s(\tau,k)^* b_{-\mathbf{k}}^{s\dagger}(\tau)\right]~,
\end{equation}
where the new annihilation operator acquires a time dependence through the Bogoliubov transformation
\begin{equation}
	b_{\mathbf{k}}^s(z)\equiv \alpha_s(z) a_{\mathbf{k}}^s+\beta_s(z)^* a_{\mathbf{k}}^{s\dagger}~.
\end{equation}
And the original vacuum annihilated by $a_{\mathbf{k}}^s$ now contains a spectrum of particles,
\begin{equation}
	\frac{\langle n^s_{\mathbf{k}}(\tau) \rangle}{V}=\frac{\langle b^{s\dagger}_{\mathbf{k}}b^s_{\mathbf{k}}(\tau) \rangle}{V}=|\beta_s(z)|^2~.
\end{equation}

Thus our aim is to solve the time dependence of the Bogoliubov coefficients $\alpha(z)$, $\beta(z)$. The solution satisfies the initial and normalization (Wronskian) conditions  as
\begin{align}
	f(z_i)\to\frac{e^{-i\lambda\int^{z}_{z_i}W^{(n)}(z_1)dz_1}}{\sqrt{2\lambda W^{(n)}(z)}} \ , \
	f{f^*}'-f^*f'=i \ .
\end{align} As pointed out in \cite{Dabrowski:2014ica,Dabrowski:2016tsx}, the constant Wronskian implies a degree of freedom of defining the derivative of $f$:
\begin{align}
	f'(z)=\left(-i\lambda W^{(n)}(z)+V(z)\right)\alpha(z)\frac{e^{-i\lambda\int^{z}_{z_i}W^{(n)}(z_1)dz_1}}{\sqrt{2\lambda W^{(n)}(z)}}+\left(i\lambda W^{(n)}(z)+V(z)\right)\beta(z)\frac{e^{i\lambda\int^{z}_{z_i}W^{(n)}(z_1)dz_1}}{\sqrt{2\lambda W^{(n)}(z)}} \ , \label{eq:fprime_form}
\end{align}
where $V(z)$ is an arbitrary real function. Choosing the time-dependent function $V(z)$ decides the evolution of the Bogoliubov coefficients
\begin{align}
	\frac{d}{dz}\begin{pmatrix}
		\alpha(z)\\
		\beta(z)
	\end{pmatrix}=\delta(z)\begin{pmatrix}
		1 & \left( \frac{\Delta(z)}{\delta(z)}+1\right)e^{2i\lambda\int^z_{z_i} W^{(n)}(z_1)dz_1} \\
		\left( \frac{\Delta(z)}{\delta(z)}-1\right)e^{-2i\lambda\int^z_{z_i} W^{(n)}(z_1)dz_1} & -1\\
	\end{pmatrix}\begin{pmatrix}
		\alpha(z)\\
		\beta(z)
	\end{pmatrix} \ ,
\end{align}
where
\begin{align}
	\delta(z)=\frac{\lambda}{2iW^{(n)}(z)}\left[w^2(z)-(W^{(n)}(z))^2+\frac{1}{\lambda^2}\left(V'(z)+V^2(z)\right)\right] \ , \ \Delta(z)=\frac{W'^{(n)}(z)}{2W^{(n)}(z)}+V(z) \ ,
\end{align}
and with the initial condition
\begin{align}
	\alpha(z_i)=\alpha_i \ , \ \beta(z_i)=\beta_i \ .
\end{align} Therefore, the appropriate choice of $V(z)$ should minimize the change of the Bogoliubov coefficients as we intends to minimize the difference between the basis function $f^{(n)}$ and the exact solution, and such a choice is $V(z)=-\frac{W'^{(n)}(z)}{2W^{(n)}(z)}$, as suggested in \cite{Berry:1990,Dabrowski:2016tsx} with different reasons. With this choice, the evolution of the Bogoliubov coefficients satisfies
\begin{align}
	\frac{d}{dz}\begin{pmatrix}
		\alpha(z)\\
		\beta(z)
	\end{pmatrix}&=\lambda\frac{(W^{(n+1)}(z))^2-(W^{(n)}(z))^2}{2iW^{(n)}(z)}\begin{pmatrix}
		1 & e^{2i\lambda\int^z_{z_i} W^{(n)}(z_1)dz_1} \\
		-e^{-2i\lambda\int^z_{z_i} W^{(n)}(z_1)dz_1} & -1\\
	\end{pmatrix}\begin{pmatrix}
		\alpha(z)\\
		\beta(z)
	\end{pmatrix}  \nonumber \\
	&=\delta(z)\begin{pmatrix}
		1 & e^{2i\lambda\int^z_{z_i} W^{(n)}(z_1)dz_1} \\
		-e^{-2i\lambda\int^z_{z_i} W^{(n)}(z_1)dz_1} & -1\\
	\end{pmatrix}\begin{pmatrix}
		\alpha(z)\\
		\beta(z)
	\end{pmatrix} \ , \label{eq:Boson_alpha_beta}
\end{align}
and the diagonal term can be removed by defining variables
\begin{align}
	\alpha(z)=e^{\int^z_{z_i}\delta(z_1)dz_1}S_+(z) \ , \ \beta(z)=e^{-\int^z_{z_i}\delta(z_1)dz_1}e^{-2i\lambda\int^{z_c}_{z_i}W^{(n)}(z_1)dz_1}S_-(z) \ , \label{eq:S_define}
\end{align}
implying that $S_\pm$ are the Stokes multipliers for the positive and negative modes respectively, so the evolution equation is simplified as
\begin{align}
	\frac{dS_{\pm}}{dF}&=\pm\frac{i\delta(z)}{2\lambda w}\exp{\left[\pm2\left(\int^z_{z_c} i\lambda W^{(n)}(z_1)dz_1-\int_{z_i}^z\delta(z_1)dz_1 \right)\right]}S_{\mp}  \nonumber \\
	&=\pm \frac{e^{\mp 2\int^z_{z_i}\delta(z_1)dz_1}}{4w\lambda^2}\frac{e^{\pm i\lambda\int^z_{z_c}W^{(n)}(z_1)dz_1}}{\sqrt{W^{(n)}}} \left[\left(\frac{e^{\pm i\lambda\int^z_{z_c}W^{(n)}(z_1)dz_1}}{\sqrt{W^{(n)}}}\right)''+\lambda^2 w^2 \frac{e^{\pm i\lambda\int^z_{z_c}W^{(n)}(z_1)dz_1}}{\sqrt{W^{(n)}}}\right]S_\mp
	\ . \label{eq:S_evolution}
\end{align} The term in the square bracket can be interpreted as the $\mathcal{O}(\lambda^{-2n-1})$ error when we approximate the EoM (\ref{eq:EoM_Boson}) with the $2n$-th order partial sum of the asymptotic series, and this can be calculated with the asymptotic series of the positive-frequency part $f_+$ 
\begin{align}
	f_+(z)=C\frac{e^{-i\lambda \int^z_{z_c}w(z_1)dz_1}}{\sqrt{2w(z)}}\sum_{n=0}^{\infty}\frac{b_n(z)}{\lambda^n} \ . \label{eq:asymptotic_f}
\end{align} Substituting this series solution into the EoM (\ref{eq:EoM_Boson}) implies
\begin{align}
	b'_{n+1}(z)=-\frac{i}{2w(z)}b''_n(z)+\frac{iw'(z)}{2w^2(z)}b'_n(z)+i\left(\frac{w''(z)}{4w^2(z)}-\frac{3w'(z)^2}{8w^3(z)}\right)b_n(z) \ . \label{eq:bn_recurrence}
\end{align} In all the scenarios that we study in Sect. \ref{TheLongTechinicalSection}, $w^2(z)$ has a simple root at $z_c$, so we approximate $w(z)$ as
\begin{align}
	w(z)= A(z-z_c)^{\frac{1}{2}}+\mathcal{O}(|z-z_c|^{\frac{3}{2}}) \ , \label{eq:w_expand}
\end{align} and (\ref{eq:bn_recurrence}) is reduced to 
\begin{align}
	b'_{n+1}(z)\approx-\frac{i}{2A(z-z_c)^{\frac{1}{2}}}b''_n(z)+\frac{i\gamma}{4A(z-z_c)^{\frac{3}{2}}}b'_n(z)-\frac{5i}{32A(z-z_c)^{\frac{5}{2}}}b_n(z) \ ,
\end{align}
and the solution of this recurrence relation with $b_0=1$ is
\begin{align}
	\frac{b_n}{\lambda^n}&\approx \frac{2^{-2 n-1} 3^n \left(\frac{i}{A}\right)^n \Gamma \left(n+\frac{1}{6}\right) \Gamma
		\left(n+\frac{5}{6}\right)}{\pi  \Gamma (n+1) (z-z_c)^\frac{3n}{2}} \nonumber \\
	&\approx \frac{\Gamma \left(n+\frac{1}{6}\right) \Gamma \left(n+\frac{5}{6}\right)}{2 \pi  n! F^n}\ . \label{eq:bn_solution}
\end{align}
Applying the asymptotic series of $f(z)$ (\ref{eq:asymptotic_f}), The term in the square bracket of (\ref{eq:S_evolution}) can be calculated explicitly
\begin{align}
	&\left[\left(\frac{e^{- i\lambda\int^z_{z_c}W^{(n)}(z_1)dz_1}}{\sqrt{W^{(n)}}}\right)''+\lambda^2 w^2 \frac{e^{- i\lambda\int^z_{z_c}W^{(n)}(z_1)dz_1}}{\sqrt{W^{(n)}}}\right]\left(\frac{e^{- i\lambda\int^z_{z_c}W^{(n)}(z_1)dz_1}}{\sqrt{W^{(n)}}}\right)^{-1} \nonumber \\
	&=\left[\left(\frac{e^{- i \lambda\int^z_{z_c}w(z_1)dz_1}}{\sqrt{w}}\sum_{j=0}^{2n} \frac{b_j}{\lambda^j}\right)''+\lambda^2w^2\left(\frac{e^{- i \lambda\int^z_{z_c}w(z_1)dz_1}}{\sqrt{w}}\sum_{j=0}^{2n} \frac{b_j}{\lambda^j}\right) \right]\left(\frac{e^{- i \lambda\int^z_{z_c}w(z_1)dz_1}}{\sqrt{w}}\right)^{-1} \ ,
\end{align}
implying that
\begin{align}
	\left(\frac{e^{- i\lambda\int^z_{z_c}W^{(n)}(z_1)dz_1}}{\sqrt{W^{(n)}}}\right)''+\lambda^2 w^2 \frac{e^{- i\lambda\int^z_{z_c}W^{(n)}(z_1)dz_1}}{\sqrt{W^{(n)}}}&=\frac{e^{- i \lambda\int^z_{z_c}W^{(n)}(z_1)dz_1}}{\lambda^{2n}\sqrt{W^{(n)}}}\left[b_{2n}''-\frac{w'b'_{2n}}{w}+\left(\frac{3w'^2}{4w^2}-\frac{w''}{2w}\right)b_{2n}\right] \nonumber \\
	&= 2iw\frac{e^{- i \lambda\int^z_{z_c}W^{(n)}(z_1)dz_1}}{\lambda^{2n}\sqrt{W^{(n)}}}b'_{2n+1} \ ,
\end{align}
where (\ref{eq:bn_recurrence}) is applied to obtain the last line. Therefore, the evolution of $S_\pm$ (\ref{eq:S_evolution}) is reduced to
\begin{align}
	\frac{d}{dF}\begin{pmatrix}
		S_+ \\
		S_-
	\end{pmatrix}&= \begin{pmatrix}
		0 & \left[\frac{d }{dF}\left(\frac{b_{2n+1}}{\lambda^{2n+1}}\right)\right]^*\frac{w }{W^{(n)}}e^{ 2 i\lambda\int^z_{z_c}W^{(n)}(z_1)dz_1} \\
		-\frac{d }{dF}\left(\frac{b_{2n+1}}{\lambda^{2n+1}}\right)\frac{w }{W^{(n)}}e^{- 2 i\lambda\int^z_{z_c}W^{(n)}(z_1)dz_1} & 0
	\end{pmatrix}\begin{pmatrix}
		S_+ \\
		S_-
	\end{pmatrix}+\mathcal{O}(\delta^2(z))  \nonumber \\
	&= \begin{pmatrix}
		0 & \left[\frac{d }{dF}\left(\frac{b_{2n+1}}{\lambda^{2n+1}}\right) \right]^*e^{-F} \\
		-\frac{d }{dF}\left(\frac{b_{2n+1}}{\lambda^{2n+1}}\right) e^F & 0
	\end{pmatrix}\begin{pmatrix}
		S_+ \\
		S_-
	\end{pmatrix}+\mathcal{O}\left(\frac{1}{\lambda^{2n+2}}\right) \ , \label{eq:simplified_S}
\end{align}
where we keep only the term with $\mathcal{O}(\lambda^{-2n-1})$. It is clear that $|dS_+/dF|\ll|dS_-/dF|$ as there is an exponential suppression $e^{-F}$ for the former, so we can solve (\ref{eq:simplified_S}) perturbatively starting from the initial values $S^i_\pm=S^{(0)}_\pm$, determined by $\alpha_i$ and $\beta_i$ through (\ref{eq:S_define}), and the leading-order change $S_-$ is an integral along the straight line with constant positive ${\rm Re}F$ in the complex $F$ plane:
\begin{align}
	S^{(0)}_-+S^{(1)}_-(F)&=S^i_--S^i_+\int^F_{{\rm Re}F+i\infty}\frac{d }{dF}\left(\frac{b_{2n+1}}{\lambda^{2n+1}}\right) e^F dF  \nonumber \\
	&=S^i_-+S^i_+\frac{R_n}{2\pi}\int^F_{{\rm Re}F+i\infty}\frac{(2n+1)!}{F^{2n+2}}e^F dF \nonumber \\
	&=S^i_-+S^i_+\frac{R_n(2n+1)!}{2\pi}\tilde{\Gamma}(-1-2n,-F) \ , \label{S0_integral}
\end{align}
where the value of prefactor 
\begin{align}
	R_n=\frac{\Gamma \left(2 n+\frac{7}{6}\right) \Gamma \left(2 n+\frac{11}{6}\right)}{(2 n+1)! \Gamma
		(2 n+1)} \ ,
\end{align} and the function $\tilde{\Gamma}(-1-2n,-F)$ is the continuous version of the incomplete Gamma function, defined as
\begin{align}
	\tilde{\Gamma}(-1-2n,-F)=\begin{cases}\Gamma(-1-2n,-F)  & {\rm Im}F\ge 0\\
		\Gamma(-1-2n,-F)+\lim_{c\to 0^+}[\Gamma(-1-2n,-{\rm Re}F-ic)-\Gamma(-1-2n,-{\rm Re}F+ic)] & {\rm Im}F<0 \ .
	\end{cases}
\end{align}
The prefactor $R_n\to 1$ when $n\gg 0$, implying that $\lim_{F\to {\rm Re}F-i\infty}S_-^{(0)}+S_-^{(1)}(F)=S^i_- -iS^i_+$. The incomplete Gamma function of (\ref{S0_integral}) can oscillate dramatically for general $n$, and we can choose an optimal truncation order $n$ such that the phase is stationary at $F={\rm Re}F$,  the moment when the integral receives dominant contribution
\begin{align}
	\left(\frac{d }{dF}\frac{b_{2n+1}}{\lambda^{2n+1}}\right)^{-1}\frac{d^2 }{dF^2}\left(\frac{b_{2n+1}}{\lambda^{2n+1}}\right)\Big|_{F={\rm Re}F}+1&=0 \ , \label{eq:stationary_phase}
\end{align}
and the solution is $n={\rm Int}\left(\frac{{\rm Re}F}{2}\right)-1$. It is noteworthy that the optimal truncation order cannot be determined by the stationary-phase condition when ${\rm Re}F<2$, and we will consider such a situation in the later part of this subsection. For simplifying the following perturbative calculation, we set the initial condition as $(\alpha_i,\beta_i)\approx(S^i_+,S^i_-)=(S^{(0)}_+,S^{(0)}_-)=(1,0)$, and the cases with general initial conditions can be obtained easily based on (\ref{S0_integral}).  Assuming that the optimal truncation is applicable with ${\rm Re}F \ge 2$, we can thus approximate the integrand as a Gaussian function around the point with stationary phase, and thus the Stokes multiplier reduces to an error function.
\begin{align}
	S^{(1)}_{-}(F)\approx -\frac{i R_n}{2}\left[1+{\rm Erf}\left(-\frac{{\rm Im}F}{\sqrt{2 {\rm Re} F}}\right)\right] \ . \label{eq:s-_Erf}
\end{align} 
We are ready to calculate the first-order term of $S_+$ by solving 
\begin{align}
	\frac{dS^{(1)}_+}{dF}&=\left[\frac{d }{dF}\left(\frac{b_{2n+1}}{\lambda^{2n+1}}\right)\right]^*\frac{w }{W^{(n)}}e^{ 2 i\lambda\int^z_{z_c}W^{(n)}(z_1)dz_1}S^{(1)}_- \nonumber \\
	&=e^{4i\lambda\int^{z_*}_{z_c}W^{(n)}(z_1)dz_1}\left[\frac{d }{dF}\left(\frac{b_{2n+1}}{\lambda^{2n+1}}\right)\frac{w }{W^{(n)}}e^{ -2 i\lambda\int^z_{z_c}W^{(n)}(z_1)dz_1}\right]^*S^{(1)}_-  \nonumber \\
	&=e^{4i\lambda\int^{z_*}_{z_c}W^{(n)}(z_1)dz_1}\frac{d{S_-^{(1)}}^*}{dF}S_-^{(1)} \ ,
\end{align}
where $z_*$ is the intersection between the Stokes line and the real $z$ axis, and the phase integral along the Stokes line $e^{4i\lambda\int^{z_*}_{z_c}W^{(n)}(z_1)dz_1}$ is real. Such a relation between $S^{(1)}_+$ and $S_-^{(1)}$ implies a much simpler form of the magnitude of $S^{(0)}_++S^{(1)}_+$
\begin{align}
	\left|S^{(0)}_++S^{(1)}_+\right|^2&\approx 1+S^{(1)}_++{S^{(1)}_+}^*+O\left(e^{8i\lambda\int^{z_*}_{z_c}W^{(n)}(z_1)dz_1}\right) \nonumber \\
	&=1+e^{4i\lambda\int^{z_*}_{z_c}W^{(n)}(z_1)dz_1}\int^{F}_{{\rm Re}F+i\infty}\left(\frac{d{S_-^{(1)}}^*}{dF}S_-^{(1)}+\frac{d{S_-^{(1)}}}{dF}{S_-^{(1)}}^*\right)dF+\mathcal{O}\left(e^{8i\lambda\int^{z_*}_{z_c}W^{(n)}(z_1)dz_1}\right) \nonumber \\
	&=1+e^{4i\lambda\int^{z_*}_{z_c}W^{(n)}(z_1)dz_1}\left|S_-^{(1)}(F)\right|^2+\mathcal{O}\left(e^{8i\lambda\int^{z_*}_{z_c}W^{(n)}(z_1)dz_1}\right) \ ,
\end{align}
and the definition of $S_{\pm}$ (\ref{eq:S_define}) implies that the normalization of the Bogoliubov coefficients preserves under the the first-order perturbation.

\begin{figure}[h!]
	\centering
	\includegraphics[width=9cm]{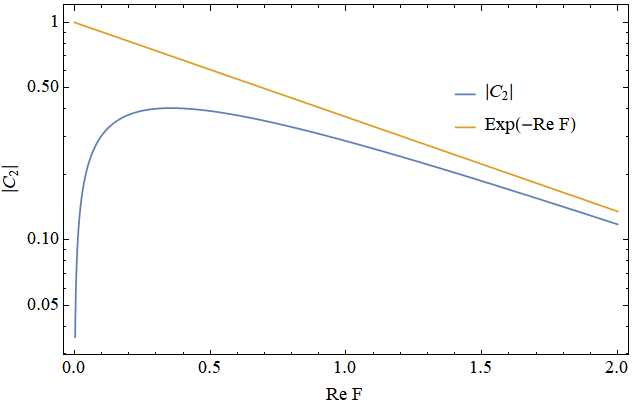}
	\caption{The comparison between $|C_2|$ and $e^{-{\rm Re}F}$ with $0<{\rm Re}F\le 2$, and the vertical axis is in logarithmic scale.}\label{fig:C2vsexp}
\end{figure}

For the situations with ${\rm Re}F<2$, we expect that higher-order perturbations are required to solve for (\ref{eq:simplified_S}), and the perturbation theory may break down when $e^{-{\rm Re}F}\to 1$, so we should analyze such situations carefully. In the cases with ${\rm Re}F<2$, we cannot choose an optimal truncation because of the failure of the stationary phase condition (\ref{eq:stationary_phase}) and the magnitudes of the terms in the asymptotic series solution (\ref{eq:asymptotic_f}) $b_n$ increase, starting from the first term. Such a divergent series defined by (\ref{eq:bn_solution}) behaves like the generalized hypergeometric function $_2 F_0(a,b;;F^{-1})$ which diverges everywhere from its original definition, but it can be defined meaningfully by applying the Borel summation:
\begin{align}
	B(F)&=\sum_{n=0}^{\infty}\frac{b_n}{\lambda^n} \nonumber \\
	&=\int^{+\infty}_0 e^{-t} \sum_{n=0}^\infty \frac{\Gamma \left(n+\frac{1}{6}\right) \Gamma \left(n+\frac{5}{6}\right)}{2 \pi  (n!)^2 F^n}t^n dt \nonumber \\
	&=\int^{+\infty}_0 e^{-t} \, _2F_1\left(\frac{1}{6},\frac{5}{6};1;\frac{t}{F}\right)dt \nonumber \\
	&=\frac{e^{-F/2} \sqrt{-F} K_{\frac{1}{3}}\left(-\frac{F}{2}\right)}{ \sqrt{\pi}} \ ,
\end{align}
where $K_a(z)$ is the modified Bessel function. The function $B(F)$ is discontinuous when it crosses the Stokes line with ${\rm Im}F=0$, implying that the exact solution after crossing the Stokes line should depend on different set of linear combination:
\begin{align}
	f(z)=\begin{cases} 
		\alpha_i\frac{e^{-i\lambda \int^z_{z_i}w(z_1)dz_1}}{\sqrt{2\lambda w(z)}} B(F(z))+\beta_i\frac{e^{i\lambda \int^z_{z_i}w(z_1)dz_1}}{\sqrt{2\lambda w(z)}} B^*(F(z))& {\rm Im}F>0 \\
		C_1\frac{e^{-i\lambda \int^z_{z_i}w(z_1)dz_1}}{\sqrt{2\lambda w(z)}} B(F(z))+C_2\frac{e^{i\lambda \int^z_{z_i}w(z_1)dz_1}}{\sqrt{2\lambda w(z)}} B^*(F(z)) & {\rm Im}F<0 
	\end{cases} \ ,
\end{align}
where the $C_1$ and $C_2$ are constants. After knowing the expression of $f(z)$, the Bogoliubov coefficients can be solved by combining (\ref{eq:superadiabatic_ansatz}) and (\ref{eq:fprime_form}), and the results can be fully recorded by the singulant $F$
\begin{align}
	\alpha(F)=\begin{cases}
		\alpha_i\left(B(F)+\frac{dB(F)}{dF}\right)-\beta_ie^{-F+{\rm Re}F}\left(\frac{dB(F)}{dF}\right)^* &  {\rm Im}F>0 \\
		C_1\left(B(F)+\frac{dB(F)}{dF}\right)-C_2e^{-F+{\rm Re}F}\left(\frac{dB(F)}{dF}\right)^* & {\rm Im}F<0 
	\end{cases}  \ ,
\end{align}
and 
\begin{align}
	\beta(F)=\begin{cases}
		-\alpha_ie^{F-{\rm Re}F} \frac{dB(F)}{dF}+\beta_i \left(B(F)+\frac{dB(F)}{dF}\right)^* & {\rm Im}F>0 \\
		-C_1e^{F-{\rm Re}F} \frac{dB(F)}{dF}+C_2 \left(B(F)+\frac{dB(F)}{dF}\right)^* & {\rm Im}F<0 
	\end{cases} \ , \label{eq:beta_resum}
\end{align}
and thus the constants $C_1$ and $C_2$ are chosen such that the Bogoliubov coefficients and their derivatives are continuous at ${\rm Im}F= 0$. To compare with the particle production in dS spacetime, we set the initial condition as $(\alpha_i,\beta_i)=(1,0)$. Since $B(F)\to 1$ when ${\rm Im}F\to\pm \infty$, we can know that $C_1$ and $C_2$ are the final values of $\alpha(F)$ and $\beta(F)$ respectively, and numerical checking confirms the normalization condition $|C_1|^2-|C_2|^2=1$. As shown in FIG. \ref{fig:C2vsexp}, $|C_2|$ agrees with the tendency of $e^{-{\rm Re}F}$ for the region with ${\rm Re}F\gtrsim 0.5$, but large deviations appear when ${\rm Re}F \to 0$. Such deviations may be partly attributed to the failure of the approximation of $w(z)$ (\ref{eq:w_expand}) when the two complex roots begin to merge when ${\rm Re}F\to 0$, whereas part of the deviations are expected. For example, the $|\beta(k)|$ in dS spacetime (\ref{spin0ProdAmountExact}) cannot be fully described as an exponential factor in some parameter ranges, and thus the behavior of large $|\beta(k)|$ depends on the details of scenarios. The universal property is that it approaches to the exponential form $e^{-{\rm Re}F}$ when $|\beta(k)|$ decreases, so we use the exponential form to describe the tendency of the production amount but not its exact value for the cases with small ${\rm Re}F$.

\begin{figure}[h!]
	\centering
	\includegraphics[width=9cm]{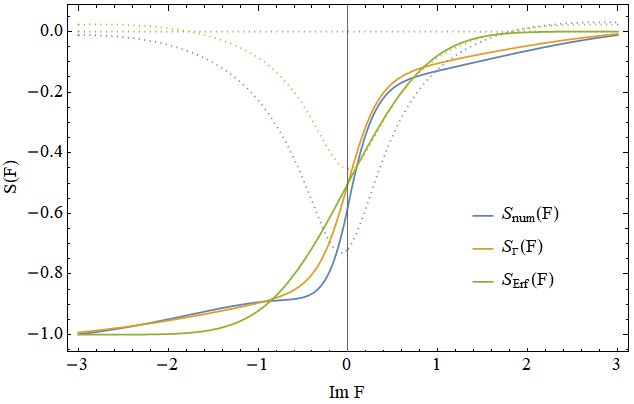}
	\caption{The comparison between $S_{\rm num}(F)$, $S_{\Gamma}(F)$ and $S_{\rm Erf}(F)$, where all of them are normalized such that the final value approaches $-i$ and ${\rm Re}F=0.5$. The solid lines represent the imaginary parts, whereas the dashed lines are the real parts.}\label{fig:Comparisions_Snum_SErf}
\end{figure}

On the other hand, we also compare the Stokes multiplier $S_{\rm num}(F)$ obtain from the numerical result (\ref{eq:beta_resum}) with the approximations utilizing the incomplete gamma function $S_{\Gamma}(F)$ with $n=0$ from the first-order perturbation (\ref{S0_integral}) and the error function $S_{\rm Erf}(F)$ (\ref{eq:s-_Erf}) respectively, as shown in FIG. \ref{fig:Comparisions_Snum_SErf}. It is clear that only the imaginary part of $\beta(k)$ remains non-zero after finishing the particle production, and different approximations have significant errors of describing the real part of $\beta(k)$ which vanishes rapidly after crossing the Stokes line, but the error function can still describe the width of the production process. 
After knowing how the production amount and width depends on the singulant $F(z)$, we can write down a simple form of $\beta(z)$ which includes the tendencies of particle production with vacuum initial condition when related parameters are changed:
\begin{empheq}[box=\widefbox]{align}
	\beta(z)\approx -\frac{ie^{-2i\int^{z_c}_{z_i}W^{(n)}(z_1)dz_1}}{2}\left[1+{\rm Erf}\left(-\frac{{\rm Im}F(z)}{\sqrt{2 {\rm Re} F}}\right)\right] \ , \label{eq:explicit_beta_boson}
\end{empheq}
where we replace $\lambda w \to w$ in (\ref{eq:EoM_Boson}) for simplicity since $\lambda$ and $w(z)$ always appear together in the final results, and the generalization to arbitrary initial conditions is straightforward based on the results (\ref{S0_integral}) and (\ref{eq:beta_resum}).

\subsection{Fermionic case}\label{StokesLineMethodReviewFermionic}
The fermionic case is logically similar to the bosonic case, but with important differences in the mathematical details. To be more specific, let us consider the action (\ref{fermionLagrangian}) for a Majorana fermion with chemical potential in an FRW background. We will use the Van der Waerden notation for two-component spinors and the conventions follow from \cite{Dreiner:2008tw}.

We begin by rewriting the mode expansion of $\psi_\alpha$ in the Van der Waerden notation,
\begin{equation}
	\psi_\alpha(\tau,\mathbf{x})=\int\frac{d^3k}{(2\pi)^3}\sum_{s=\pm}\left[u_s(\tau,k)e^{i\mathbf{k}\cdot\mathbf{x}}h^s_\alpha(\mathbf{\hat{k}})b_{\mathbf{k}}^s+v_s(\tau,k)^*e^{-i\mathbf{k}\cdot\mathbf{x}}\epsilon_{\alpha\beta} h^{s\dagger}_{\dot{\gamma}}(\mathbf{\hat{k}})\bar{\sigma}^{0\dot{\gamma}\beta} b_{\mathbf{k}}^{s\dagger}\right]~.
\end{equation}
Notice that it is sometimes customary to omit the zeroth Pauli matrix since it is an identity matrix in component form. However, for the sake of balancing the indices, we will keep them explicit here. The eigenvalue equation for the helicity basis can be written in a number of different equivalent forms:
\begin{subequations}
	\begin{eqnarray}
		-\hat{k}^i\sigma^{0}_{\alpha\dot{\beta}}\bar{\sigma}^{i\dot{\beta}\gamma}h^s_\gamma(\mathbf{\hat{k}})&=&s h^s_\alpha(\mathbf{\hat{k}})\\
		h^{s\dagger}_{\dot{\gamma}}(\mathbf{\hat{k}})\bar{\sigma}^{0\dot{\gamma}\beta}\sigma^i_{\beta\dot{\alpha}}\hat{k}^i&=&s h^{s\dagger}_{\dot{\alpha}}(\mathbf{\hat{k}})~.
	\end{eqnarray}
\end{subequations}
It is sometimes useful to choose an explicit component form of the helicity basis,
\begin{equation}
	h^+_\alpha(\mathbf{\hat k})=\left(\begin{array}{ccc}
		\cos\frac{\theta}{2}\\
		e^{i\phi}\sin\frac{\theta}{2}
	\end{array}\right)_\alpha~~,~~~h^-_\alpha(\mathbf{\hat k})=\left(\begin{array}{ccc}
		-e^{-i\phi}\sin\frac{\theta}{2}\\
		\cos\frac{\theta}{2}
	\end{array}\right)_\alpha~,\label{helicityBasisComponentForm}
\end{equation}
with $\mathbf{\hat k}$ pointing toward the $(\theta,\phi)$ direction in spherical coordinates. 

Substituting the mode expansion into the equation of $\psi_\alpha$ obtained from varying the action (\ref{fermionLagrangian}), we obtain the EoM of the mode functions,
\begin{equation}
	i\frac{\partial}{\partial \tau}
	\left(\begin{array}{ccc}
		u_s\\
		v_s
	\end{array}\right)=\left(\begin{array}{ccc}
		-s k-a\kappa & a m\\
		a m & s k+a\kappa
	\end{array}\right)\left(\begin{array}{ccc}
		u_s\\
		v_s
	\end{array}\right)~.\label{fermionEoM2}
\end{equation}
This EoM preserves the combination $|u_s|^2+|v_s|^2$, with the normalization constant determined by the canonical quantization condition $\{\psi_\alpha(\tau,\mathbf{x}),\frac{\delta S_{1/2}}{\delta\partial_\tau\psi_{\beta}(\tau,\mathbf{x}')}\}=i\delta_\alpha^\beta\delta^3(\mathbf{x}-\mathbf{x}')$. After plugging in the mode expansion, this is reduced to a c-number equation
\begin{equation}
	\delta_\alpha^\beta=\sum_{s=\pm}\left(|u_s|^2 h^s_\alpha h^{s\dagger}_{\dot{\beta}}\bar{\sigma}^{0\dot{\beta}\beta}+|v_s|^2 \sigma^{0}_{\alpha\dot{\beta}}h^{-s\dagger\dot{\beta}}h^{-s\beta}\right)~.
\end{equation}
Taking the trace and the determinant of the above equation yields
\begin{equation}
	\sum_{s=\pm}(|u_s|^2+|v_s|^2)=2~~,\text{and}~~~\prod_{s=\pm}(|u_s|^2+|v_s|^2)=1~,
\end{equation}
thus fixing the normalization condition $|u_s|^2+|v_s|^2=1$ separately for different helicities.

As mentioned in Sect.~\ref{fermionGeneralities}, the EoM (\ref{fermionEoM2}) can be interpreted as describing the transition of a two-level system with a Hamiltonian
\begin{equation}
	H(\tau)=\left(\begin{array}{ccc}
		Z(\tau) & X(\tau)\\
		X(\tau) & -Z(\tau)
	\end{array}\right)= E(\tau)\left(\begin{array}{ccc}
		C(\tau) & S(\tau)\\
		S(\tau) & -C(\tau)
	\end{array}\right)~,\label{eq:two_level_H}
\end{equation}
where 
\begin{equation}
	E\equiv\sqrt{Z^2+X^2}=k^2+2s\kappa a+(m^2+\kappa^2)a^2
\end{equation}
and $C\equiv\frac{Z}{E}$, $S\equiv\frac{X}{E}$. The instantaneous eigenstates of $H(\tau)$ are given by
\begin{equation}
	H\left(\begin{array}{ccc}
		C\\
		S
	\end{array}\right)=E\left(\begin{array}{ccc}
		C\\
		S
	\end{array}\right)~,~~~H\left(\begin{array}{ccc}
		S\\
		-C
	\end{array}\right)=-E\left(\begin{array}{ccc}
		S\\
		-C
	\end{array}\right)~.\label{eq:eigenstates}
\end{equation}
Therefore, an ansatz of the solution of (\ref{fermionEoM2}) can be constructed as
\begin{equation}
	\left(\begin{array}{ccc}
		u_s\\
		v_s
	\end{array}\right)=\alpha_s e^{-i\int E_s d\tau}\left(\begin{array}{ccc}
		\tilde C_s\\
		\tilde S_s
	\end{array}\right)+\beta_s e^{i\int E_s d\tau}\left(\begin{array}{ccc}
		\tilde S_s^*\\
		-\tilde C_s^*
	\end{array}\right)~.\label{eq:ansatz_CS}
\end{equation}
where $\tilde C_s,\tilde S_s$ are slowly varying functions whose detailed form as a super-adiabatic basis will be computed later. If we choose $|\tilde{C}_s|^2+|\tilde{S}_s|^2=1$, the coefficient functions $\alpha_s,\beta_s$ will satisfy the normalization $|\alpha_s|^2+|\beta_s|^2=1$, as required by the normalization of $u_s, v_s$ and unitarity.

Now we can insert the ansatz back into the mode expansion of $\psi_\alpha$,
\begin{eqnarray}
	\psi_\alpha(\tau,\mathbf{x})&=&\int\frac{d^3k}{(2\pi)^3}\sum_{s=\pm}\Bigg[\left(\alpha_s e^{-i\int E_s d\tau}\tilde C_s+\beta_s e^{i\int E_s d\tau}\tilde S_s^*\right)e^{i\mathbf{k}\cdot\mathbf{x}}h^s_\alpha(\mathbf{\hat{k}})b_{\mathbf{k}}^s\\
	&&~~~~~~~~~~~~~~~~+\left(\alpha_s^* e^{i\int E_s d\tau}\tilde S_s^*-\beta_s^* e^{-i\int E_s d\tau}\tilde C_s\right)e^{-i\mathbf{k}\cdot\mathbf{x}}\epsilon_{\alpha\beta} h^{s\dagger}_{\dot{\gamma}}(\mathbf{\hat{k}})\bar{\sigma}^{0\dot{\gamma}\beta} b_{\mathbf{k}}^{s\dagger}\Bigg]~.
\end{eqnarray}
The time-dependent creation/annihilation operators are selected according to the instantaneous negative/positive frequency parts of $\psi_\alpha$. Therefore, we can regroup the terms according to the dynamical phase $e^{\mp i\int E_s d\tau}$. First, we note the relation
\begin{equation}
	\epsilon_{\alpha\beta} h^{s\dagger}_{\dot{\gamma}}(-\mathbf{\hat{k}})\bar{\sigma}^{0\dot{\gamma}\beta}\equiv \eta^s(\mathbf{\hat{k}})h^{s}_{\alpha}(\mathbf{\hat{k}})~,\label{helicityBasisPhase}
\end{equation}
where $\eta^s(\mathbf{\hat{k}})$ is a phase factor satisfying
\begin{equation}
	\eta^s(-\mathbf{\hat{k}})=-\eta^s(\mathbf{\hat{k}})~~.
\end{equation}
This can be seen directly from left-multiplying (\ref{helicityBasisPhase}) by $h^{s\dagger}_{\dot{\alpha}}\bar{\sigma}^{0\dot{\alpha}\alpha}$ and solving out $\eta^s(\mathbf{\hat{k}})$, or from directly inspecting the component form (\ref{helicityBasisComponentForm}). After applying (\ref{helicityBasisPhase}), we arrive at an alternative expansion,
\begin{equation}
	\psi_\alpha(\tau,\mathbf{x})=\int\frac{d^3k}{(2\pi)^3}\sum_{s=\pm}\left[\tilde C_s e^{-i\int E_s d\tau}e^{i\mathbf{k}\cdot\mathbf{x}}h^s_\alpha(\mathbf{\hat{k}})d_{\mathbf{k}}^s+ \tilde S_s^* e^{i\int E_s d\tau} e^{-i\mathbf{k}\cdot\mathbf{x}}\epsilon_{\alpha\beta} h^{s\dagger}_{\dot{\gamma}}(\mathbf{\hat{k}})\bar{\sigma}^{0\dot{\gamma}\beta}d_{\mathbf{k}}^{s\dagger}\right]~,
\end{equation}
where the new time-dependent annihilation operator is obtained as a Bogoliubov transformation \cite{Landete:2013axa,Adshead:2015kza},
\begin{equation}
	d_{\mathbf{k}}^s(\tau)\equiv \alpha_s(\tau,k) b_{\mathbf{k}}^s-\beta_s^*(\tau,k)\eta^s(\mathbf{\hat{k}})b_{\mathbf{k}}^{s\dagger}~.
\end{equation}
The anti-commutation relation is preserved:
\begin{equation}
	\{d^s_{\mathbf{k}},d^{s'\dagger}_{\mathbf{k}'}\}=\left(|\alpha_s|^2+|\beta_s|^2\right)\{b^s_{\mathbf{k}},b^{s'\dagger}_{\mathbf{k}'}\}=(2\pi)^3\delta^{ss'}\delta^3(\mathbf{k}-\mathbf{k}')~.
\end{equation}
Therefore, the vacuum annihilated by the original operator $b^s_{\mathbf{k}}$ now contains a spectrum of particles with comoving number density
\begin{equation}
	\frac{\langle n^s_{\mathbf{k}}(\tau) \rangle}{V}=\frac{\langle d^{s\dagger}_{\mathbf{k}}d^s_{\mathbf{k}}(\tau) \rangle}{V}=|\beta_s(\tau,k)|^2~,
\end{equation}
where $V$ is the comoving volume. Here rotational symmetry demands the isotropy of particle production spectrum, as $\beta_s$ only depends on the magnitude of the momentum.

The structure of the EoM of fermion (\ref{fermionEoM2}) is similar to the two-state systems in quantum mechanics, and thus we adopt the framework of analyzing the quantum transition histories of such systems \cite{Berry:1990histories}. To calculate the particle production, it is convenient to use the bra-ket notation. Similar to the case of boson, we rewrite the equation of motion with the asymptotic parameter $\lambda$
\begin{align}
	i\frac{d}{dz}|\psi\rangle=\lambda H |\psi\rangle \ , \label{eq:EoM_bra_ket}
\end{align}
where we use $\psi\rangle$ to denote the two-component mode function, and $z=k\tau$. For the ansatz (\ref{eq:ansatz_CS}) in the bra-ket notation
\begin{align}
	|\psi\rangle=\alpha e^{-i\lambda\int^z_{z_i}E(z_1)dz_1}|\psi_\alpha\rangle+\beta e^{i\lambda\int^z_{z_i}E(z_1)dz_1}|\psi_\beta\rangle \ , \label{eq:fermion_exact_form}
\end{align} the left hand side of (\ref{eq:EoM_bra_ket}) is
\begin{align}
	i\frac{d}{dz}|\psi\rangle=i\left[(\alpha'-i\lambda E\alpha)|\psi_\alpha\rangle+\alpha |\psi_\alpha\rangle'\right]e^{-i\lambda\int^z_{z_i}E(z_1)dz_1}+i\left[(\beta'+i\lambda E\beta)|\psi_\beta\rangle+\beta |\psi_\beta\rangle'\right]e^{i\lambda\int^z_{z_i}E(z_1)dz_1} \ , \label{eq:LHS_fermion}
\end{align}
whereas the right hand side is
\begin{align}
	\lambda H |\psi\rangle=\lambda\left(\alpha e^{-i\lambda\int^z_{z_i}E(z_1)dz_1} H|\psi_\alpha\rangle+\beta e^{i\lambda\int^z_{z_i}E(z_1)dz_1} H|\psi_\beta\rangle\right) \ . \label{eq:RHS_fermion}
\end{align}
If $|\psi_\alpha \rangle$ and $|\psi_\beta \rangle$ are the two exact solutions, the positive and negative modes evolve independently with constant $\alpha$ and $\beta$, implying that
\begin{align}
	-i\lambda(H-E)|\psi_\alpha\rangle-|\psi_\alpha\rangle'&=0\nonumber \\
	-i\lambda(H+E)|\psi_\beta\rangle-|\psi_\beta\rangle'&=0 \ . \label{eq:psi_alphabeta}
\end{align}
Similar to the case of boson, we approximate the solutions with the asymptotic series
\begin{align}
	|\psi_\alpha\rangle&=\sum_{j=0}^\infty\frac{c_j(z)|\psi^{(0)}_\alpha\rangle+d_j(z)|\psi^{(0)}_\beta\rangle}{\lambda^j} \nonumber \\
	|\psi_\beta\rangle&=-\sum_{j=0}^\infty\frac{d_j^*(z)|\psi^{(0)}_\alpha\rangle-c_j^*(z)|\psi^{(0)}_\beta\rangle}{\lambda^j} \ , \label{eq:psi_asymptotic}
\end{align}
where $|\psi^{(0)}_\alpha\rangle$ and $|\psi^{(0)}_\beta\rangle$ are the instantaneous eigenstates found in (\ref{eq:eigenstates}), and they satisfy
\begin{align}
	|\psi_\alpha^{(0)}\rangle'&=-\frac{\theta'}{2}|\psi_\beta^{(0)}\rangle \nonumber \\
	|\psi_\beta^{(0)}\rangle'&=\frac{\theta'}{2}|\psi_\alpha^{(0)}\rangle \ , \label{eq:psi0_derivative}
\end{align}
where $\theta(z)=\arctan\left(\frac{X(z)}{Z(z)}\right)$, defined in (\ref{eq:two_level_H}). Substitute (\ref{eq:psi_asymptotic}) and (\ref{eq:psi0_derivative}) to (\ref{eq:psi_alphabeta}), we have 
\begin{align}
	c_j'+\frac{d_j}{2}\theta'&=0 \nonumber \\
	2iE(z) d_{j+1}-d_j'+\frac{c_j}{2}\theta' &=0\ . \label{eq:c_d_equation}
\end{align}
Similar to the case of boson, we solve for $c_j(z)$ and $d_j(z)$ near the complex root $z_c$ of $E^2(z)=X^2(z)+Z^2(z)=0$. In all the scenarios that we study in Sect. \ref{TheLongTechinicalSection}, $z_c$ is a first-order root, implying that
\begin{align}
	X(z)&\approx X_c+X'_c(z-z_c)+\mathcal{O}(|z-z_c|^2)\nonumber \\
	Z(z)&\approx iX_c+Z'_c(z-z_c)+\mathcal{O}(|z-z_c|^2) \ , \label{eq:approximate_XZ}
\end{align}
with $X_c^2+Z_c^2=0$, and
\begin{align}
	\theta'(z)&=\frac{Z^2(z)}{X^2(z)+Z^2(z)}\left(\frac{X(z)}{Z(z)}\right)' \nonumber \\
	&\approx \frac{i}{2(z-z_c)} \ .
\end{align}
With this approximation of $\theta'(z)$, we obtain the recurrence relation of $c_j$ from (\ref{eq:c_d_equation})
\begin{align}
	c_{j+1}'(z)=-\frac{i}{2A(z-z_c)^{\frac{1}{2}}}c_j''(z)-\frac{i}{2A(z-z_c)^{\frac{3}{2}}}c_j'(z)+\frac{i}{32A(z-z_c)^{\frac{5}{2}}}c_j(z) \ ,
\end{align}
where $A$ is defined by $E\approx A(z-z_c)^{1/2}$, and the solutions of $c_j(z)$ and $d_j(z)$ with $c_0=1$ and $d_0=0$ are
\begin{align}
	\frac{c_j(z)}{\lambda^j}&=-\frac{3^{j-2} 4^{-j-1} \left(\frac{i}{A}\right)^j \left(\frac{5}{6}\right)_{j-1}
		\left(\frac{7}{6}\right)_{j-1}}{\Gamma (j+1)(z-z_c)^{3j/2}} \nonumber \\
	&=-\frac{\left(\frac{5}{6}\right)_{j-1} \left(\frac{7}{6}\right)_{j-1}}{36 \Gamma (j+1)F^j}\ ,
\end{align}
and 
\begin{align}
	\frac{d_j(z)}{\lambda^j}=\frac{i  \Gamma \left(j-\frac{1}{6}\right) \Gamma \left(j+\frac{1}{6}\right)}{2 \pi 
		\Gamma (j)F^j} \ ,
\end{align}
respectively, where $(a)_k=\Gamma(a+k)/\Gamma(a)$ is the Pochhammer symbol, and the singulant is
\begin{align}
	F(z)=-2i\int^z_{z_c}E(z_1)dz_1 \ .
\end{align} Clearly $c_j(z)$ and $d_j(z)$ are divergent asymptotic series, and thus we truncate them at the order of $n$ and denote the partial sums of (\ref{eq:psi_asymptotic}) as $|\psi_\alpha^{(n)}\rangle$ and $|\psi_\beta^{(n)}\rangle$. Equating (\ref{eq:LHS_fermion}) and (\ref{eq:RHS_fermion}) with the truncated asymptotic series can derive the coupled differential equations of $\alpha$ and $\beta$
\begin{align}
	&\frac{d}{dz}\begin{pmatrix} 
		\alpha(z) \\
		\beta(z)
	\end{pmatrix} \nonumber \\
	&=\begin{pmatrix}
		i\lambda E(z)-\langle \psi^{(n)}_\alpha|\psi^{(n)}_\alpha\rangle'-i\lambda \langle \psi^{(n)}_\alpha|H|\psi^{(n)}_\alpha\rangle & -\left(\langle \psi^{(n)}_\alpha|\psi^{(n)}_\beta\rangle'+i\lambda \langle \psi^{(n)}_\alpha |H| \psi^{(n)}_\beta\rangle\right)e^{2i\lambda\int^z_{z_i}E(z_1)dz_1} \\
		-\left(\langle \psi^{(n)}_\beta|\psi^{(n)}_\alpha\rangle'+i\lambda \langle \psi^{(n)}_\beta |H| \psi^{(n)}_\alpha\rangle\right)e^{-2i\lambda\int^z_{z_i}E(z_1)dz_1}& -i\lambda E(z)-\langle \psi^{(n)}_\beta|\psi^{(n)}_\beta\rangle'-i\lambda \langle \psi^{(n)}_\beta|H|\psi^{(n)}_\beta\rangle
	\end{pmatrix} \begin{pmatrix} 
		\alpha(z) \\
		\beta(z)
	\end{pmatrix} \ .
\end{align}
The values of matrix elements can be obtained by evaluating (\ref{eq:psi_alphabeta}) with the truncated series
\begin{align}
	-i\lambda(H-E)|\psi^{(n)}_\alpha\rangle-|\psi^{(n)}_\alpha\rangle'&=\frac{1}{\lambda^n}\left(\frac{c_n(z)}{2}\theta'(z)-d_n'(z)\right)|\psi_\beta^{(0)}\rangle \nonumber \\
	-i\lambda(H+E)|\psi^{(n)}_\beta\rangle-|\psi^{(n)}_\beta\rangle'&= \frac{1}{\lambda^n}\left(\frac{c_n(z)}{2}\theta'(z)-d_n'(z)\right)^* |\psi_\alpha^{(0)}\rangle\ ,
\end{align} and thus
\begin{align}
	\frac{d}{dz}\begin{pmatrix} 
		\alpha (z) \\
		\beta (z)
	\end{pmatrix}=\begin{pmatrix}
		\delta_2(z) & -\Delta^*_2(z)e^{2i\lambda\int^z_{z_i}E(z_1)dz_1}\\
		\Delta_2(z)e^{-2i\lambda\int^z_{z_i}E(z_1)dz_1} & \delta_2^*(z) 
	\end{pmatrix} \begin{pmatrix}
		\alpha(z) \\
		\beta(z)
	\end{pmatrix} \ ,
\end{align}
where 
\begin{align}
	\delta_2(z)&=\frac{1}{\lambda^n}\left(\frac{c_n(z)}{2}\theta'(z)-d_n'(z)\right)\sum_{j=1}^n\frac{d^*_j(z)}{\lambda^j} \nonumber \\
	\Delta_2(z)&=\frac{1}{\lambda^n}\left(\frac{c_n(z)}{2}\theta'(z)-d_n'(z)\right)\sum_{j=0}^n \frac{c_j(z)}{\lambda^j}
\end{align}
which has the same structure as vector boson (\ref{eq:Boson_alpha_beta}). We remove the diagonal term by defining the Stokes multipliers
\begin{align}
	\alpha(z)=e^{\int^z_{z_i}\delta_2(z_1)dz_1} S_+(z) \ , \ \beta(z)=e^{\int^z_{z_i}\delta_2^*(z_1)dz_1}e^{-2i\lambda\int^{z_c}_{z_i}E(z_1)dz_1}S_-(z) \ ,
\end{align}
and
\begin{align}
	\frac{d}{dF}\begin{pmatrix}
		S_+ \\
		S_-
	\end{pmatrix} &= \begin{pmatrix}
		0 & \frac{\Delta_2^*}{2i\lambda E}e^{-F+\int^z_{z_i}\delta^*(z_1)-\delta(z_1)dz_1} \\
		-\frac{\Delta_2}{2i\lambda E} e^{F+\int^z_{z_i}\delta(z_1)-\delta^*(z_1)dz_1}& 0
	\end{pmatrix}\begin{pmatrix}
		S_+ \\
		S_-
	\end{pmatrix} \nonumber \\
	&=\begin{pmatrix}
		0 & -iT_n\left(\frac{n!}{2\pi F^{n+1}}\right)^*e^{-F}\\
		iT_n\frac{n!}{2\pi F^{n+1}}e^F &
	\end{pmatrix}\begin{pmatrix}
		S_+ \\
		S_-
	\end{pmatrix} +\mathcal{O}\left(\frac{1}{\lambda^{n+1}}\right)\ , \label{eq:Spm_fermion}
\end{align}
where we only keep the $\mathcal{O}(\lambda^{-n})$ term, and the prefactor
\begin{align}
	T_n=\frac{\Gamma \left(n+\frac{5}{6}\right) \Gamma \left(n+\frac{7}{6}\right)}{\Gamma (n+1)^2} \ ,
\end{align}
which converges to $1$ for $n\to\infty$. Similar to the bosonic case (\ref{eq:simplified_S}), (\ref{eq:Spm_fermion}) can be solved perturbatively starting with the initial condition $(S^i_+,S^i_-)=(S^{(0)}_+,S^{(0)}_-)$:
\begin{align}
	S_-^{(0)}+S_-^{(1)}(F)&=S^i_-+iS^i_+T_n\int^F_{{\rm Re}F+i\infty}\frac{n!}{2\pi F^{n+1}}e^F dF \nonumber \\
	&=S^i_-+S^i_+T_n\frac{i(-1)^{n+1} n!}{2\pi}\tilde{\Gamma}(-n,-F) \ . \label{eq:S0_fermion}
\end{align}
By setting the vacuum initial condition $(S^i_+,S^i_-)=(S^{(0)}_+,S^{(0)}_-)=(1,0)$, the first-order perturbation of $S_+$ implies that
\begin{align}
	\left|S^{(0)}_++S^{(1)}_+ \right|^2 &=1-e^{-2{\rm Re}F}\int^F_{{\rm Re}F+i\infty} \frac{{dS_-^{(1)}}^*}{dF}S_-^{(1)}+\frac{dS_-^{(1)}}{dF}{S_-^{(1)}}^* dF+\mathcal{O}\left(e^{-4{\rm Re}F}\right) \nonumber \\
	&=1-e^{-2{\rm Re}F}\left|S_-^{(1)}(F)\right|^2+\mathcal{O}\left(e^{-4{\rm Re}F}\right) \ ,
\end{align}
which agrees with the normalization of the Bogoliubov coefficients. By choosing the the optimal truncation order as $n={\rm Re}F-1$, the integrand is stationary at $F={\rm Re}F$, and thus $S_-^{(1)}$ can be approximated as
\begin{align}
	S_-^{(1)}(z)\approx \frac{T_n}{2}\left[1+{\rm Erf}\left(-\frac{{\rm Im}F(z)}{\sqrt{2 {\rm Re} F}}\right)\right] \ . \label{eq:SErf_fermion}
\end{align}

\begin{figure}[h!]
	\centering
	\includegraphics[width=9cm]{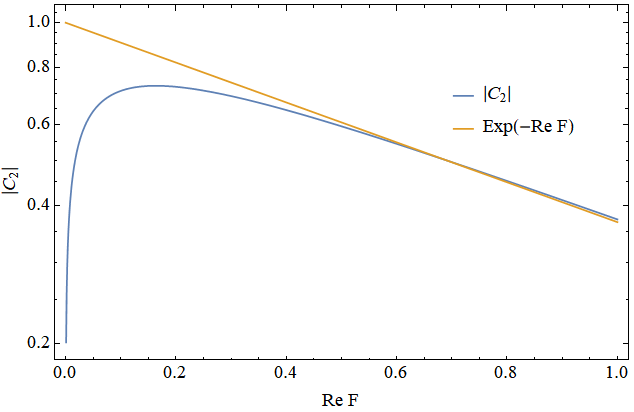}
	\caption{The comparison between $|C_2|$ and $e^{-{\rm Re}F}$ with $0<{\rm Re}F\le 1$, and the vertical axis is in logarithmic scale.}\label{fig:C2vsexp_fermion}
\end{figure}

\begin{figure}[h!]
	\centering
	\includegraphics[width=9cm]{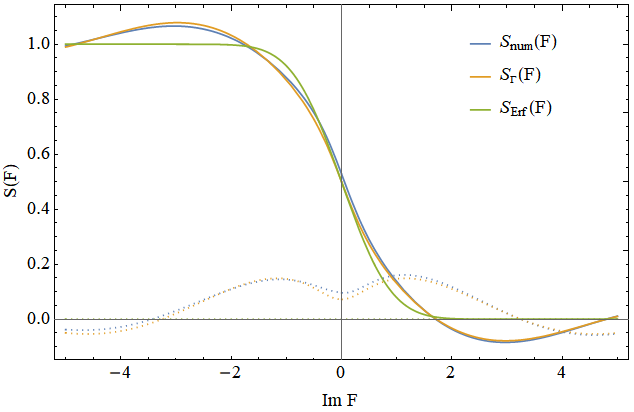}
	\caption{The comparison between $S_{\rm num}(F)$, $S_{\Gamma}(F)$ and $S_{\rm Erf}(F)$, where all of them are normalized such that the final value approaches $1$ and ${\rm Re}F=0.5$. The solid lines represent the real parts, whereas the dashed lines are the imaginary parts.}\label{fig:Comparisions_Snum_SGamma_SErf_fermion}
\end{figure}

Similar to the bosonic case, the situations with ${\rm Re}F<1$ implies the failure of choosing an optimal truncation for the asymptotic series (\ref{eq:psi_asymptotic}), and we may apply the Borel sum to evaluate such divergent series:
\begin{align}
	I(F)&=\sum_{j=0}^\infty\frac{c_j}{\lambda^j}\nonumber \\
	&=\frac{U\left(-\frac{1}{6},\frac{2}{3},-F\right)}{\sqrt[6]{-F}} \ , \nonumber \\
	J(F)&=\sum_{j=0}^\infty\frac{d_j}{\lambda^j} \nonumber \\
	&=6iF \frac{dI(F)}{dF}\ ,
\end{align}
where $U(a,b,z)$ is the confluent hypergeometric function, and the last line is obtained from the relation between $c_j$ and $d_j$ (\ref{eq:c_d_equation}). Since $I(F)$ and $J(F)$ are discontinuous at ${\rm Im}F=0$, we rewrite the exact solution (\ref{eq:fermion_exact_form}) into two parts:
\begin{align}
	|\psi\rangle=\begin{cases}
		\alpha_ie^{-i\lambda\int^z_{z_i}E(z_1)dz_1}\left(I(F)|\psi^{(0)}_\alpha\rangle+J(F)|\psi^{(0)}_\beta\rangle\right)\\
		~~~-\beta_i e^{i\lambda\int^z_{z_i}E(z_1)dz_1}\left(J^*(F)|\psi^{(0)}_\alpha\rangle-I^*(F)|\psi^{(0)}_\beta\rangle\right) &,~ {\rm Im}F>0 \\
		C_1e^{-i\lambda\int^z_{z_i}E(z_1)dz_1}\left(I(F)|\psi^{(0)}_\alpha\rangle+J(F)|\psi^{(0)}_\beta\rangle\right)\\
		~~~+C_2 e^{i\lambda\int^z_{z_i}E(z_1)dz_1}\left(J^*(F)|\psi^{(0)}_\alpha\rangle-I^*(F)|\psi^{(0)}_\beta\rangle\right)&,~ {\rm Im}F<0 \ ,
	\end{cases}
\end{align}
where $\alpha_i$ and $\beta_i$ are the initial Bogoliubov coefficients, $C_1$ and $C_2$ are constants to let $|\psi\rangle$ and its derivative continuous at ${\rm Im}F=0$. The time-dependent Bogoliubov coefficients defined with respect to the eigenstates $|\psi^{(0)}_\alpha\rangle$ and $|\psi^{(0)}_\beta\rangle$ are thus:

\begin{align}
	\alpha(F)&=\begin{cases}
		\alpha_iI(F)-\beta_ie^{-F+{\rm Re}F}J^*(F)&,~ {\rm Im}F>0\\
		C_1I(F)+C_2e^{-F+{\rm Re}F}J^*(F)&,~ {\rm Im}F<0 
	\end{cases} \nonumber \\
	\beta(F)&=\begin{cases}
		\alpha_ie^{F-{\rm Re}F} J(F)+\beta_i I^*(F) &,~ {\rm Im}F>0\\
		C_1e^{F-{\rm Re}F} J(F)-C_2 I^*(F)&,~ {\rm Im}F<0 
	\end{cases}\ . \label{eq:beta_resum_fermion}
\end{align}
With the vacuum initial condition $(\alpha_i,\beta_i)=(1,0)$, the two unknown constants are determined by matching the two parts. Since $I(F)\to 1$ and $J(F)\to 0$ for ${\Im F} \to \pm \infty$, it is clear that $C_1$ and $-C_2$ are the final values of $\alpha(k)$ and $\beta(k)$ respectively, and numerical checking confirms the normalization $|C_1|^2+|C_2|^2=1$. As shown in FIG. \ref{fig:C2vsexp_fermion}, the amount of particle production fits the tendency of $e^{-{\rm Re}F}$ for ${\rm Re}F\gtrsim 0.2$, but large deviations appear when ${\rm Re}F\to 0$, similar to the case of boson.
We also compare the Stokes multiplier $S_{\rm num}(F)$ obtain from the numerical result (\ref{eq:beta_resum_fermion}) with the approximations utilizing the incomplete gamma function $S_{\Gamma}(F)$ with $n=0$ from the first-order perturbation (\ref{eq:S0_fermion}) and the error function $S_{\rm Erf}(F)$ (\ref{eq:SErf_fermion}) respectively, as shown in FIG. \ref{fig:Comparisions_Snum_SGamma_SErf_fermion}.

Similar to the bosonic case, the asymptotic parameter $\lambda$ always appear with the instantaneous eigenvalue $E(z)$, so we may set $\lambda H \to H$ in (\ref{eq:EoM_bra_ket}). We summarize a simple form of $\beta(z)$ which reflects the tendency of the particle production starting from vacuum initial condition
\begin{empheq}[box=\widefbox]{align}
	\beta(z)\approx \frac{e^{-F(z_i)}}{2}\left[1+{\rm Erf}\left(-\frac{{\rm Im}F(z)}{\sqrt{2 {\rm Re} F}}\right)\right] \ ,\label{eq:explicit_beta_fermion}
\end{empheq}
where it is noteworthy that there is no additional prefactor $-i$ compared to the bosonic case (\ref{eq:explicit_beta_boson}), and the generalization to arbitrary initial conditions can be easily done base on (\ref{eq:S0_fermion}) and (\ref{eq:beta_resum_fermion}). 

Note that the production histories solely depend on the singulant $F$, and the definition of the singulant $F$ in the fermion case is similar to the boson case, with 
\begin{equation}
	E(z)=\sqrt{1+\frac{2s\kappa a}{k}+\frac{(m^2+\kappa^2)a^2}{k^2}}
\end{equation}
in replacement of 
\begin{equation}
	W(z)\approx w(z)=\sqrt{1+\frac{2s\kappa a}{k}+\frac{m^2a^2}{k^2}}~.
\end{equation}
Therefore, combining this observation with (\ref{eq:explicit_beta_boson}) and (\ref{eq:explicit_beta_fermion}), we arrive at a simple replacement rule for the production histories ($i.e.,~|\beta(z)|^2$) of vector bosons and fermions:
\begin{empheq}[box=\widefbox]{align}
	\nonumber m^2&\leftrightarrow m^2+\kappa^2\\
	\text{(bosons)}&~~~~~\text{(fermions)}~.
\end{empheq}

\section{Analysis of particle production in various spacetimes}\label{TheLongTechinicalSection}
Armed with these powerful mathematical tools, we are now in a position to compute the fine-grained particle production histories of both massive vector bosons and Majorana fermions in various setups.

In this section, we have in mind that the chemical potential is provided as a external source by, for instance, a rolling scalar field. The backreaction to the external field that generates the chemical potential is also assumed to be negligible. In particular, we will assume the chemical potential $\kappa$ is a constant in spacetime. The reason for such a choice is three-fold. First of all, this is indeed true in some cases. For example, the Hubble friction during inflation drives a rolling scalar to an attractor phase with constant speed $\dot{\phi}$, which corresponds to a constant $\kappa$ when coupled to vectors or fermions. In a radiation/matter-dominated universe, specifically chosen scalar potentials also give rise to constant rolling speeds. Second, physically speaking, for any slowly-varying $\kappa(\tau)\neq 0$, a constant chemical potential is always a leading order approximation. As long as the typical time scale of $\kappa(\tau)$ is longer than the particle production time scale, this approximation will be valid. Third, mathematically speaking, a constant $\kappa$ leads to simple and analytical results that already contain lots of information in the general cases, which can always be dealt with using numerical methods.

We will focus on five familiar types of FRW spacetimes whose singulant integrals are exactly computable. The resulting production amount, time and width are given explicitly as analytical expressions. Some of these results are exact while others are approximate or empirical with percent-level error in most parameter regimes. To distinguish them from each other, we will use $=$ when the result is exact. We use $\simeq$ for results which are easily computable to any desired precision but which are shown with finite accuracy. And $\approx$ will be used for empirical results whose relative error is at percent-level.

Due to the replacement rule mentioned in Sect.~\ref{StokesLineMethodReview}, we will only work out the spin-1 case with $|\kappa|<m$, and obtain the spin-1/2 results for all parameter regions by simple substitutions. Also because different helicities are related by a sign flip of $\kappa$, we will focus on the negative helicity state, whose production is enhanced if $\kappa$ is positive. Throughout this section, we will be working in comoving coordinates and using conformal time rather than cosmic time\footnote{This choice has interesting implications for dS. See Sect.~\ref{dSAnalysis} and Appendix~\ref{OneFourthPuzzle} for more details.}. Tilde variables will be used to define dimensionless parameters measured in units of a certain Hubble scale, $e.g.$, $\tilde{m}\equiv\frac{m}{H}$, $\tilde{\kappa}\equiv\frac{\kappa}{H}$, etc. And we will typically expand quantities in powers of $\frac{\tilde{\kappa}}{\tilde{m}}=\frac{\kappa}{m}$.

\subsection{dS}\label{dSAnalysis}
In an exact dS spacetime, the scale factor has a time dependence $a(\tau)=-\frac{1}{H\tau}=e^{H t}$. The EoM for spin-1 particles with chemical potential can be written in terms of a dimensionless variable $z=-k\tau=\frac{k}{a H}$,
\begin{equation}
	\frac{d^2 f(z)}{dz^2}+w^2(z)f(z)=0~,~w^2(z)=1-\frac{2\tilde\kappa}{z} +\frac{\tilde m^2}{z^2}~.\label{dSAnalysisSpin1EoM}
\end{equation}
Notice that the variable $z$ now runs from the right to the left, and the physical region is the positive real axis $z>0$. After analytical continuation, one can define $w(z)$ on the whole complex plane. The two roots of $w^2(z)=0$ lie at $z_c$ and $z_c^*$, with
\begin{equation}
	z_c=\tilde{\kappa}+ i\sqrt{\tilde{m}^2-\tilde{\kappa}^2}~.
\end{equation}
Notice that $z_c$ lies on the upper half complex plane since the original $\tau_c$ is on the lower half complex plane and they differ by a sign. The absence of tachyonic instability requires $\tilde{m}>|\tilde{\kappa}|$. Therefore, the two complex turning points lie in a symmetric fashion across the real axis. Starting from $z_c$ and $z_c^*$ are two branch cuts that meet their ends at the pole at $z=0$.

\begin{figure}[h!]
	\centering
	\includegraphics[width=15cm]{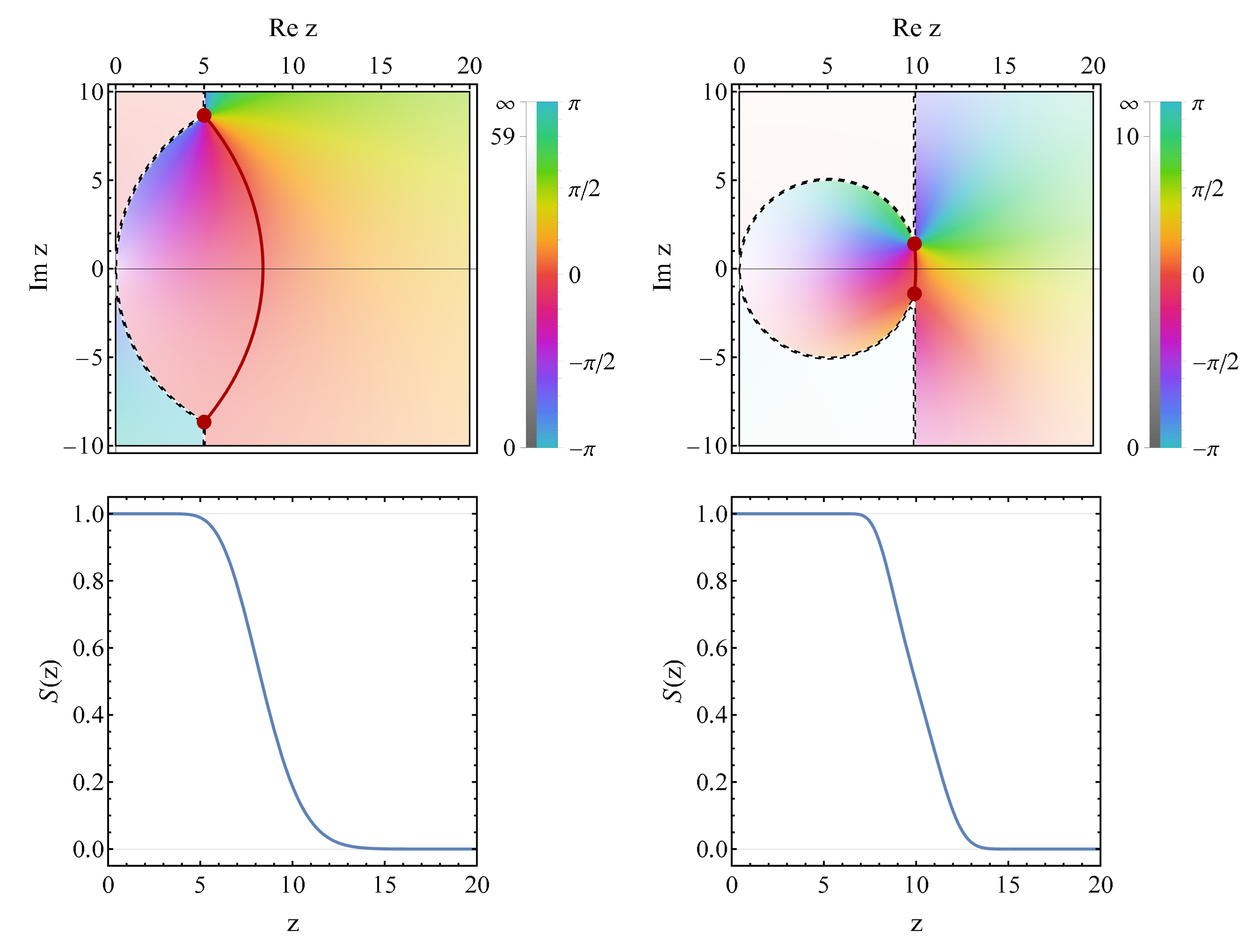}~~~~~~
	\caption{The super-adiabatic singulant $F(z)$ in dS for $\tilde\kappa=5$ (upper left panel) and $\tilde{\kappa}=9.9$ (upper right panel), with the dS-corrected mass $\mu=10$. The hue represents the phase $\arg F(z)$ while the brightness represents the modulus $|F(z)|$. In both panels, the dark red line is the Stokes line joining the complex turning points. 
	The lower two panels show the Stokes multiplier $S(z)$ corresponding to the parameters chosen above.
	}\label{dS01Singulant}
\end{figure}

In the $z$-domain, the singulant is evaluated as
\begin{equation}
	F(z)=2i\int_{z_c}^{z}w(z)dz~.
\end{equation}
Notice the sign change compared to (\ref{eq:singulant_F}). The phase integral is exactly solvable:
\begin{eqnarray}
	\nonumber F(z)&=&2i\left[\sqrt{\tilde{m}^2-2 z \tilde{\kappa }+z^2}+\tilde{m} \tanh ^{-1}\left(\frac{z \tilde{\kappa
		}-\tilde{m}^2}{\tilde{m} \sqrt{\tilde{m}^2-2 z \tilde{\kappa }+z^2}}\right)+\tilde{\kappa } \tanh
	^{-1}\left(\frac{\tilde{\kappa }-z}{\sqrt{\tilde{m}^2+z \left(z-2 \tilde{\kappa }\right)}}\right)\right]\\
	&&+\pi (\tilde{m}-\tilde{\kappa})~,
\end{eqnarray}
where the first line is purely imaginary for $z$ lying on the positive real axis. Its behavior on the right-half $z$-plane as well as the Stokes lines are shown in FIG.~\ref{dS01Singulant}.
\begin{enumerate}
	\item[$\bullet$] Production amount. The production amount is straightforwardly given by the formula
	\begin{equation}
		|\beta|^2=e^{-2\Re F(z_i)}=e^{-2\pi(\tilde{m}-\tilde{\kappa})}~,\label{dSSpin1ProdAmountBeforeResum}
	\end{equation}
	with $z_i\gg \tilde{m}$ taken on the real axis. 
	
	\item[$\bullet$] Production time. The crossing time $z_*$ lies on the real axis, and therefore satisfies a real-numbered equation:
	\begin{equation}
		0=\sqrt{\tilde{m}^2-2 z_* \tilde{\kappa }+z_*^2}+\tilde{m} \tanh ^{-1}\left(\frac{z_* \tilde{\kappa
			}-\tilde{m}^2}{\tilde{m} \sqrt{\tilde{m}^2-2 z_* \tilde{\kappa }+z_*^2}}\right)+\tilde{\kappa } \tanh
		^{-1}\left(\frac{\tilde{\kappa }-z_*}{\sqrt{\tilde{m}^2+z_* \left(z_*-2 \tilde{\kappa }\right)}}\right)~.\label{dSProdTimeEqn}
	\end{equation}
	This equation can be solved using a perturbative expansion in powers of $\tilde{\kappa}$. Plug in the ansatz
	\begin{equation}
		z_*(\tilde{m},\tilde{\kappa})=\sum_{n=0}^\infty \frac{b_n\tilde{\kappa}^n}{\tilde{m}^{n-1}}
	\end{equation}
	into the equation (\ref{dSProdTimeEqn}) and collect the terms order-by-order in $\frac{\tilde{\kappa}}{\tilde{m}}$, we are able to solve the coefficients iteratively,
	\begin{subequations}
		\begin{eqnarray}
			1&=&\frac{1}{\sqrt{1+b_0^2}}\tanh ^{-1}\left(\frac{1}{\sqrt{1+b_0^2}}\right)\\
			b_1&=&\frac{b_0}{\sqrt{1+b_0^2}}\tanh ^{-1}\left(\frac{b_0}{\sqrt{1+b_0^2}}\right)\\
			b_2&=&\frac{b_1+2 b_1 b_0^2-b_0^2}{2 b_0+2 b_0^3}\\
			b_3&=&\frac{\left(b_1+b_0^2\right) \left(-2 b_1+6 b_0 b_2+b_0^2 \left(2-3
				b_1\right) b_1+6 b_0^3 b_2+b_0^4\right)}{6 b_0^2
				\left(1+b_0^2\right){}^2}\\
			\nonumber&\cdots&~.
		\end{eqnarray}
	\end{subequations}
	In this way, we obtain the particle production time as
	\begin{equation}
		z_*(\tilde{m},\tilde{\kappa})\simeq 0.6627 \tilde{m}+0.3435\tilde{\kappa}-0.0102\frac{\tilde{\kappa}^2}{\tilde{m}}+0.0064\frac{\tilde{\kappa}^3}{\tilde{m}^2}+\cdots~.\label{dSSpin1ProdTimeBeforeResum}
	\end{equation}
	Clearly, with a larger effective mass and a larger positive chemical potential, the production time becomes earlier.
	
	\item[$\bullet$] Production width. The derivative of the singulant is none other than the frequency itself:
	\begin{equation}
		\Im F'(z_*)=2\Re w(z_*)=2\sqrt{1-\frac{2\tilde\kappa}{z_*} +\frac{\tilde m^2}{z_*^2}}~.
	\end{equation}
	Thus the production width in the $z$-domain is
	\begin{equation}
		\Delta z_*=\frac{2\sqrt{2|\Re F(z_*)|}}{|\Im F'(z_*)|}=\sqrt{\frac{2\pi\left(\tilde{m}-\tilde{\kappa}\right)}{1-\frac{2\tilde\kappa}{z_*(\tilde{m},\tilde{\kappa})} +\frac{\tilde m^2}{z_*^2(\tilde{m},\tilde{\kappa})}}}~.
	\end{equation}
	We can also translate the production width into the $t$-domain by $\Delta t_*=\frac{\Delta z_*}{H z_*}$. Another useful measure of production width is the e-folding numbers during which the production is complete. In terms of a power series in $\tilde\kappa/\tilde m$, we have
	\begin{equation}
		\Delta N_*=H\Delta t_*=\frac{\Delta z_*}{z_*}\simeq\frac{2.0895}{\sqrt{\tilde{m}}}-\frac{0.4131
			\tilde{\kappa }}{\tilde{m}^{3/2}}+\frac{0.1323
			\tilde{\kappa
			}^2}{\tilde{m}^{5/2}}-\frac{0.0523
			\tilde{\kappa
			}^3}{\tilde{m}^{7/2}}+O\left(\frac{\tilde{\kappa
		}^4}{\tilde{m}^{9/2}}\right)~.\label{dSSpin1ProdWidthBeforeResumey}
	\end{equation}
\end{enumerate}

Now the alert readers may find an inconsistency here. If one compares the production amount (\ref{dSSpin1ProdAmountBeforeResum}) computed from the Stokes-line method with that of the exact result, namely (\ref{spin0ProdAmountExact}) or (\ref{spin0ProdAmountExactLargeMass}), one finds that there is a mismatch of mass: The true result should contain the dS-corrected mass $\mu=\sqrt{\tilde{m}^2-\frac{1}{4}}$ instead of $\tilde{m}$ as given by the Stokes-line method. This mismatch implies that the results obtained by the naive application of Stokes-line method are subjected to a relative error $\mathcal{O}(\tilde{m}^{-2})$, which can be important if $\tilde{m}$ is small.

To trace the origin of this $1/4$ puzzle, let us go back to the super-adiabatic basis for the EoM (\ref{dSAnalysisSpin1EoM}),
\begin{equation}
	f(z)=\frac{1}{\sqrt{2W(z)}}e^{i\int^z_{z_i} W(z')dz'}-\frac{iS(z)e^{-F(z_i)}}{\sqrt{2W(z)}}e^{-i\int^z_{z_i} W(z')dz'}~.\label{SuperAdiabaticSpin1}
\end{equation}
where $S(z)$ is the Stokes multiplier. $W(z)$ is solved order-by-order as specified in Sect.~\ref{StokesLineMethodReviewBosonic}. The leading order reads $W^{(0)}(z)=w(z)$ and
\begin{equation}
	f^{(0)}(z)=\frac{1}{\sqrt{2w(z)}}e^{i\int^z_{z_i} w(z')dz'}-\frac{iS(z)e^{-F(z_i)}}{\sqrt{2w(z)}}e^{-i\int^z_{z_i} w(z')dz'}~.\label{naiveWKB}
\end{equation}
This leading order solution is sometimes called WKB approximation. The late-time behavior of the frequency function is
\begin{equation}
	W^{(0)}(z)=w(z)\xrightarrow{z\to 0}\frac{\tilde{m}}{z}+\mathcal{O}(z^0)~.
\end{equation}
Upon integration over $z$, the phase of the positive-frequency mode becomes linearly increasing with cosmic time $t$:
\begin{equation}
	\int^z_{z_i} w(z')dz'=\int^z_{z_i}\left(\frac{\tilde{m}}{z'}+\mathcal{O}(z'^0)\right)dz'=\tilde{m}\ln z+\text{const}=-m t+\text{const}~,
\end{equation}
where we have used $z=-k\tau=-\frac{k}{H}e^{-Ht}$. Hence the late time behavior of the WKB basis is $f^{(0)}\sim e^{\mp imt}$. This, however, corresponds to a \textit{wrong} oscillation frequency for spin-1 vector particles. The correct frequency can be easily obtained by inspecting the late-time behavior of the EoM itself. Namely we can plug in the ansatz $f\sim z^{\Delta-1}$ and expand (\ref{dSAnalysisSpin1EoM}) to leading order in $z$,
\begin{equation}
	((\Delta-1)(\Delta-2)+\tilde{m})z^{\Delta-3}+\mathcal{O}(z^{\Delta-2})=0~.
\end{equation}
This gives $\Delta_\pm=\frac{3}{2}\pm i\mu$ and the correct IR behavior $f\sim z^{(1/2\pm i\mu)t}\propto e^{\mp i\mu Ht}\neq e^{\mp imt}$. As a result, the leading order super-adiabatic solution does not capture the correct IR oscillation frequency, which is dictated by dS symmetries. Particles in dS are classified according to the unitary irreducible representation of the dS group \cite{Thomas:1941uir,Newton:1950nrds} and a massive spin-$S$ particle in the principal series has a conformal weight \cite{Lee:2016vti}
\begin{equation}
	\Delta_\pm^{(S)}=\frac{3}{2}\pm i\mu_S~,~~\mu_S=\sqrt{\frac{m^2}{H^2}-\left(S-\frac{1}{2}\right)^2}~,~~S\geqslant 1~.
\end{equation}
In other words, the geometry of dS modifies the effective mass of spinning particles in the IR (small $z$), and this fact is not taken into account by the naive WKB approximation (\ref{naiveWKB}).

Fortunately, the advantage of the smoothed Stokes-line method is that the higher-order terms in the super-adiabatic basis can, and actually do, give essential corrections to the leading order solution. The first order correction to $W$ is
\begin{equation}
	\delta W^{(1)}(z)=\frac{3 w'(z)^2-2 w(z) w''(z)}{8 w(z)^3}=\frac{6 z \tilde{m}^2 \left(\tilde{\kappa
		}-z\right)-\tilde{m}^4+z^2 \tilde{\kappa } \left(4 z-3 \tilde{\kappa }\right)}{8 \left(z^2-2 \tilde{\kappa }z+\tilde{m}^2\right)^3}\sqrt{1-\frac{2 \tilde{\kappa }}{z}+\frac{\tilde{m}^2}{z^2}}~.
\end{equation}
This function has two third-order poles at $z_c$ and $z_c^*$. The branch cuts brought by $w(z)$ are still present and they connect the third-order poles to a simple pole at $z=0$,
\begin{equation}
	\delta W^{(1)}(z)=-\frac{1}{8 \tilde{m}z}+\mathcal{O}(z^0)~,~~~\Re z>0~.
\end{equation}
The effect of this pole at the origin is exactly to give an $\mathcal{O}(\tilde{m}^{-2})$ correction to the IR oscillation frequency:
\begin{equation}
	W^{(1)}(z)=W^{(0)}(z)+\delta W^{(1)}(z)=\left(\tilde{m}-\frac{1}{8 \tilde{m}}\right)\frac{1}{z}+\mathcal{O}(z^0)~.
\end{equation}
Including the higher order corrections in the super-adiabatic series, we recover the correct IR oscillation frequency,
\begin{eqnarray}
	\nonumber W(z)&=&W^{(0)}(z)+\delta W^{(1)}(z)+\delta W^{(2)}(z)+\cdots=\left(\tilde{m}-\frac{1}{8 \tilde{m}}-\frac{1}{128\tilde{m}^3}+\cdots\right)\frac{1}{z}+\mathcal{O}(z^0)\\
	&=&\sqrt{\tilde{m}^2-\frac{1}{4}}\times\frac{1}{z}+\mathcal{O}(z^0)~.
\end{eqnarray}

\begin{figure}[h!]
	\centering
	\includegraphics[width=8cm]{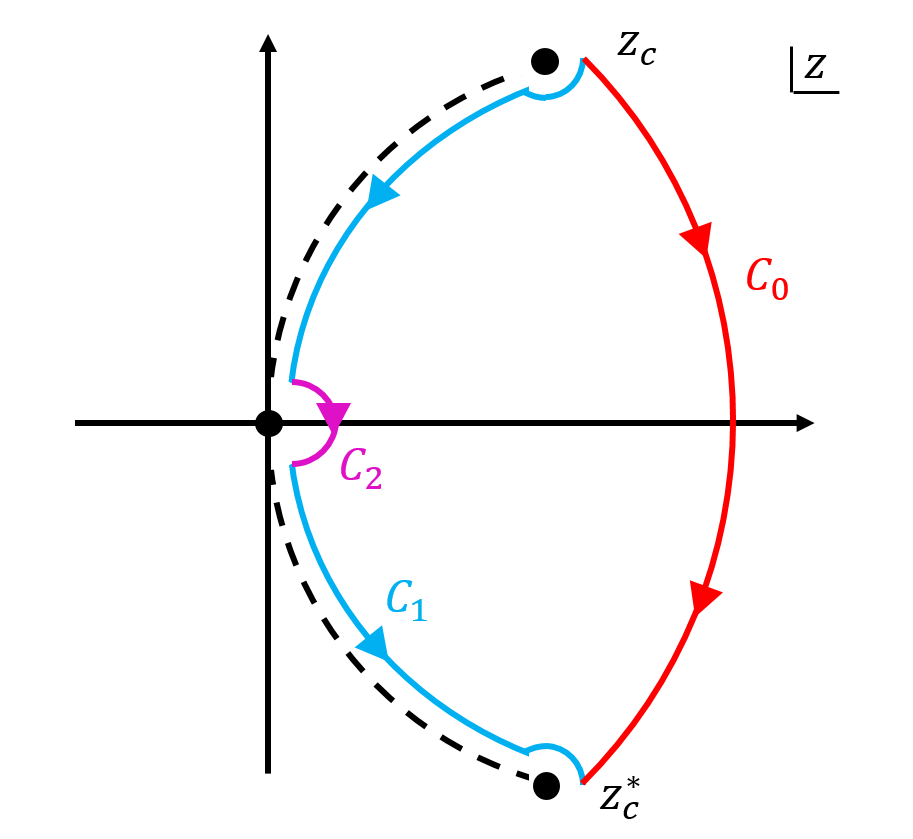}\\
	\caption{The integration contour can be deformed from along the Stokes line ($\mathcal{C}_0$) to along the branch cut ($\mathcal{C}_1\cup \mathcal{C}_2$), collecting half of the residue at $z=0^+$. The black dashed line indicates the branch cut.}\label{contour1}
\end{figure}

Therefore, the $1/4$ puzzle can be resolved by taking into account the full super-adiabatic basis and resumming the higher-order corrections to the mass. We can recover the correct production amount by redefining the singulant as an integral of $W$ instead of $W^{(0)}=w$. Technically, since this integral is ill-defined around the complex turning points where $\delta W^{(n)}(z)$ diverges, we need to manually impose a principal value prescription. We deform the integration contour to lie along the branch cut and tour along a semi-circle around the pole at origin (see FIG.~\ref{contour1} for illustration). Then by some arguments of complex analysis, the only non-zero contribution comes from the semi-circle at the origin, where the IR frequency correction is at work. Thus the corrected production amount can be written as half of the residue of $W(z)$ at $z=0^+$, subtracting half of the residue at $z=\infty$,
\begin{equation}
	\ln|\beta^2|=-2i\int_{z_c}^{z_c^*}W(z)dz=-2\pi\left(\Res_{z\to 0^+} W(z)-\Res_{z\to \infty} W(z)\right)=-2\pi(\mu-\kappa)~.\label{dSBetaResummed}
\end{equation}

For a detailed mathematical discussion of these arguments, we refer the readers to Appendix~\ref{OneFourthPuzzle}. In summary, the super-adiabatic corrections completely fix the mass mismatch and we only need to replace $\tilde{m}\to\mu$ in (\ref{dSSpin1ProdAmountBeforeResum}), (\ref{dSSpin1ProdTimeBeforeResum}) and (\ref{dSSpin1ProdWidthBeforeResumey}) to recover the true results. We list them below for the sake of clarity.
\begin{enumerate}
	\item[$\bullet$] Production amount
	\begin{equation}
		|\beta|^2=e^{-2\pi(\mu-\tilde{\kappa})}~,\label{dSSpin1ProdAmountAfterResum}
	\end{equation}
	\item[$\bullet$] Production time
	\begin{equation}
		z_*(\mu,\tilde{\kappa})\simeq 0.6627 \mu+0.3435\tilde{\kappa}-0.0102\frac{\tilde{\kappa}^2}{\mu}+0.0064\frac{\tilde{\kappa}^3}{\mu^2}+\cdots~.\label{dSSpin1ProdTimeAfterResum}
	\end{equation}
	\item[$\bullet$] Production width
	\begin{equation}
		\Delta N_*=H\Delta t_*=\frac{\Delta z_*}{z_*}\simeq\frac{2.0895}{\sqrt{\mu}}-\frac{0.4131
			\tilde{\kappa }}{\mu^{3/2}}+\frac{0.1323
			\tilde{\kappa
			}^2}{\mu^{5/2}}-\frac{0.0523
			\tilde{\kappa
			}^3}{\mu^{7/2}}+O\left(\frac{\tilde{\kappa
			}^4}{\mu^{9/2}}\right)~.\label{dSSpin1ProdWidthAfterResumey}
	\end{equation}
\end{enumerate}

\begin{figure}[h!]
	\centering
	\includegraphics[width=15cm]{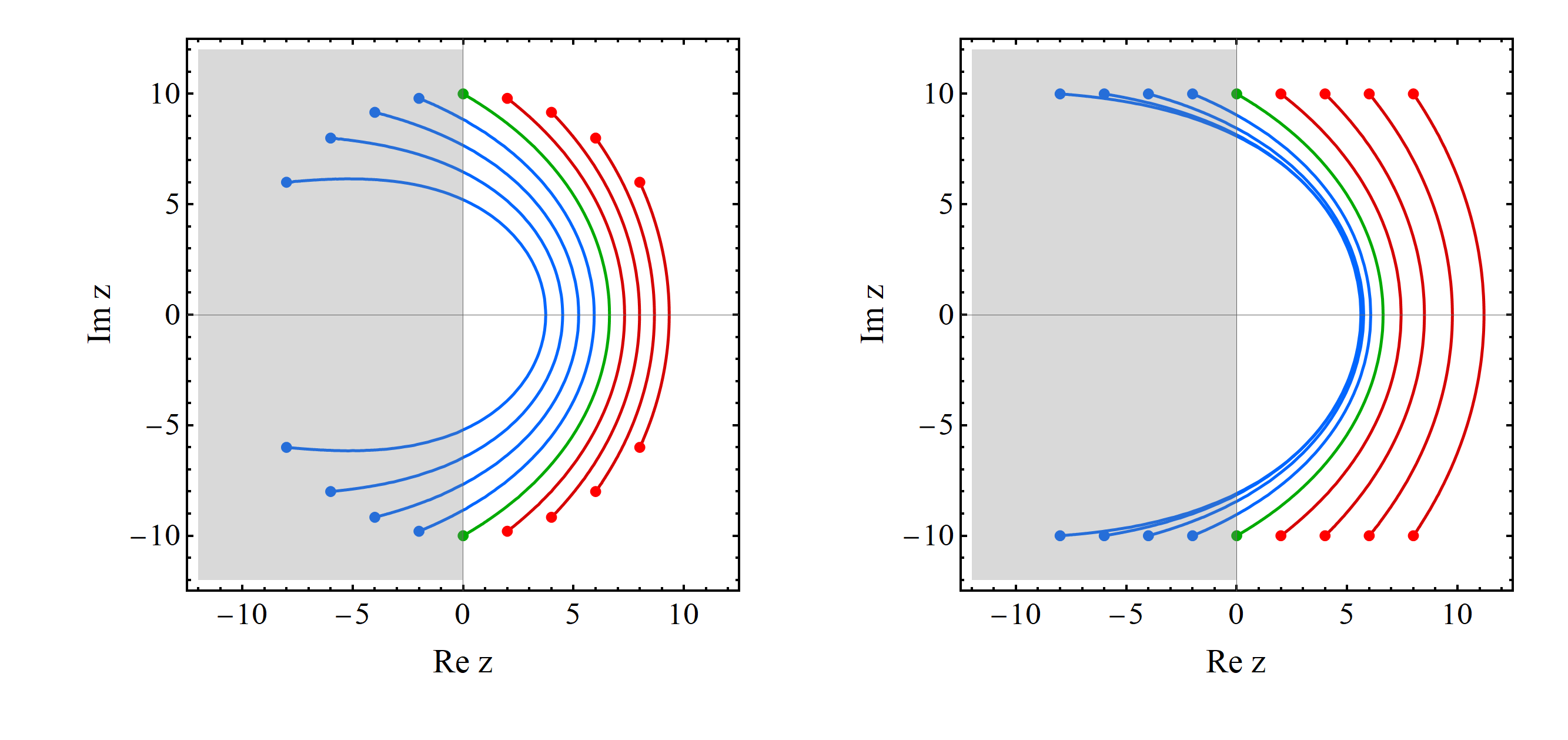}~~~~~~~~~~~~~~~~~~~~\\
	\caption{The Stokes lines for spin-1 bosons (left panel) and spin-1/2 fermions (right panel). In both panels, the chemical potential ranges over $\tilde{\kappa}=-8,-6,\cdots,6,8$ (from left to right, and the green lines correspond to the case without chemical potential), with spin-1 boson mass and spin-1/2 fermion mass $\mu=\tilde{m}=10$. The gray region with $\Re z<0$ cannot be reached physically in dS.
	}\label{dS01StokesLines}
\end{figure}
For spin-1/2 fermions, however, there is no such a problem, since the leading order WKB result already capture the correct IR behavior. The instantaneous eigenvalue of the Hamiltonian for its EoM (\ref{fermionEoM}) written in $z$-domain is 
\begin{equation}
	E_\pm(z)=\sqrt{1\pm\frac{2\tilde{\kappa}}{z}+\frac{\tilde{m}^2+\tilde{\kappa}^2}{z^2}}=\frac{\sqrt{\tilde{m}^2+\tilde{\kappa}^2}}{z}+\mathcal{O}(z^0)~.
\end{equation}
Thus the mode functions behave as $u, v\sim e^{\pm i\int E_\pm (z)dz}\sim e^{\mp i\sqrt{\tilde{m}^2+\tilde{\kappa}^2}t}$. This oscillation frequency indeed agrees with the late-time behavior of (\ref{fermionEoM}). Therefore, higher orders in the super-adiabatic basis do not offer any $\mathcal{O}(\tilde{m}^{-n})$ corrections to the oscillation phase, hence to the production history. The production amount, time and width can simply be obtained by replacing $\mu\to \sqrt{\tilde{m}^2+\tilde{\kappa}^2}$ in (\ref{dSSpin1ProdAmountAfterResum}-\ref{dSSpin1ProdWidthAfterResumey}). Its Stokes lines are shown together with spin-1 particles in FIG.~\ref{dS01StokesLines}.

\begin{figure}[h!]
	\centering
	\includegraphics[width=17cm]{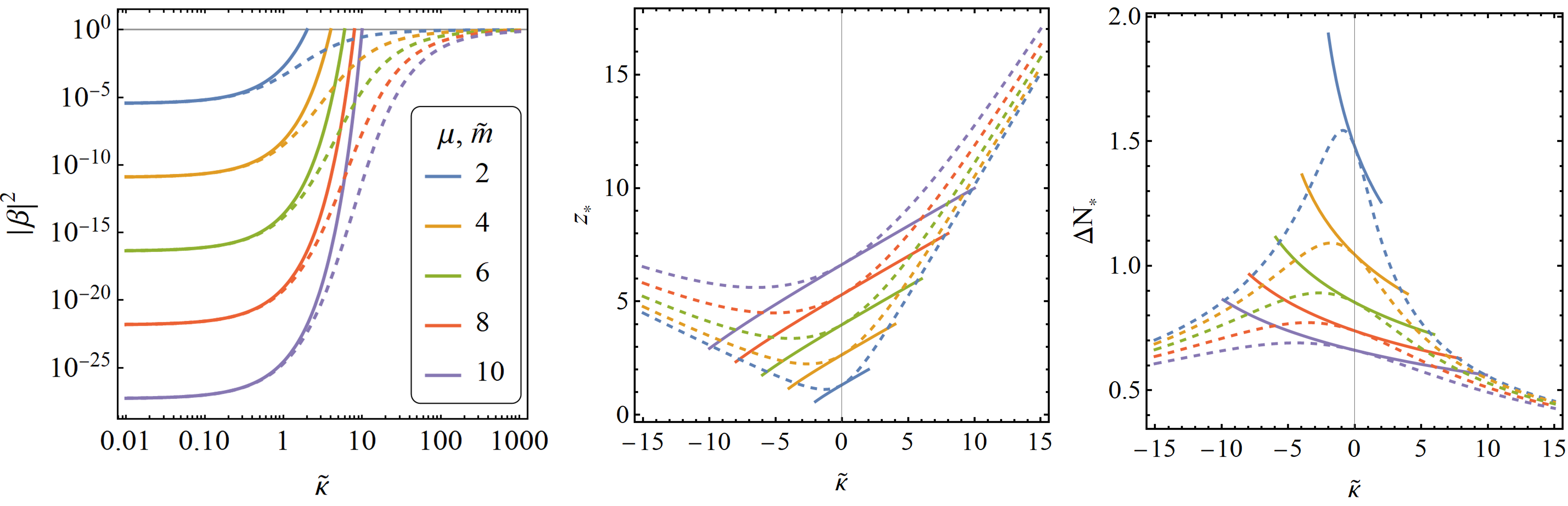}\\
	\caption{The production histories in exact dS. Left panel: Production amount as a function of the dimensionless chemical potential for different particle masses. Middle panel: The $z$-domain production time dependence on chemical potential and mass. Right panel: Production width measured in e-folding numbers. In all three plots, solid lines represent spin-1 bosons while dashed lines represent spin-1/2 fermions. Particles with different masses are distinguished by the colors of the lines according to the legend in the left panel. For bosons, we limit the range of chemical potential to be smaller than the mass, so that no tachyonic instability is induced. For fermions, the chemical potential is not restricted and we allow it to take arbitrarily large values.}\label{dS01ProdHistory}
\end{figure}

Finally, we plot the parameter dependence of production histories for both bosons and fermions in FIG.~\ref{dS01ProdHistory}. From the plots, one can see that bosons and fermions share similar production histories when the chemical potential is small. However, they begin to depart from each other as $|\tilde{\kappa}|$ becomes large, and in the end their large chemical-potential limits are drastically different: Bosons enter the tachyonic regime, while fermions saturate and approach an asymptotic limit where $z_*\propto |\kappa|$ and $\Delta N_*\propto |\kappa|^{-1/2}$. Note that for bosons, both $z_*$ and $\Delta N_*$ are monotonic functions of the chemical potential, whereas neither are monotonic for fermions.

\subsection{Deviation from dS of the $\epsilon$-type}
The exact and rigid dS spacetime is maximally symmetric with a simple time dependence in the scale factor. The mode functions for free fields are exactly solvable, even though knowing the production time and width still requires the technique of smoothed Stokes phenomenon. However, dS only works as a leading order approximation to certain stages of cosmic evolution such as inflation or dark-energy dominated era. The actual evolution can deviate from that of dS in different ways, with distinctive impacts on the particle production history that we are after. In this subsection and the next, we will focus on two simplest ways to deform the dS geometry, namely with a constant $\epsilon$ parameter and with a constant $\eta$ parameter. We will call these deviations the $\epsilon$-type and the $\eta$-type.

The simplest kind of deformation is to introduce a (small) constant $\epsilon$ parameter,
\begin{equation}
	\epsilon(\tau)=-\frac{H'(\tau)}{a(\tau)H^2(\tau)}=\epsilon=\text{const}~,~~~H(\tau)=\frac{a'(\tau)}{a^2(\tau)}~.
\end{equation}
Integrating over $\tau$ yields a scale factor
\begin{equation}
	a(\tau)=\left(-H_p\tau\right)^{-\frac{1}{1-\epsilon}}~,
\end{equation}
where $H_p\equiv H(\tau_p)$ is the Hubble parameter evaluated at the time $\tau_p$ when the scale factor is $a(\tau_p)=(1-\epsilon)^{-1/\epsilon}$. Because the Hubble parameter is decreasing with conformal time as a power-law, modes that exit the horizon experience a slightly different gravitational background. This soft breaking of scale invariance, as we will see, manifests itself in the scale dependence of production history.

\begin{figure}[h!]
	\centering
	\includegraphics[width=15cm]{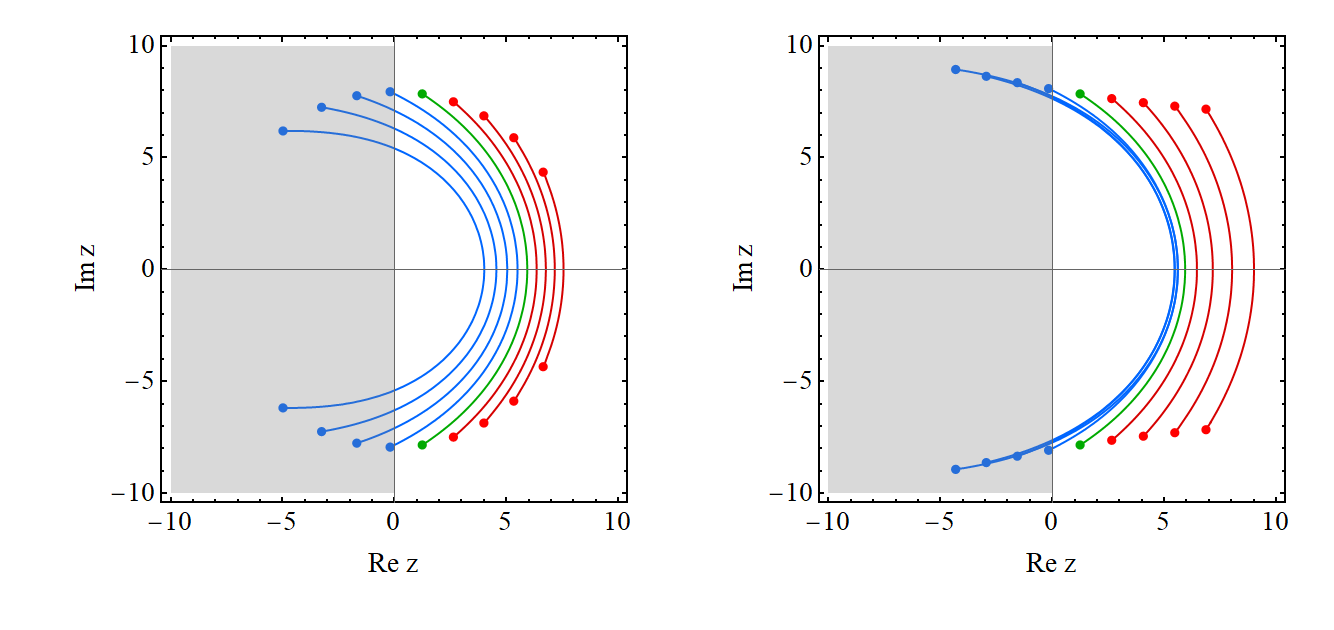}~~~~~~~~~~~~~~~~~~~~\\
	\caption{The Stokes lines for spin-1 bosons (left panel) and spin-1/2 fermions (right panel) in dS with $\epsilon$-type deviation and $\epsilon=0.1$. In both panels, the momentum $k$ is fixed to have a dimensionless chemical potential ranging over $\tilde{\kappa}_p=-8,-6,\cdots,6,8$ (from left to right, and the green lines correspond to the case without chemical potential), with spin-1 boson mass and spin-1/2 fermion mass $\mu_p=\tilde{m}_p=10$. The gray region with $\Re z<0$ cannot be reached physically.
	}\label{dSEpsStokesLines}
\end{figure}

Defining the dimensionless variable $z=-k\tau$, the EoM of a massive spin-1 particle reads
\begin{equation}
	\frac{d^2 f(z)}{dz^2}+w^2(z,k)f(z)=0~,~w^2(z,k)=1-\frac{2\tilde\kappa_p(k)}{z^{\frac{1}{1-\epsilon}}} +\frac{\tilde m_p^2(k)}{z^{\frac{2}{1-\epsilon}}}~,
\end{equation}
where we have denoted the scale-dependent dimensionless mass and chemical potential as
\begin{equation}
	\tilde\kappa_p(k)=\frac{\kappa}{H_p}\left(\frac{k}{H_p}\right)^{\frac{\epsilon}{1-\epsilon}},~~~\tilde m_p(k)=\frac{m}{H_p}\left(\frac{k}{H_p}\right)^{\frac{\epsilon}{1-\epsilon}}~.\label{scaleDepMassAndChemPtl}
\end{equation}
The complex turning points lie at $z_c$, $z_c^*$, with
\begin{equation}
	z_c=\left(\tilde{\kappa}_p+ i\sqrt{\tilde{m}_p^2-\tilde{\kappa}_p^2}\right)^{1-\epsilon}~.
\end{equation}
The phase integral is
\begin{eqnarray}
	\nonumber\int w(z,k) dz&=&
	(1-\epsilon)\int \frac{dz'}{z'^\epsilon}\sqrt{1-\frac{2\tilde\kappa_p}{z'} +\frac{\tilde m_p^2}{z'^2}}\\
	\nonumber&=&-\frac{z (1-\epsilon )}{\epsilon }\sqrt{1-\frac{2\tilde\kappa_p}{z^{\frac{1}{1-\epsilon}}} +\frac{\tilde m_p^2}{z^{\frac{2}{1-\epsilon}}}}\\
	\nonumber&&+\frac{z}{\epsilon }F_1\left(-1+\epsilon ;\frac{1}{2},\frac{1}{2};\epsilon ;z^{\frac{1}{-1+\epsilon }} \left(\tilde{\kappa
	}_p+i \sqrt{\tilde{m}_p^2-\tilde{\kappa }_p^2}\right),z^{\frac{1}{-1+\epsilon }} \left(\tilde{\kappa }_p-i
	\sqrt{\tilde{m}_p^2-\tilde{\kappa }_p^2}\right)\right)\\
	\nonumber&&+\frac{z^{\frac{\epsilon }{-1+\epsilon }} (1-\epsilon )  \tilde{\kappa }_p}{\epsilon ^2}F_1\left(\epsilon ;\frac{1}{2},\frac{1}{2};1+\epsilon
	;z^{\frac{1}{-1+\epsilon }} \left(\tilde{\kappa }_p+i \sqrt{\tilde{m}_p^2-\tilde{\kappa
		}_p^2}\right),z^{\frac{1}{-1+\epsilon }} \left(\tilde{\kappa }_p-i \sqrt{\tilde{m}_p^2-\tilde{\kappa
		}_p^2}\right)\right)~.\\
	&& 
\end{eqnarray}
Here $F_1(a;b_1,b_2;c;x,y)$ is the Appell hypergeometric function defined by the double series
\begin{equation}
	F_1(a;b_1,b_2;c;x,y)=\sum _{m=0}^{\infty } \sum _{n=0}^{\infty }
	\frac{ (a)_{m+n} (b_1)_m
		(b_2)_n}{(c)_{m+n} }\frac{x^m y^n}{m! n!}~,
\end{equation}
where $(a)_n=\frac{\Gamma(a+n)}{\Gamma(a)}$ is the Pochhammer symbol. The Stokes lines are plotted in FIG.~\ref{dSEpsStokesLines}. Comparing to FIG.~\ref{dS01StokesLines}, one finds that with a non-zero $\epsilon$, the distribution of turning points becomes asymmetric around the imaginary axis and the Stokes lines become more squeezed with a larger positive chemical potential.

\begin{enumerate}
	\item[$\bullet$] Production amount. For $z_i$ lying on the real axis, the above integral is real, suggesting no contribution to the particle production amount $e^{-2\Re F(z_i)},z_i\in \mathbb{R}$. Therefore, the particle production amount only receives contribution at the lower end $z_c$. Converting this to an integral along the whole Stokes line, we obtain
	\begin{eqnarray}
		\nonumber 2\Re F(z_i)&=&2i\int_{z_c}^{z_c^*}w(z,k)dz\\
		\nonumber&=&-2 \sqrt{\pi } (1-\epsilon ) \Gamma (-\epsilon ) \tilde{m}_p^{1-2 \epsilon }\\
		&&\times\Im\Bigg[\left(\tilde{\kappa }_p+i
		\sqrt{\tilde{m}_p^2-\tilde{\kappa }_p^2}\right){}^{\epsilon } \, _2\tilde{F}_1\left(-\frac{1}{2},-\epsilon
		;\frac{3}{2}-\epsilon ;-1+\frac{2 \tilde{\kappa }_p \left(\tilde{\kappa }_p-i \sqrt{\tilde{m}_p^2-\tilde{\kappa
				}_p^2}\right)}{\tilde{m}_p^2}\right)\Bigg]~.\label{dSEpsProdAmountFull}
	\end{eqnarray}
	For $\frac{k}{H_p}$ not far away from unity, a small dS-deformation parameter $\epsilon\ll 1$ can be used as an expansion parameter. After a further expansion in powers of $\tilde{\kappa}_p$ and resummation, we find
	\begin{equation}
		\nonumber 2\Re F(z_i)=2 \pi  \left[\tilde{m}_p-\tilde{\kappa }_p+\epsilon  \left(\left(\tilde{m}_p+\tilde{\kappa
		}_p\right) \ln \frac{\tilde{m}_p+\tilde{\kappa }_p}{2}-2 \tilde{m}_p \ln\tilde{m}_p\right)+\mathcal{O}(\epsilon^2)\right]~.
	\end{equation}
	Therefore, the particle production amount is now scale dependent:
	\begin{eqnarray}
		\nonumber|\beta(k)|^2&=&e^{ -2 \pi  \left[\mu_p-\tilde{\kappa }_p+\epsilon  \left(\left(\tilde{m}_p+\tilde{\kappa
			}_p\right) \ln \frac{\tilde{m}_p+\tilde{\kappa }_p}{2}-2 \tilde{m}_p \ln\tilde{m}_p+\mathcal{O}(\tilde{m}_p^{-1})\right)+\mathcal{O}(\epsilon^2)\right]}\\
		&=&\exp\left[-\frac{2\pi}{H_p}\left(\sqrt{m^2-\frac{H_p^2}{4}}-\kappa\right) \left(1+\epsilon  \ln
		\frac{k}{m}+\epsilon\frac{m+\kappa}{m-\kappa}\ln
		\frac{m+\kappa}{2 m}+\mathcal{O}\left(\frac{\epsilon H_p^2}{m^2},\epsilon^2\right)\right)\right]~.\label{dSEpsProdAmount}
	\end{eqnarray}
	where we have resummed the $\mathcal{O}(\epsilon^0)$ super-adiabatic corrections to the mass and replaced
	\begin{equation}
		\tilde{m}_p(k)\to\mu_p(k)=\sqrt{\frac{m^2}{H_p^2}-\frac{1}{4}}\left(\frac{k}{H_p}\right)^{\frac{\epsilon}{1-\epsilon}}~.
	\end{equation}
	The $\mathcal{O}(\epsilon)$ order, however, cannot be treated in the same way as in (\ref{dSBetaResummed}), because of the presence of a branch cut extending from the origin all the way to infinity. Therefore, considering the fact that the adiabatic parameter is $\frac{w'(z)}{w(z)^2}\sim \tilde{m}_p^{-2}$, we expect that the relative error of (\ref{dSEpsProdAmount}) is of order $\frac{\epsilon H_p^2}{m^2}$, which is negligible if the mass $m$ is large.
	
	For a positive $\epsilon$, the production amount necessarily drops with scale $k$, because the effective mass in the Boltzmann factor is measured in units of the time-dependent Hubble parameter, which is decreasing during inflation. 
	\item[$\bullet$] Production time. This can only be solved numerically from $\Im F(z_*)=0$. However, by an educated guess, we found an empirical formula that describes the $\mathcal{O}(\epsilon)$ contribution to $z_*$ very well, with an error of 2\% on average (see FIG.~\ref{dSEpsErrorDisplay} for example). Namely,
	\begin{eqnarray}
		\nonumber z_*(\tilde{m}_p,\tilde{\kappa}_p,\epsilon)&\equiv&z_*^{(0)}+z_*^{(1)}+\cdots\\
		\nonumber&\approx&0.6627 \mu_p+0.3435 \tilde{\kappa }_p-\frac{0.0102 \tilde{\kappa }_p^2}{\mu_p}+\frac{0.0064 \tilde{\kappa }_p^3}{\mu_p^2}+\cdots\\
		\nonumber &&+\epsilon 
		\left[0.82\tilde{m}_p-0.42\tilde{\kappa}_p-0.32\tilde{m}_p\ln\tilde{m}_p-\left(0.33\tilde{m}_p+0.30\tilde{\kappa}_p\right)\ln(\tilde{m}_p+\tilde{\kappa}_p)+\mathcal{O}\left(\tilde{m}_p^{-1}\right)\right]\\
		&&+\mathcal{O}(\epsilon^2)~,\label{dSEpsProdTime}
	\end{eqnarray}
	Here the first line ($z_*^{(0)}$) has the same form as the exact solution found in dS spacetime and the second line ($z_*^{(1)}$) represents the leading-order slow-roll correction as an empirical formula, with an uncertainty due to the unresummed super-adiabatic corrections.
	\begin{figure}[h!]
		\centering
		\includegraphics[width=10cm]{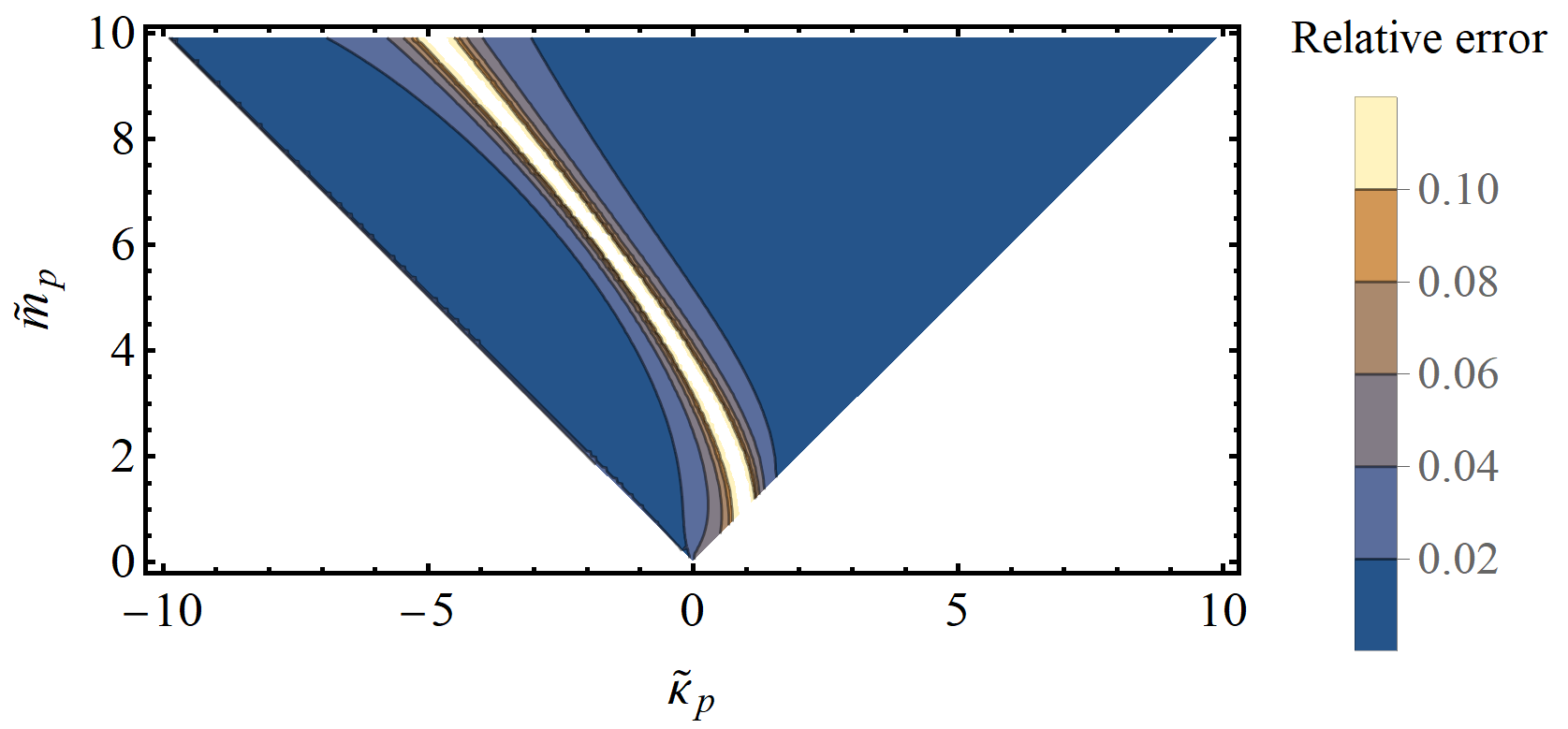}\\
		\caption{The relative error of the $\mathcal{O}(\epsilon)$ empirical formula (\ref{dSEpsProdTime}) as compared to the numeric result. The bright band cutting across the plot is where the $\mathcal{O}(\epsilon)$ correction crosses zero, hence the large relative error.}\label{dSEpsErrorDisplay}
	\end{figure}
	
	\item[$\bullet$] Production width. The production width can also be expressed partially analytically. On the real axis, the singulant function accumulates no real parts and therefore $\Re F(z_*)=\Re F(z_i)$. Plugging in the expression for $\Im F'(z_*)$, we obtain
	\begin{equation}
		\Delta z_*=\frac{2\sqrt{2|\Re F(z_*)|}}{|\Im F'(z_*)|}=\sqrt{\frac{2\pi\left[\mu_p-\tilde{\kappa }_p+\epsilon  \left(\left(\tilde{m}_p+\tilde{\kappa
				}_p\right) \ln \frac{\tilde{m}_p+\tilde{\kappa }_p}{2}-2 \tilde{m}_p \ln\tilde{m}_p\right)\right]}{1-\frac{2\tilde\kappa_p}{z_*^{\frac{1}{1-\epsilon}}} +\frac{\tilde m^2_p}{z_*^{\frac{2}{1-\epsilon}}}}}~.
	\end{equation}
	When expressed in units of e-folding numbers and expanded into powers of $\frac{\kappa}{m}$, we have the empirical expression
	\begin{eqnarray}
		\nonumber&& \Delta N_*(k;m,\kappa,\epsilon)\\
		\nonumber&=&\frac{1}{1-\epsilon}\frac{\Delta z_*}{z_*}\\
		\nonumber&\equiv&\Delta N_*^{(0)}+\Delta N_*^{(1)}+\cdots\\
		\nonumber&\approx&\frac{1}{(m^2/H_p^2-1/4)^{1/4}}\Bigg\{2.0895-\frac{0.4131
			\kappa}{(m^2-H_p^2/4)^{1/2}}+\frac{0.1323
			\kappa^2}{m^2-H_p^2/4}-\frac{0.0523
			\kappa^3}{(m^2-H_p^2/4)^{3/2}}+\cdots\\
		\nonumber&&~~~~~~~~~~~~~~~~~~~~~~~~~+\epsilon \left[-\frac{1}{2}\left(2.0895-\cdots\right) \ln\frac{k}{m}+\frac{1.4\kappa}{m}-\frac{0.70 \kappa ^2}{m^2}+\frac{0.32\kappa^3}{m^3}+\mathcal{O}\left(\frac{H_p^2}{m^2}\right)\right]\\
		&&~~~~~~~~~~~~~~~~~~~~~~~~+\mathcal{O}(\epsilon^2)\Bigg\}~.\label{dSEpsProdWidth}
	\end{eqnarray}
	Again, the first line ($=\Delta N_*^{(0)}$) is exact while the second line ($=\Delta N_*^{(1)}$) is approximate, with the exception of the coefficient before the running term $\ln\frac{k}{m}$, which inherits its exactness from the first line.

\end{enumerate}

\begin{figure}[h!]
	\centering
	\includegraphics[width=17cm]{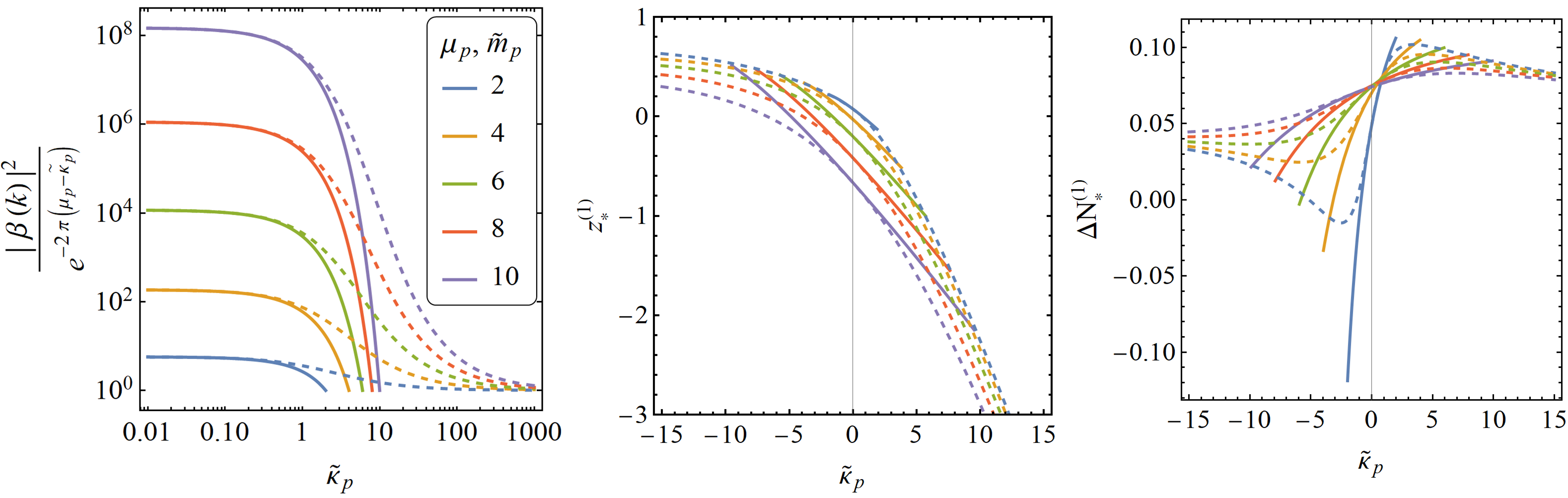}\\
	\caption{The $\mathcal{O}(\epsilon)$ corrections to production histories in deformed dS with $\epsilon=0.1$ . Left panel: Production amount excluding zeroth-order contribution as a function of the dimensionless chemical potential for different particle masses. Middle panel: The correction to $z$-domain production time dependence on chemical potential and mass. Right panel: The correction to production width measured in e-folding numbers. In all three plots, solid lines represent spin-1 bosons while dashed lines represent spin-1/2 fermions. Particles with different masses are distinguished by the colors of the lines according to the legend in the left panel. For bosons, we limit the range of chemical potential to be smaller than the mass, so that no tachyonic instability is induced. For fermions, the chemical potential is not restricted and we allow it to take arbitrarily large values.}\label{dSEpsProdHistory}
\end{figure}

The fermionic case is then easily obtained by applying the replacement $\mu_p,\tilde{m}_p\to\sqrt{\tilde{m}_p^2+\tilde{\kappa}_p^2}$ to the above results. Since the zeroth order in $\epsilon$ is the same as the exact dS result with a Hubble constant $H_p$, we focus on the $\mathcal{O}(\epsilon)$ corrections due to the deformation and plot them in FIG.~\ref{dSEpsProdHistory}. As shown in the plots, the behavior of bosons and fermions are again different for large chemical potentials. 

Before ending this subsection, we note that the validity of our perturbative expansion is actually controlled by $\epsilon \ln\max\{\tilde{m}_p,\tilde{\kappa}_p\}$ instead of just $\epsilon$. Thus, for particles with extremely large mass or chemical potential, $i.e.$, $\tilde{m}_p,\tilde{\kappa}_p\gtrsim e^{1/\epsilon}$, perturbation theory fails and one would have to rely on the full result (\ref{dSEpsProdAmountFull}) for production amount and numerically solve production time and width. For parameters chosen in FIG.~\ref{dSEpsProdHistory}, the corrections are small compared to the zeroth order dS results, suggesting the perturbative expansion is valid.

\subsection{Deviation from dS of the $\eta$-type}
The second type of deformation of dS is obtained from introducing a weak time dependence in $\epsilon(\tau)$ described by a non-zero $\eta$ parameter,
\begin{equation}
	\eta(\tau)=\frac{\epsilon'(\tau)}{a(\tau)H(\tau)\epsilon(\tau)}~.
\end{equation} 
A further simplification appears if $\epsilon(\tau)\ll \eta(\tau)$ for $\tau$ lying in the range of interest. We will call this type of deviation the $\eta$-type. It is reasonable to analyze particle production in such a scenario since for inflation, the Planck 2018 data \cite{Akrami:2018odb} favors a smaller first slow-roll parameter compared to the second slow-roll parameter. 

\begin{figure}[h!]
	\centering
	\includegraphics[width=15cm]{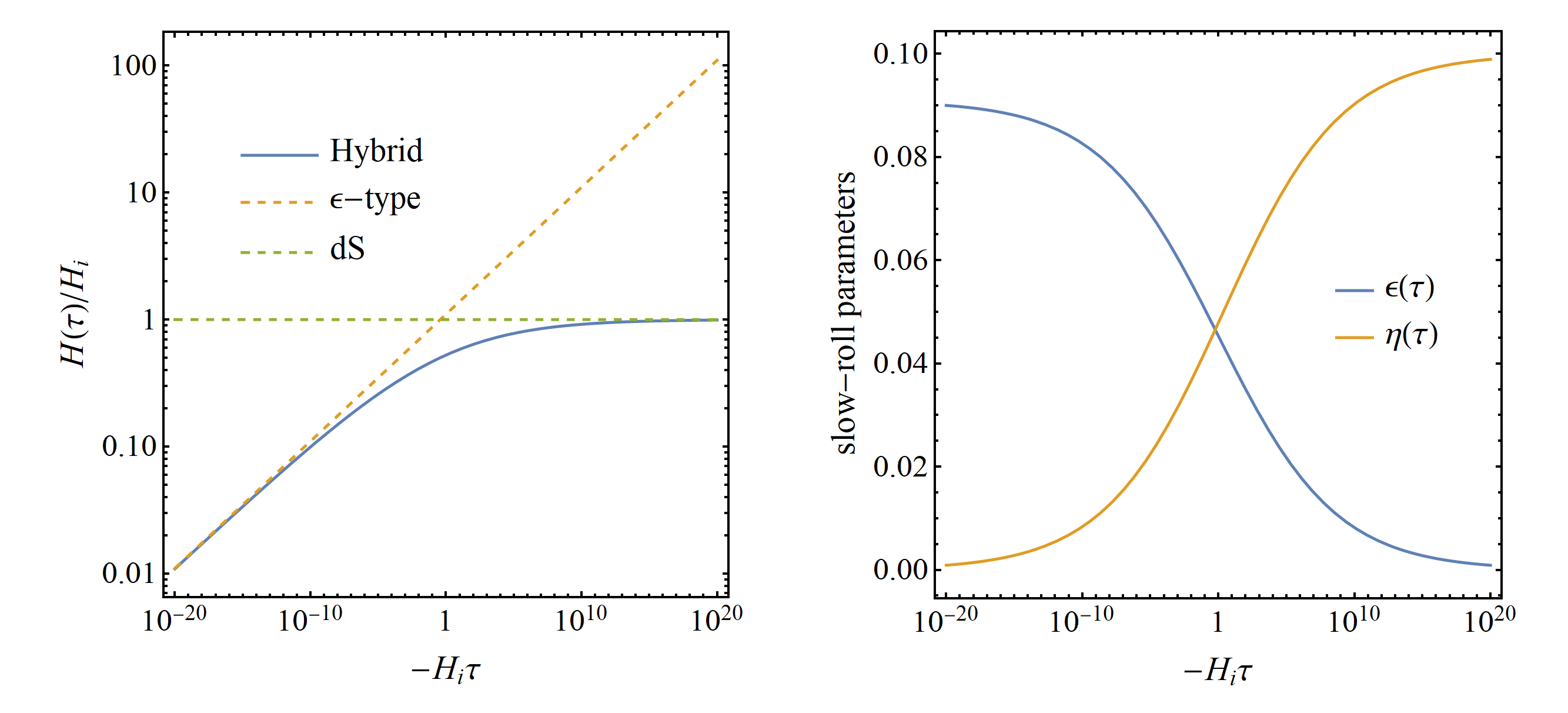}\\
	\caption{The Hubble parameter and slow-roll parameters of the hybrid model (\ref{etaTypeDef}). Here we have chosen $\eta_i=0.1$ for illustration. Clearly, the hybrid model is separated into two stages, with early time dominated by $\eta$-type and late time dominated by $\epsilon$-type.}
	\label{hybridIllustration}
\end{figure}

To study $\eta$-type deviation from dS, we suppose the scale factor can be expressed as a series with variable $-H_i \tau$, where $H_i$ has the dimension of Hubble. So the leading-order deviation can be approximated as
\begin{align}
	a(\tau)=-\frac{1}{H_i \tau}+\frac{1}{(-H_i \tau)^{1+\eta_i}} \ ,`\label{etaTypeDef}
\end{align}
with $\eta_i>0$ and $\tau<0$. With this scale factor, the Hubble parameter is 
\begin{align}
	H(\tau)=\frac{H_i (-H_i \tau )^{\eta_i} \left[1+\eta_i+(-H_i \tau )^{\eta_i}\right]}{\left[1+(-H_i \tau )^{\eta_i}\right]^2} \ ,
\end{align} and therefore the physical meaning of $H_i$ is the Hubble parameter at $\tau\to -\infty$. On the other hand, the first and second slow-roll parameters are
\begin{align}
	\epsilon(\tau)&=\eta_i \frac{1+\eta_i+(1-\eta_i)(-H_i\tau)^{\eta_i}}{\left[1+\eta_i+(-H_i \tau)^{\eta_i}\right]^2} \nonumber \\
	\eta(\tau)&=\eta_i\frac{(-H_i \tau)^{\eta_i}\left[1+(-H_i \tau)^{\eta_i}\right]\left[(1+\eta_i)^2+(1-\eta_i)(-H_i \tau)^{\eta_i}\right]}{\left[1+\eta_i+(-H_i \tau)^{\eta_i}\right]^2\left[1+\eta_i+(1-\eta_i)(-H_i \tau)^{\eta_i}\right]} \ .
\end{align}
The asymptotic behaviors of the slow-roll parameters in the early-/late- time limit are 
\begin{eqnarray}
	(\epsilon(\tau),\eta(\tau))\to\begin{cases}
		(0,\eta_i)&,~(-H_i \tau)^{\eta_i}\gg 1\\
		\left(\frac{\eta_i}{1+\eta_i},0\right)&,~(-H_i \tau)^{\eta_i}\ll 1~.
	\end{cases}
\end{eqnarray}
Thus (\ref{etaTypeDef}) describes a hybrid spacetime in which the deviation starts as $\eta$-type and then becomes $\epsilon$-type (see FIG.~\ref{hybridIllustration} for illustration). In this section, we will be interested in the early stage where $\eta$-type deviation dominates.

We first evaluate the phase integrals in the limit with $(-H_i \tau)^{\eta_i}\gg 1$. With the variable $z\equiv-k\tau$, the time-dependent frequency is
\begin{align}
	w^2(z,k)=1-2\frac{\tilde{\kappa}}{z}\left(1+\frac{c(k)}{z^{\eta_i}}\right)+\frac{\tilde{m}^2}{z^2}\left(1+\frac{c(k)}{z^{\eta_i}}\right)^2 \ , \label{eq:omega2_eta}
\end{align}
where $\tilde{\kappa}\equiv\frac{\kappa}{H_i}$, $\tilde{m}\equiv\frac{m}{H_i}$ and $c(k)\equiv(\frac{k}{H_i})^{\eta_i}$. The phase integral with this frequency cannot be evaluated analytically. However, for the modes which cross the horizon at $z=-k\tau\sim 1$ early, when $i.e.$, $(-H_i\tau)^{-\eta_i}=c(k)z^{-\eta_i}\sim c(k)\ll 1$, we can solve perturbatively. This suggests expanding the frequency in the series of $c(k)$,
\begin{align}
	w(z,k)=\sqrt{1-\frac{2\tilde{\kappa}}{z}+\frac{\tilde{m}^2}{z^2}}
	+\frac{c(k) z^{-\eta_i-1} \left(\tilde{m}^2-\tilde{\kappa}  z\right)}{\sqrt{\tilde{m}^2+z (z-2 \tilde{\kappa}
			)}}+\mathcal{O}\left((c(k)z^{-\eta_i})^2\right) \ .
\end{align}
Thus the phase integral reads
\begin{align}
	\int w(z,k)dz=&\sqrt{\tilde{m}^2+z^2-2 \tilde{\kappa}  z}+m \tanh ^{-1}\left(\frac{\tilde{\kappa}  z-\tilde{m}^2}{\tilde{m} \sqrt{\tilde{m}^2+z^2-2
			\tilde{\kappa}  z}}\right)+\tilde{\kappa}  \tanh ^{-1}\left(\frac{\tilde{\kappa} -z}{\sqrt{\tilde{m}^2+z (z-2 \tilde{\kappa}
			)}}\right) \nonumber \\
	&+\frac{c(k) z^{-\eta _i}}{\tilde{m}}  \Bigg[\frac{z^2 F_1\left(2-\eta _i;\frac{1}{2},\frac{1}{2};3-\eta
		_i;\frac{z \left(\tilde{\kappa}+i \sqrt{\tilde{m}^2-\tilde{\kappa
				}^2}\right)}{\tilde{m}^2},\frac{z \left(\tilde{\kappa}-i
			\sqrt{\tilde{m}^2-\tilde{\kappa}^2}\right)}{\tilde{m}^2}\right)}{\eta
		_i-2} \nonumber \\
	&~~~~~~~~~~~~~~~~~-\frac{z \tilde{\kappa} F_1\left(1-\eta _i;\frac{1}{2},\frac{1}{2};2-\eta
		_i;\frac{z \left(\tilde{\kappa}+i \sqrt{\tilde{m}^2-\tilde{\kappa
				}^2}\right)}{\tilde{m}^2},\frac{z \left(\tilde{\kappa}-i
			\sqrt{\tilde{m}^2-\tilde{\kappa}^2}\right)}{\tilde{m}^2}\right)}{\eta
		_i-1} \nonumber \\
	&~~~~~~~~~~~~~~~~~-\frac{\tilde{m}^2 F_1\left(-\eta _i;-\frac{1}{2},-\frac{1}{2};1-\eta
		_i;\frac{z \left(\tilde{\kappa}+i \sqrt{\tilde{m}^2-\tilde{\kappa
				}^2}\right)}{\tilde{m}^2},\frac{z \left(\tilde{\kappa}-i
			\sqrt{\tilde{m}^2-\tilde{\kappa}^2}\right)}{\tilde{m}^2}\right)}{\eta
		_i}\Bigg]\\
	&+\mathcal{O}\left(c(k)^2\right) \ .
\end{align} To calculate $z_c$, we expand it in the order of $c(k)$:
\begin{align}
	z_c= z^{(0)}_c+c(k)z^{(1)}_c+\mathcal{O}(c(k)^2) \ ,
\end{align}
and the roots of (\ref{eq:omega2_eta}) satisfy
\begin{align}
	\frac{1}{z_c}\left(1+\frac{c(k)}{z_c^{\eta_i}}\right)=\frac{\tilde{\kappa}^2-i\sqrt{\tilde{m}^2-\kappa^2}}{\tilde{m}^2} \ ,
\end{align}
and thus
\begin{align}
	z_c=\tilde{\kappa}+i\sqrt{\tilde{m}^2-\kappa^2}+c(k)\left(\tilde{\kappa}+i\sqrt{\tilde{m}^2-\kappa^2}\right)^{1-\eta_i}+\mathcal{O}(c(k)^2) \ .
\end{align}

\begin{figure}[h!]
	\centering
	\includegraphics[width=15cm]{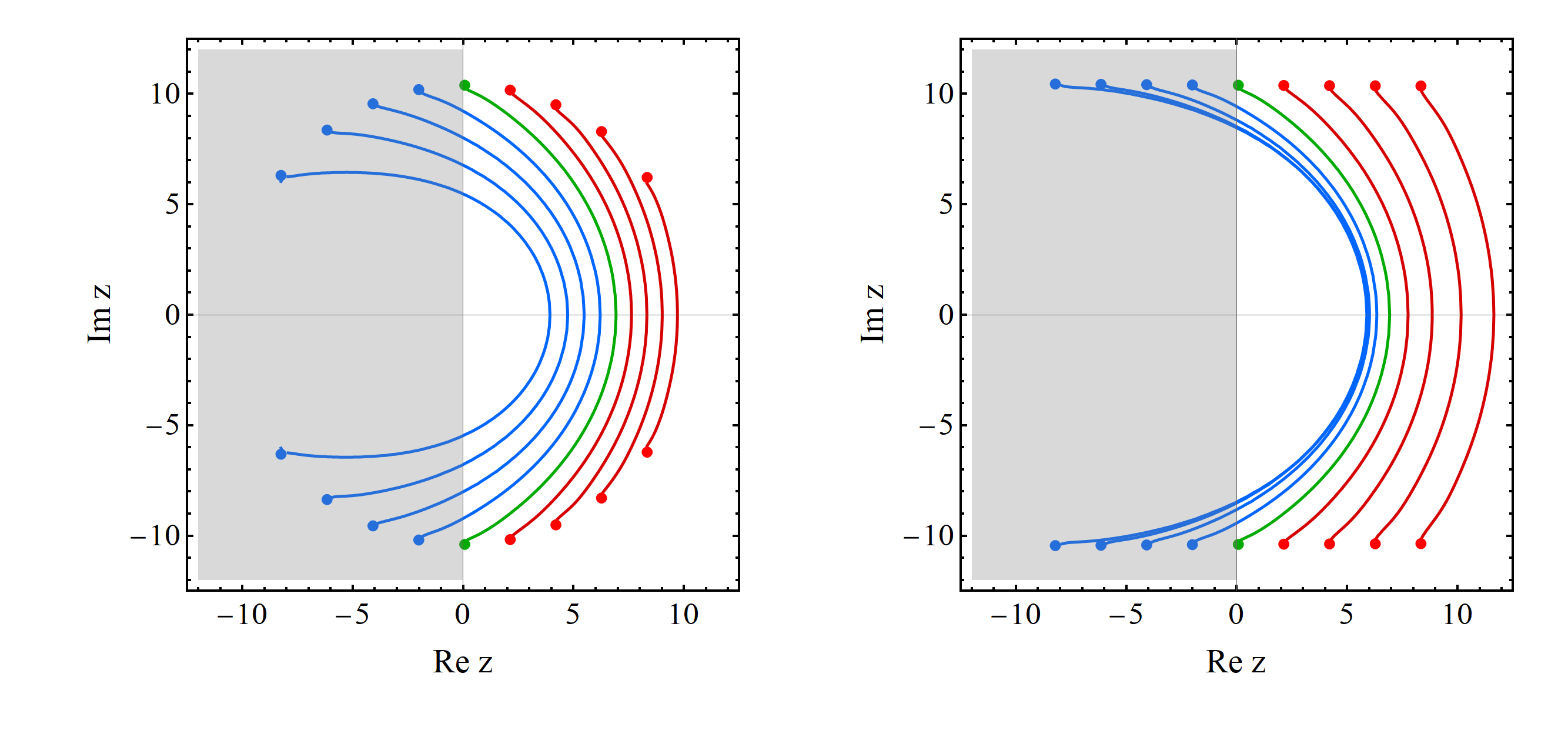}~~~~~~~~~~~~~~~~~~~~\\
	\caption{The Stokes lines for spin-1 bosons (left panel) and spin-1/2 fermions (right panel) in dS with $\eta$-type deviation and $\eta=0.1$, $c(k)=0.05$. In both panels, the dimensionless chemical potential ranging over $\tilde{\kappa}=-8,-6,\cdots,6,8$ (from left to right, and the green lines correspond to the case without chemical potential), with spin-1 boson mass and spin-1/2 fermion mass $\mu=\tilde{m}=10$. The gray region with $\Re z<0$ cannot be reached physically.
	}\label{dSEtaStokesLines}
\end{figure}

The singulant can now be solved up to the first order in $c(k)$:
\begin{align}
	F(z)&=2i\left[\int^{z}_{{z}^{(0)}_c+c(k){z}^{(1)}_c}\left(w^{(0)}(z,k)+c(k)w^{(1)}(z,k)\right)dz\right]+\mathcal{O}(c(k)^2) \nonumber \\
	&=2i\left[\int^{z}_{{z}^{(0)}_c}\left(w^{(0)}(z,k)+c(k)w^{(1)}(z,k)\right)dz-w^{(0)}(z^{(0)}_c,k)c(k){z}^{(1)}_c\right]+\mathcal{O}(c(k)^2) \nonumber \\
	&=2i\Bigg\{
	\sqrt{\tilde{m}^2+z^2-2 \tilde{\kappa}  z}+m \tanh ^{-1}\left(\frac{\tilde{\kappa}  z-\tilde{m}^2}{\tilde{m} \sqrt{\tilde{m}^2+z^2-2
			\tilde{\kappa}  z}}\right)+\tilde{\kappa}  \tanh ^{-1}\left(\frac{\tilde{\kappa} -z}{\sqrt{\tilde{m}^2+z (z-2 \tilde{\kappa}
			)}}\right) \nonumber \\
	&~~~~~~~~~~+\frac{c(k) z^{-\eta _i}}{\tilde{m}}  \Bigg[\frac{z^2 F_1\left(2-\eta _i;\frac{1}{2},\frac{1}{2};3-\eta
		_i;\frac{z \left(\tilde{\kappa}+i \sqrt{\tilde{m}^2-\tilde{\kappa
				}^2}\right)}{\tilde{m}^2},\frac{z \left(\tilde{\kappa}-i
			\sqrt{\tilde{m}^2-\tilde{\kappa}^2}\right)}{\tilde{m}^2}\right)}{\eta
		_i-2} \nonumber \\
	&~~~~~~~~~~~~~~~~~~~~~~~~~~~-\frac{z \tilde{\kappa} F_1\left(1-\eta _i;\frac{1}{2},\frac{1}{2};2-\eta
		_i;\frac{z \left(\tilde{\kappa}+i \sqrt{\tilde{m}^2-\tilde{\kappa
				}^2}\right)}{\tilde{m}^2},\frac{z \left(\tilde{\kappa}-i
			\sqrt{\tilde{m}^2-\tilde{\kappa}^2}\right)}{\tilde{m}^2}\right)}{\eta
		_i-1} \nonumber \\
	&~~~~~~~~~~~~~~~~~~~~~~~~~~~-\frac{\tilde{m}^2 F_1\left(-\eta _i;-\frac{1}{2},-\frac{1}{2};1-\eta
		_i;\frac{z \left(\tilde{\kappa}+i \sqrt{\tilde{m}^2-\tilde{\kappa
				}^2}\right)}{\tilde{m}^2},\frac{z \left(\tilde{\kappa}-i
			\sqrt{\tilde{m}^2-\tilde{\kappa}^2}\right)}{\tilde{m}^2}\right)}{\eta
		_i}\Bigg] \nonumber \\
	&~~~~~~~~~~\nonumber+\frac{\sqrt{\pi } c(k) \tilde{m} \Gamma \left(2-\eta _i\right) \left(\tilde{\kappa }+i \sqrt{\tilde{m}^2-\tilde{\kappa }^2}\right)^{-\eta _i} \, _2\tilde{F}_1\left(-\frac{1}{2},-\eta
		_i;\frac{3}{2}-\eta _i;\frac{\left(\tilde{\kappa }+i \sqrt{\tilde{m}^2-\tilde{\kappa }^2}\right)^2}{\tilde{m}^2}\right)}{2 \eta _i}
	\Bigg\}+\pi(\tilde{m}-\tilde{\kappa})\\
	&~~~+\mathcal{O}(c(k)^2) \ .
\end{align}
The Stokes lines are shown in FIG.~\ref{dSEtaStokesLines}. The distribution of Stokes lines on the complex plane is very similar to that of dS, except for a slightly decreased Hubble parameter.

\begin{enumerate}
	\item[$\bullet$] Production amount. The singulant evaluated at $z_i$ gives
	\begin{align}
		2{\rm Re}F(z_i)&=2i\left[\int^{{z^*}^{(0)}_c+c(k){z^*}^{(1)}_c}_{{z}^{(0)}_c+c(k){z}^{(1)}_c}\left(w^{(0)}(z,k)+c(k)w^{(1)}(z,k)\right)dz\right]+\mathcal{O}(c(k)^2) \nonumber \\
		&=2i\left[\int^{{z^*}^{(0)}_c}_{{z}^{(0)}_c}\left(w^{(0)}(z,k)+c(k)w^{(1)}(z,k)\right)dz+w^{(0)}({z^*}^{(0)}_c,k)c(k){z^*}^{(1)}_c-w^{(0)}(z^{(0)}_c,k)c(k){z}^{(1)}_c\right]\nonumber \\
		&~~~+\mathcal{O}(c(k)^2)\nonumber \\
		&=2\pi(\tilde{m}-\tilde{\kappa}) \nonumber \\
		&~~~~-\frac{2 \sqrt{\pi } c(k) \tilde{m} \Gamma \left(2-\eta _i\right)
		}{\eta _i}{\rm Im}\left[\left(\tilde{\kappa}+i \sqrt{\tilde{m}^2-\tilde{\kappa}^2}\right){}^{-\eta _i} \, _2\tilde{F}_1\left(-\frac{1}{2},-\eta_i;\frac{3}{2}-\eta _i;\frac{\left(\tilde{\kappa}+i
			\sqrt{\tilde{m}^2-\tilde{\kappa}^2}\right){}^2}{\tilde{m}^2}\right)\right] \nonumber \\
		&~~~~+\mathcal{O}(c(k)^2) \ .
	\end{align}
	Thus, the production amount of particles that exit the horizon during the early $\eta$-phase is computable to the linear order in $c(k)\ll 1$:
	\begin{eqnarray}
		\nonumber&&|\beta(k)|^2\\
		\nonumber&&=\exp\left\{-2\pi\Bigg[\mu-\tilde{\kappa}-\frac{ c(k) \tilde{m} \Gamma \left(2-\eta _i\right)
		}{\sqrt{\pi }\eta _i}{\rm Im}\frac{_2\tilde{F}_1\left(-\frac{1}{2},-\eta_i;\frac{3}{2}-\eta _i;\frac{\left(\tilde{\kappa}+i
				\sqrt{\tilde{m}^2-\tilde{\kappa}^2}\right){}^2}{\tilde{m}^2}\right)}{\left(\tilde{\kappa}+i \sqrt{\tilde{m}^2-\tilde{\kappa}^2}\right)^{\eta _i}} +\mathcal{O}\left(\frac{c(k)}{\tilde{m}},c(k)^2\right)\Bigg]\right\}~, \label{dSetaspin1ProdAmount}\\
	\end{eqnarray}
	where we have resummed the $\mathcal{O}(c(k)^0)$ super-adiabatic corrections to the mass and replaced
	\begin{equation}
		\tilde{m}\to\mu=\sqrt{\frac{m^2}{H_i^2}-\frac{1}{4}}~.
	\end{equation}
	The $\mathcal{O}(c(k)^1)$ super-adiabatic corrections are complicated by the branch cut from $z=0$ to $z=-\infty$ and we are not able to analyze them. Therefore, as in the $\epsilon$-deviation case, we have indicated in (\ref{dSetaspin1ProdAmount}) the presence of an $\mathcal{O}(\tilde{m}^{-2})$ relative error.
	
	\item[$\bullet$] Production time. The equation $\Im F(z_*)=0$ can only be solved numerically. However, as in the $\epsilon$-type deviation case, we have found a useful empirical formula to the first order in $\eta_i$, whose relative error is less than $2\%$ within most parameter regions. Up to $\mathcal{O}(c(k),\eta_i)$, we found
	\begin{eqnarray}
		\nonumber z_*(\tilde{m},\tilde{\kappa},\eta_i;k)&\equiv&z_*^{(0,0)}+z_*^{(1,0)}+z_*^{(1,1)}+\cdots\\
		\nonumber&\approx&(1+c(k))\left(0.6627 \mu+0.3435 \tilde{\kappa }-\frac{0.0102 \tilde{\kappa }^2}{\mu}+\frac{0.0064 \tilde{\kappa }^3}{\mu^2}+\cdots\right)\\
		\nonumber&&+c(k)\eta_i\Big(0.84 \tilde{m}-0.37 \tilde{\kappa }-\left(0.35 \tilde{m}+0.33 \tilde{\kappa }\right) \ln \left(\tilde{\kappa
		}+\tilde{m}\right)-0.30 \tilde{m} \ln\tilde{m}+\mathcal{O}(\tilde{m}^{-1})\Big)\\
		&&+\mathcal{O}(c(k)\eta_i^2,c(k)^2)~.\label{dSetaspin1ProdTime}
	\end{eqnarray}
	Here the first line ($=z_*^{(0,0)}+z_*^{(1,0)}$) is exact while the second line ($=z_*^{(1,1)}$) is approximate\footnote{Note that in this subsection, we will denote $X^{(m,n)}$ as the $\mathcal{O}(c(k)^m\eta_i^n)$ correction to a certain quantity $X$. Since $X^{(1,0)}$ follows from an exact dS spacetime with a rescaled Hubble constant, it can always be trivially obtained by rescaling $H_i$ in $X^{(0,0)}$. In contrast, $X^{(1,1)}$ represents the leading order nontrivial $\eta$-corrections.}.
	\begin{figure}[h!]
		\centering
		\includegraphics[width=10cm]{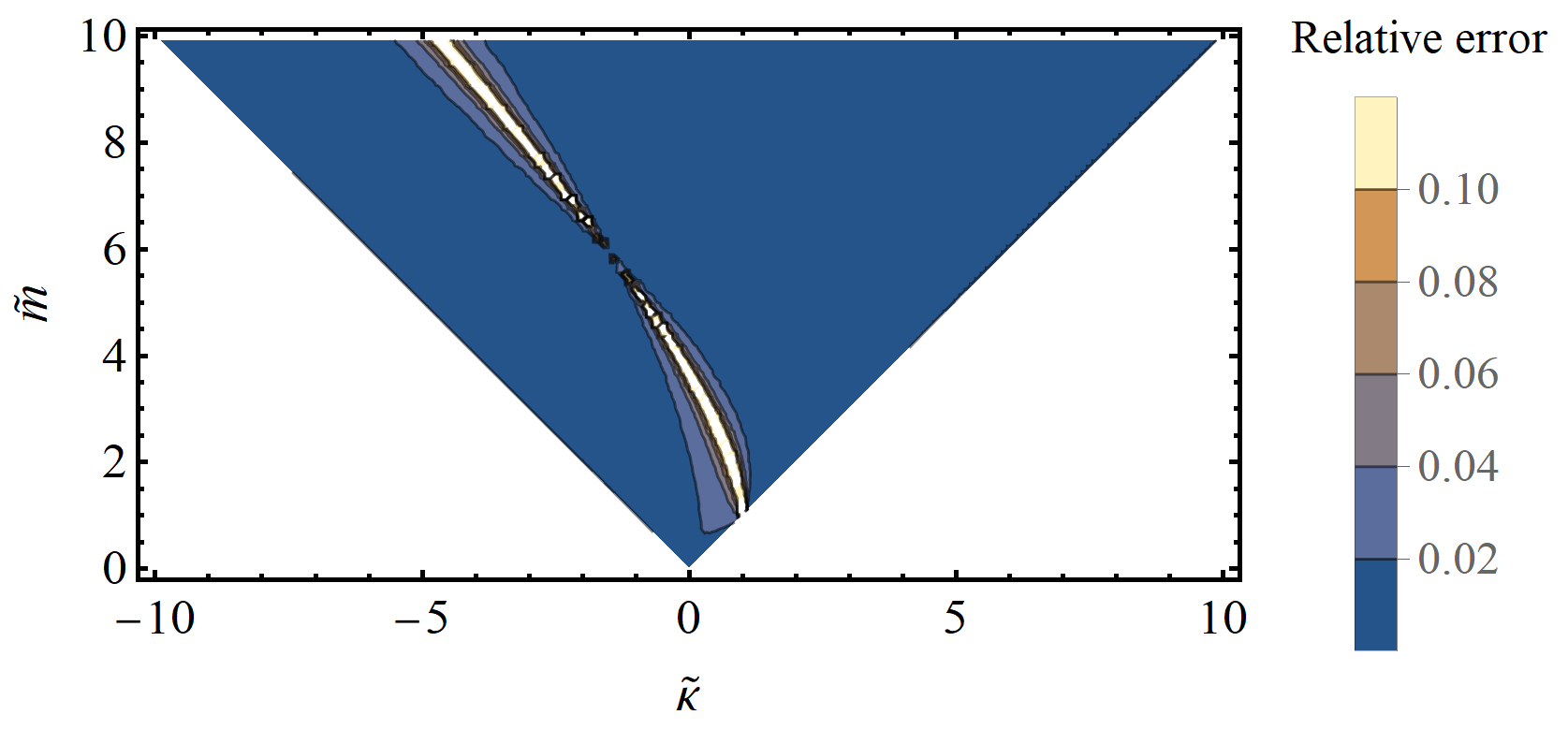}\\
		\caption{The relative error of the empirical formula (\ref{dSetaspin1ProdTime}) as compared to the numeric result. The bright band cutting across the plot is where the $z_*^{(1,1)}$ crosses zero, hence the large relative error.}
	\end{figure}
	
	\item[$\bullet$] Production width. Using the empirical formula (\ref{dSetaspin1ProdTime}), this can also be obtained approximately as a truncated power series,
	\begin{eqnarray}
		\nonumber\frac{\Delta z_*(\tilde{m},\tilde{\kappa},\eta_i;k)}{\sqrt{\tilde{m}}}&\approx&1.385+\frac{0.444 \tilde{\kappa }}{\tilde{m}}-\frac{0.07554 \tilde{\kappa }^2}{\tilde{m}^2}+\frac{0.02839
			\tilde{\kappa }^3}{\tilde{m}^3}+\cdots\\
		\nonumber&&+c(k) \Bigg\{\frac{1}{2}\left(1.385+\cdots\right)+\eta _i \Bigg[0.34-\frac{1.38
			\tilde{\kappa }}{\tilde{m}}+\frac{0.19 \tilde{\kappa }^2}{\tilde{m}^2}-\frac{0.05 \tilde{\kappa
			}^3}{\tilde{m}^3}\\
		\nonumber&&~~~~~~~~~~~~~~~~~~~~~~~~~~~~~~~~~~~~~+\left(-0.67-\frac{0.21 \tilde{\kappa }}{\tilde{m}}+\frac{0.02 \tilde{\kappa
			}^2}{\tilde{m}^2}\right) \ln\tilde{m}\Bigg]\\
		\nonumber&&~~~~~~~~~~~~~~~~~~~~~~~~~~~~~+\mathcal{O}(\eta_i^2)\Bigg\}\\
		&&+\mathcal{O}(c(k)^2)~.
	\end{eqnarray}
	In terms of e-folding numbers, the width is
	\begin{eqnarray}
		\nonumber&&\Delta N_*(m,\kappa,\eta_i;k)\\
		\nonumber&\equiv&\Delta N_*^{(0,0)}+\Delta N_*^{(1,0)}+\Delta N_*^{(1,1)}+\cdots\\
		\nonumber&\approx&\Delta z_*\times\Bigg|\frac{d}{dz}\ln\frac{1+c(k)z^{-\eta_i}}{z}\Bigg|_{z=z_*}\\
		\nonumber&\approx&\frac{1}{(\frac{m^2}{H_i^2}-\frac{1}{4})^{1/4}} \Bigg(2.089-\frac{0.4131 \kappa }{(m^2-H_i^2/4)^{1/2}}+\frac{0.1323 \kappa ^2}{m^2-H_i^2/4}-\frac{0.05226 \kappa
			^3}{(m^2-H_i^2/4)^{3/2}}\cdots\\
		\nonumber&&~~~~~~~~~~~~~+c(k) \Bigg\{-\frac{1}{2}\left(2.089-\cdots\right)\\
		\nonumber&&~~~~~~~~~~~~~~~~~~~~~~+\eta _i \Bigg[-0.04+\frac{1.4 \kappa }{m}-\frac{0.68 \kappa ^2}{m^2}+\frac{0.30 \kappa
			^3}{m^3}+\left(1.0-\frac{0.21 \kappa }{m}+\frac{0.08 \kappa ^2}{m^2}-\frac{0.04\kappa ^3}{m^3}\right)
		\ln \frac{m}{H_i}\Bigg]\\
		\nonumber&&~~~~~~~~~~~~~~~~~~~~~~+\mathcal{O}\left(\frac{\eta_i H_i^2}{m^2},\eta_i^2\right)\Bigg\}\\
		&&~~~~~~~~~~~~~~+\mathcal{O}(c(k)^2)\Bigg)~.\label{dSetaspin1ProdWidth}
	\end{eqnarray}
	Again, the first line ($=\Delta N_*^{(0,0)}$) and the second line ($=\Delta N_*^{(1,0)}$) are exact, whereas the third line ($=\Delta N_*^{(1,1)}$) is approximate.
\end{enumerate}

\begin{figure}[h!]
	\centering
	\includegraphics[width=17cm]{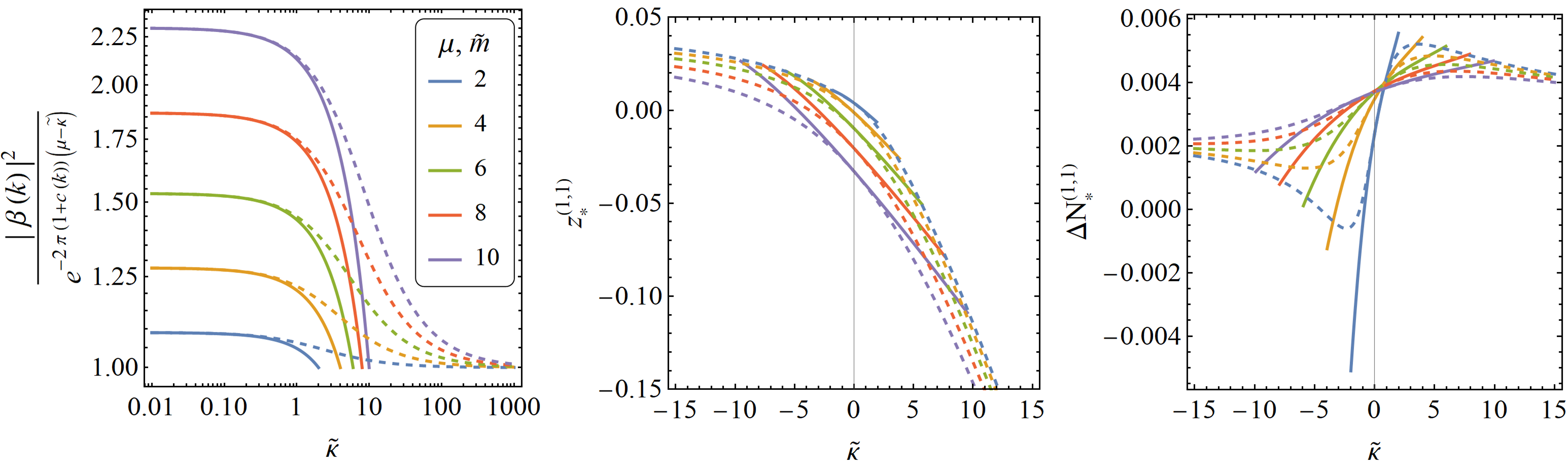}\\
	\caption{The $\mathcal{O}(c(k)\eta_i)$ corrections to production histories in deformed dS with $c(k)=0.05$ and $\eta_i=0.1$. Left panel: Production amount excluding $\mathcal{O}(\eta_i^0)$ contribution as a function of the dimensionless chemical potential for different particle masses. Middle panel: The correction to $z$-domain production time dependence on chemical potential and mass. Right panel: The correction to production width measured in e-folding numbers. In all three plots, solid lines represent spin-1 bosons while dashed lines represent spin-1/2 fermions. Particles with different masses are distinguished by the colors of the lines according to the legend in the left panel. For bosons, we limit the range of chemical potential to be smaller than the mass, so that no tachyonic instability is induced. For fermions, the chemical potential is not restricted and we allow it to take arbitrarily large values.}\label{dSEtaProdHistory}
\end{figure}

The results for fermions are again obtained via the corresponding replacement. The $\mathcal{O}(c(k)\eta_i)$ corrections to production histories in $\eta$-type deformed dS are plotted in FIG.~\ref{dSEtaProdHistory}. Comparing to FIG.~\ref{dSEpsProdHistory}, one can see that the overall behavior is similar to that of $\epsilon$-type deformed dS. Indeed, if we choose $\epsilon \sim c(k)\eta_i$, the corrections roughly match in size. This interesting fact will be discussed below. Nevertheless, we notice that there are important differences in the scale dependence of various production history parameters.

\subsubsection*{Comparison between $\epsilon$-type and $\eta$-type}
\begin{figure}[h!]
	\centering
	\includegraphics[width=12cm]{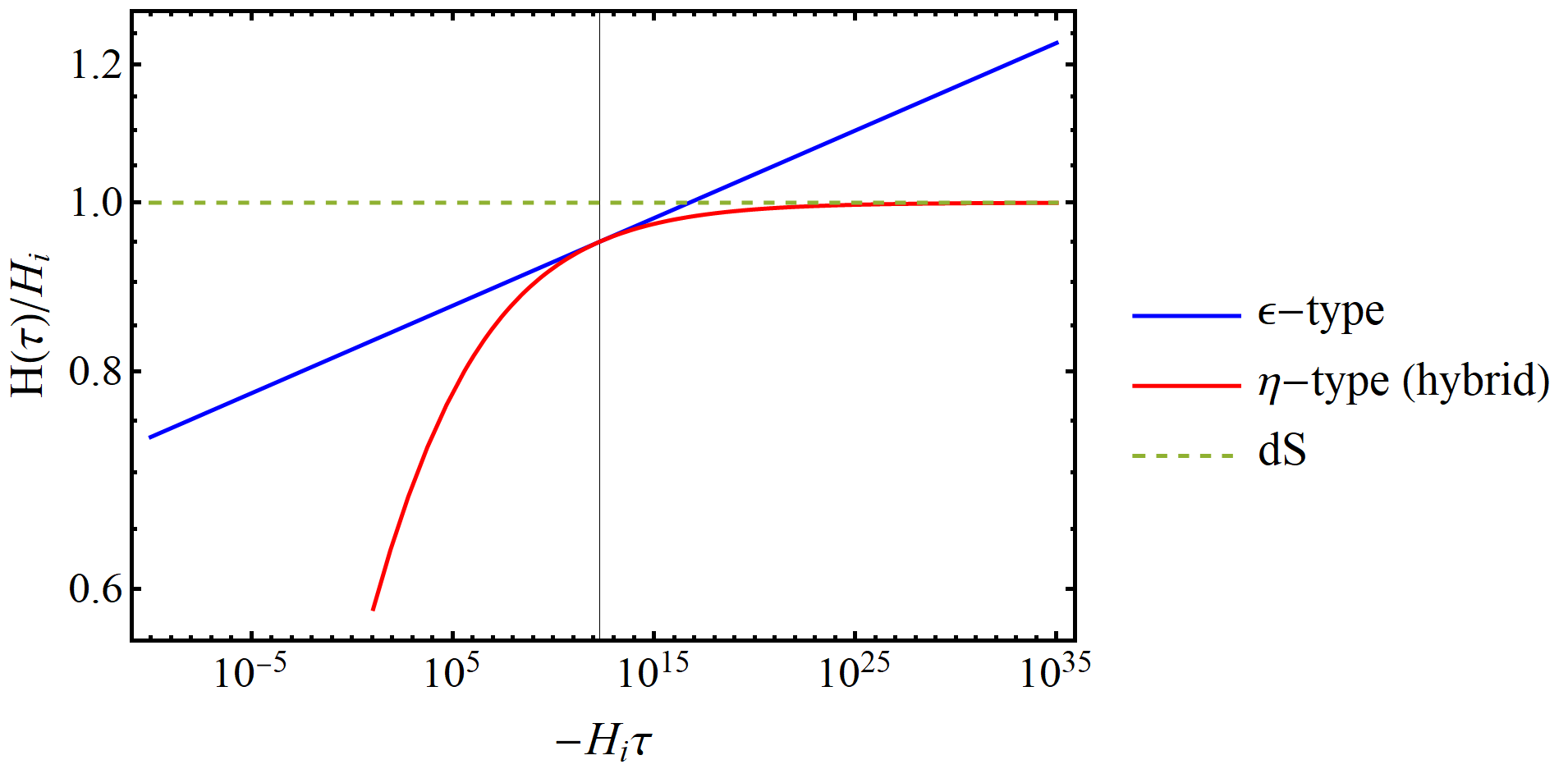}\\
	\caption{The Hubble parameter for two spacetimes are chosen to be tangent to each other at the time $\tau_0$, which is indicated by the vertical black line.}\label{Tangent}
\end{figure}

\begin{figure}[h!]
	\centering
	\includegraphics[width=15cm]{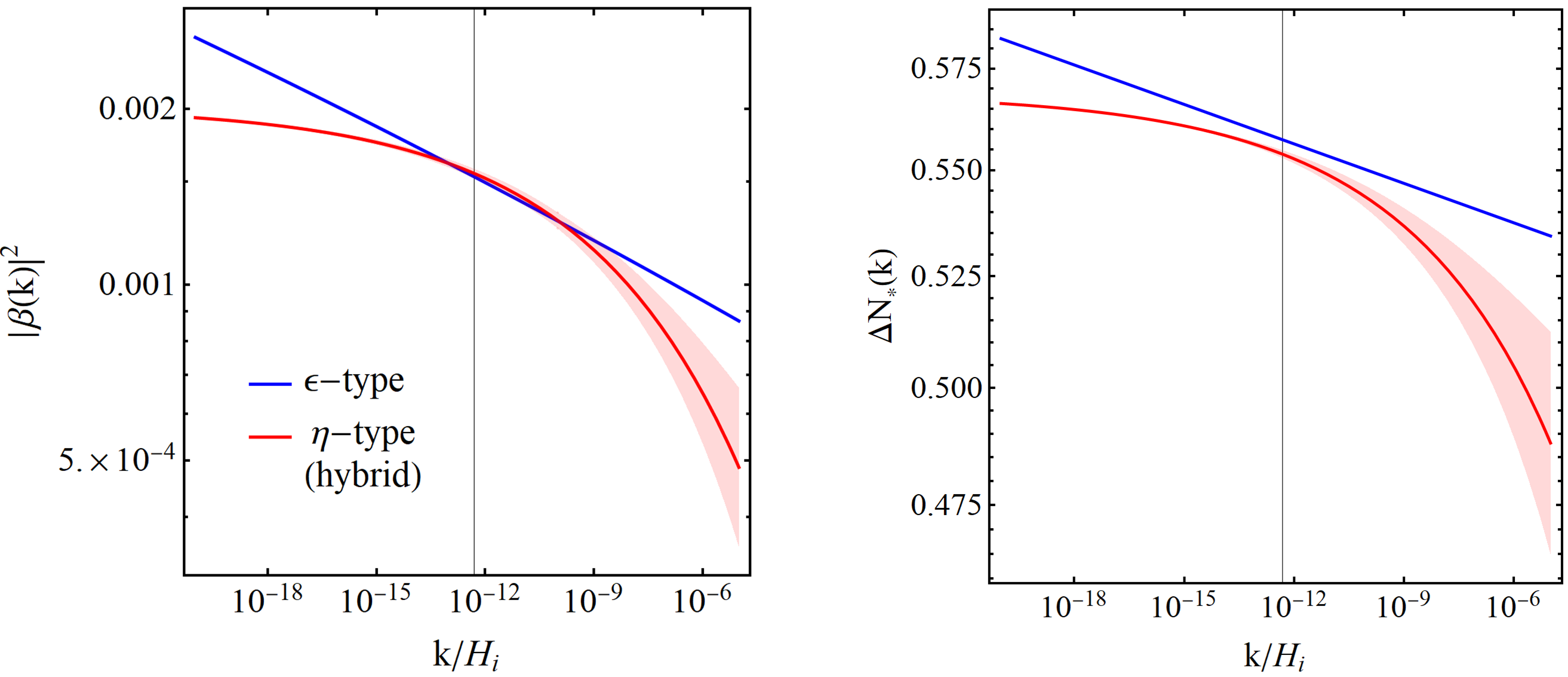}\\
	\caption{The scale dependence of production amount and width for the two spacetimes in FIG.~\ref{Tangent}. The opaque bands stand for the higher-order errors present in (\ref{dSEpsProdAmount}), (\ref{dSEpsProdWidth}), (\ref{dSetaspin1ProdAmount}) and (\ref{dSetaspin1ProdWidth}). Notice that the error band for $\epsilon$-type is too thin to be observed. The vertical black line indicates the mode $k_0=-\tau_0^{-1}$. The other parameters are chosen to be $m=10H_i$, $\kappa=9 H_i$.}
	\label{EpsVsEta}
\end{figure}

Gravitational particle production histories are crucially influenced by the Hubble parameter $H(\tau)$, as it directly enters the expression of dimensionless mass and chemical potential. Two spacetimes with different $H(\tau)$ are intrinsically different for any process that is non-local in time, including the smoothed Stokes phenomenon. Thus to test this intrinsic difference of particle production for the two types of dS deviations, we carefully select their Hubble parameters to be tangent to each other at time $\tau_0$, so that both the Hubble and the first slow-roll parameter are equal,
\begin{eqnarray}
	H^{(\epsilon)}(\tau_0,H_p,\epsilon)&=&H^{(\eta)}(\tau_0,H_i,\eta_i)\\
	\epsilon^{(\epsilon)}(\tau_0,H_p,\epsilon)&=&\epsilon^{(\eta)}(\tau_0,H_i,\eta_i)
\end{eqnarray}
For the purpose of demonstration, we will choose the following solution,
\begin{equation}
	\epsilon=0.005~,~~H_p\approxeq 0.8204 H_i~,~~\eta_i=0.1~,~~\tau_0=-2.042\times 10^{12} H_i^{-1}~.\label{tangentChoice}
\end{equation}
The Hubble parameter for the choice (\ref{tangentChoice}) is shown in FIG.~\ref{Tangent}.
The production amount and width can be computed using formula given above. Their scale dependence is shown in FIG.~\ref{EpsVsEta}. As shown in this figure, particles created near the tangent time $\tau_0$ approximately have the same production amount and width, with  mismatches of the same order as the higher-order errors in (\ref{dSetaspin1ProdAmount}) and (\ref{dSetaspin1ProdWidth}). 

As a result, one can approximate the $\eta$-deformed dS by a tangential $\epsilon$-deformed dS locally in time, and obtain the details of particles produced then, up to some higher-order errors. However, this can only be done mode-by-mode, since the scale dependence for these two types of deviations is different. Conversely, if we wish to observationally distinguish the two scenarios, we can either accurately measure the production amount/width for a single mode up to some higher orders ($e.g.$, $\mathcal{O}(c(k)^2)$), or probe the scale dependence by looking at different modes.

\subsection{Radiation-domination era}\label{RDsubsection}
Now we turn to other completely different FRW backgrounds, namely those describing the post-inflationary evolution of the universe. Roughly speaking, the scale factor evolves as $a(\tau)\propto \tau^{\frac{2}{3w+1}}$, where $w$ is the equation of state of the dominating component. In this section and the next, we will consider radiation domination era with $w=\frac{1}{3}$ and matter domination era with $w=0$, respectively.

The radiation-dominated universe has a scale factor linearly dependent on the conformal time, $a(\tau)=c_r\tau$, where $c_r>0$ and $\tau$ runs from $0$ to $+\infty$. The Hubble parameter is $H(\tau)=\frac{1}{c_r\tau^2}$. At the origin lies the Big Bang Singularity, $H(0^+)\to \infty$. This singularity can be removed by continuously deforming the spacetime to other geometries such as that of inflationary \cite{Guth:1980zm,Starobinsky:1980te,Linde:1981mu,Albrecht:1982wi}, ekpyrotic \cite{Khoury:2001wf,Buchbinder:2007ad}, bouncing \cite{Wands:1998yp,Finelli:2001sr} and string gas cosmology \cite{Brandenberger:1988aj,Nayeri:2005ck}. These physical continuations provide a cutoff time $\tau_i>0$ with $H(\tau_i)\ll M_p$. As we are interested in particle production during the later stage of the radiation-dominated era, we will limit ourselves to the modes with $k\tau_i\ll 1$ and assume that they have a vacuum initial condition at $\tau=\tau_i\approx0^+$ prepared by the earlier evolution history\footnote{If the initial condition is non-trivial with a non-zero particle number density, one can model this initial particle population by a non-zero $\beta_i$. The phase of $\beta_i$ may be fixed if the initial particles are prepared coherently, whereas it is random for thermally prepared particles. In such cases, one can return to (\ref{S0_integral}),(\ref{eq:beta_resum}),(\ref{eq:S0_fermion}),(\ref{eq:beta_resum_fermion}), and add to $\beta(\tau)$ a term proportional to $\beta_i$ and then compute the change of particle number by taking the square and performing an additional ensemble average over the phase of $\beta_i$ if it is thermally prepared.}. Thus for $\tau\gg\tau_i$, any particle mode with comoving momentum $k$ has a time-dependent physical momentum $\frac{k}{a(\tau)}=\frac{k}{c_r \tau}$. This scale is to be compared with $H(\tau)$, $m$ and $\kappa$.

\begin{figure}[h!]
	\centering
	\includegraphics[width=15cm]{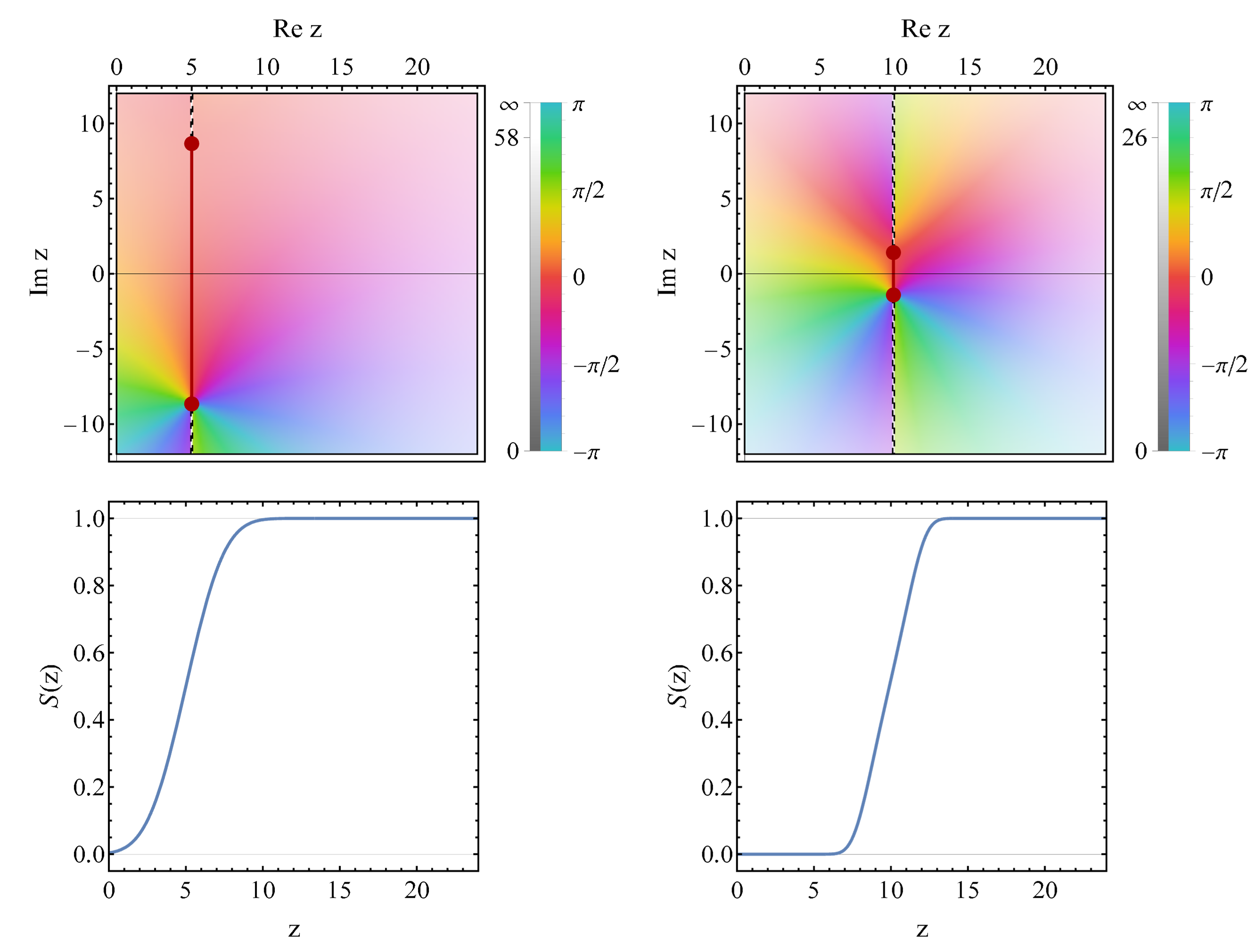}~~~~~~
	\caption{The super-adiabatic singulant $F(z)$ in radiation-dominated era for $\tilde\kappa=0.05$ (upper left panel) and $\tilde{\kappa}=0.099$ (upper right panel), with the dimensionless mass $\tilde m=0.1$. The hue represents the phase $\arg F(z)$ while the brightness represents the modulus $|F(z)|$. In both panels, the dark red line is the Stokes line joining the complex turning points. The lower two panels show the Stokes multiplier $S(z)$ corresponding to the parameters chosen above.		
	}\label{RadDomSpin1Singulant}
\end{figure}

Defining $z=k\tau$, the EoM now reads
\begin{equation}
	\frac{d^2 f(z)}{dz^2}+w^2(z,k)f(z)=0~,~w^2(z,k)=1-2\tilde\kappa(k)z +\tilde{m}^2(k) z^2~,\label{RadDomSpin1}
\end{equation}
with scale-dependent effective chemical potential and mass
\begin{equation}
	\tilde{\kappa}(k)=\frac{\kappa c_r}{k^2}~,~~~\tilde{m}(k)=\frac{m c_r}{k^2}~.
\end{equation}
Their physical meaning is the mass and chemical potential measured in units of Hubble parameter at horizon re-entry. The lower zero of $w(z;k)$ lies at
\begin{equation}
	z_c=\frac{\tilde{\kappa}-i\sqrt{\tilde{m}^2-\tilde{\kappa}^2}}{\tilde{m}^2}~.
\end{equation}
The singulant integral reads
\begin{eqnarray}
	\nonumber F(z)&=&-2i\int_{z_c}^z dz' w(z';k)\\
	&=&i\left[\frac{\tilde{\kappa }-\tilde{m}^2 z}{\tilde{m}^2}\sqrt{1-2 \tilde{\kappa }z+ \tilde{m}^2z^2}+\frac{\tilde{m}^2-\tilde{\kappa }^2}{\tilde{m}^3} \ln\frac{\sqrt{\tilde{m}^2-\tilde{\kappa
			}^2}}{\tilde{m} \left(\sqrt{1-2 \tilde{\kappa }z+ \tilde{m}^2z^2}+z \tilde{m}\right)-\tilde{\kappa }}\right]+\frac{ \pi  \left(\tilde{m}^2-\tilde{\kappa }^2\right)}{2\tilde{m}^3}~.~~~
\end{eqnarray}
The super-adiabatic corrections to $W(z)$ yield no simple pole at the origin or at infinity:
\begin{equation}
	\delta W^{(n)}(z)\xrightarrow{z\to 0}\mathcal{O}(z^0)~,~~~\delta W^{(n)}(z)\xrightarrow{z\to \infty}\mathcal{O}(z^{1-4n}),~~~n\geqslant 1~.
\end{equation}
Therefore, unlike the case of dS in conformal time coordinates, there is no super-adiabatic mass correction and hence no need for resummation. The behavior of the singulant as well as the Stokes multiplier are shown in FIG.~\ref{RadDomSpin1Singulant}.

The details of particle production can be easily obtained as follows.
\begin{enumerate}
	\item[$\bullet$] Production amount. Taking the imaginary part of $F$ on the  real axis, we have 
	\begin{equation}
		|\beta(k)|^2=e^{-\frac{\pi}{\tilde{m}^3}\left(\tilde{m}^2-\tilde{\kappa }^2\right)}=\exp\left[-\frac{\pi k^2}{c_r m^3}\left(m^2-\kappa^2\right)\right]~.\label{RadDomSpin1ProdAmount}
	\end{equation}
	There are two interesting aspects in this formula. First, $|\beta(k)|^2$ is symmetric under the flip $\kappa\leftrightarrow-\kappa$ (equivalent to a parity transformation that flips helicities), seemingly suggesting an equal enhancement for both helicities. However, this turns out to be superfluous as we will see from inspecting the production history below. Second, for a given mode $k$, the production amount does not seem to increase with mass monotonically. This phenomenon may be somewhat counterintuitive, as heavier particles naively should be more difficult to produce. This puzzle is resolved when one recalls that the radiation-dominated universe does not have a constant background temperature like dS. It effective ``temperature" is likened to the decreasing Hubble parameter $H(\tau)=\frac{1}{c_r\tau^2}$. Raising the mass may lead to two competing effects, one being increasing the difficulty of producing a real particle, the other being pushing the production time earlier, when the effective ``temperature" is higher. We will see later that for the radiation-dominated universe, the first effect dominates the applicable range of our method and heavier particles come with a smaller production amount. However, in the matter-dominated universe, this is not the case. Finally, the production amount sharply drops to zero for $k\gtrsim \sqrt{\frac{2c_r m^3}{m^2-\kappa ^2}}$, for which the ``temperature" is too low to support any real particles.
	
	\item[$\bullet$] Production time. Since the Stokes lines are vertical lines on the complex plane (see FIG.~\ref{RadDomStokesLines}), the crossing time solved from $\Im F(z_*)=0$ is simply
	\begin{equation}
		z_*=\frac{\tilde{\kappa}}{\tilde{m}^2}~,~~~\text{or}~~~\tau_*=\frac{k\kappa}{c_r m^2}~.\label{RadDomSpin1ProdTime}
	\end{equation}

	Now let us recall that the physical time domain is $0<z<+\infty$, with $z=0$ being the Big Bang singularity. The EoM (\ref{RadDomSpin1}) by itself, however, is regular at $z=0$ and admits a straightforward continuation to the \textit{unphysical} region $-\infty<z<0$. This mathematically continued EoM enjoys a $Z_2$ symmetry that is well-defined at the origin\footnote{In the dS case, this $Z_2$ symmetry is also present in the EoM, but then it is not well-defined at the dS boundary $z=0$, which is a singularity for the EoM (not a singularity for the spacetime). The lack of a smooth continuation to $z<0$ breaks the $Z_2$ symmetry spontaneously, and therefore leads to the parity asymmetry in particle production.}: $\kappa\leftrightarrow -\kappa,z\leftrightarrow-z$. This is the cause of the apparent parity symmetry in $|\beta(k)|^2$. In fact, the production time $z_*$ is also in the unphysical region for $\kappa<0$. Thus for negative $\kappa$, both (\ref{RadDomSpin1ProdAmount}) and (\ref{RadDomSpin1ProdTime}) must be understood in the analytically-continued sense. Namely, only if the initial condition at $z=0^+$ is prepared so as to match the solution of the analytically-continued EoM with Bunch-Davies initial condition at $z\to-\infty$, the Stokes-line method results are valid. For $\kappa>0$ and $0<\frac{\Delta z_*}{2}\lesssim z_*$, these results agree with that of the usual vacuum initial condition at $z=0^+$ since the Stokes line is far right to the origin and particles do not get produced until a late time. Otherwise, a direct application of the Stokes-line method may be inaccurate. In that case, the particle production history depends on the actual initial condition set at $z_i=k\tau_i\ll 1$ by an earlier cosmic evolution, which is a \textit{physical} continuation to the region $z<0$.
	
	\begin{figure}[h!]
		\centering
		\includegraphics[width=16cm]{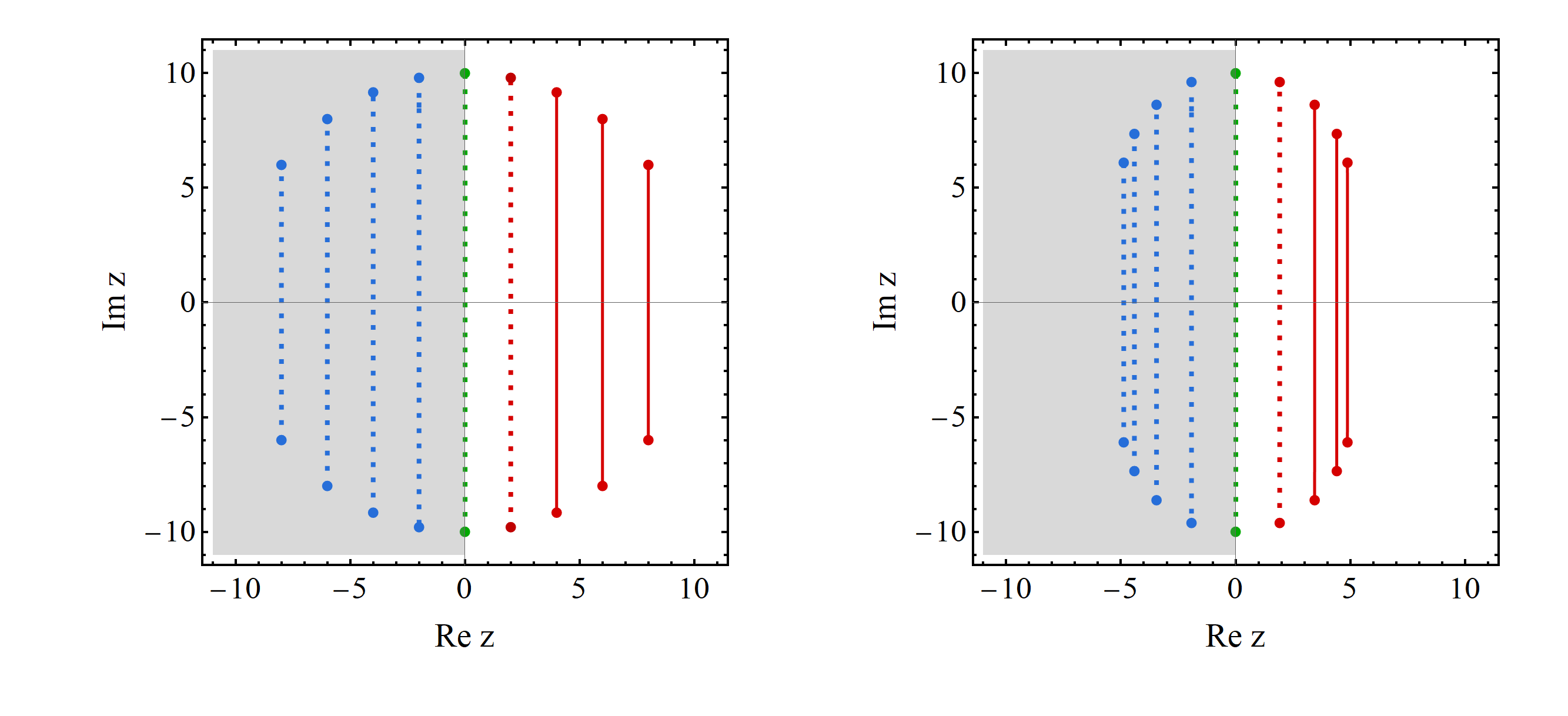}~~~~~~~~~~~~~~~~~~~~\\
		\caption{The Stokes lines for spin-1 bosons (left panel) and spin-1/2 fermions (right panel) in radiation-dominated universe. In both panels, the mass is set to $\tilde{m}=0.1$, and the momentum $k$ is fixed to have a dimensionless chemical potential ranging over $\tilde{\kappa}=-0.08,-0.06,\cdots,0.06,0.08$ (from left to right, and the green lines correspond to the case without chemical potential). The gray region with $\Re z<0$ cannot be reached physically. Solid lines represent scenarios where the Ginzburg criterion (\ref{GinzburgCriterion}) is fulfilled and vacuum initial condition is satisfied, while dotted lines can only be understood in the analytically continued sense.
		}\label{RadDomStokesLines}
	\end{figure}

	\item[$\bullet$] Production width. The production width in $z$-domain is simply calculated as
	\begin{equation}
		\Delta z_*=\frac{2\sqrt{2|\Re F(z_*)|}}{|\Im F'(z_*)|}=\sqrt{\frac{\pi}{\tilde{m}}}~,~~~\text{or}~~~\Delta\tau_*=\sqrt{\frac{\pi}{c_r m}}~.\label{RadDomSpin1ProdWidth}
	\end{equation}
	Interestingly, the production width in the $z$-domain for spin-1 boson does not depend on the chemical potential $\kappa$. The parameter region where our continuation interpretation matches that of the vacuum initial condition at $z=0^+$ is where the Ginzburg criterion is satisfied,
	\begin{equation}
		0<\frac{\Delta z_*}{2}\lesssim z_*\Rightarrow \frac{\sqrt{\pi}}{2} \tilde{m}^{3/2}\lesssim\tilde{\kappa}<\tilde{m}~,~~~\text{or}~~~\frac{\sqrt{\pi c_r} m^{3/2}}{2k}\lesssim\kappa<m~.\label{GinzburgCriterion}
	\end{equation}
	These conditions actually limits $\tilde{m}<\frac{4}{\pi}$. The parameter region satisfying the Ginzburg criterion is shown in FIG.~\ref{RadGinzburgRegion}, and one can see that both $\tilde{\kappa}$ and $\tilde{m}$ are bounded from above, $i.e.$, $\tilde{\kappa},\tilde{m}\lesssim \mathcal{O}(1)$. This suggests that the modes are still relativistic at horizon re-entry, and becomes non-relativistic only after the production time $z_*$.
	\begin{figure}[h!]
		\centering
		\includegraphics[width=9cm]{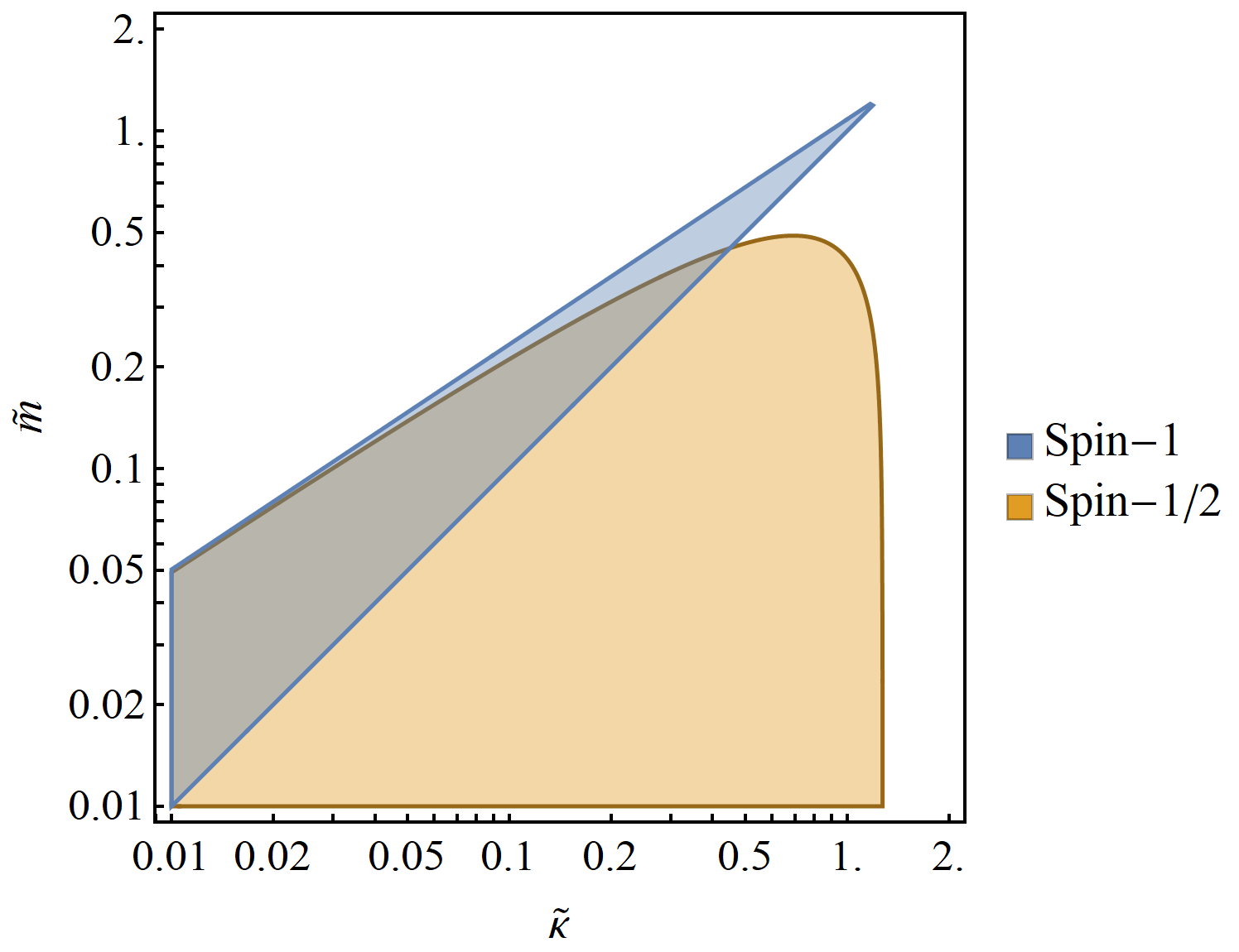}\\
		\caption{The parameter region where the Ginzburg criterion (\ref{GinzburgCriterion}) is satisfied. The blue region stands for spin-1 bosons while the yellow region stands for spin-1/2 fermions.}\label{RadGinzburgRegion}
	\end{figure}
	
\end{enumerate}

The fermion case is again obtained by a simple substitution $\tilde{m}\to\sqrt{\tilde{m}^2+\tilde{\kappa}^2}$. Here one useful check is to go to the large-chemical-potential limit with $\kappa\gg m$. There the production amount of fermions reduces to
\begin{equation}
	|\beta(k)|^2=e^{-\frac{\pi \tilde{m}^2}{\left(\tilde{m}^2+\tilde{\kappa}^2\right)^{3/2}}}\xrightarrow{\kappa\gg m}e^{-\frac{\pi  k^2 m^2}{c_r \kappa^3}}=e^{-\frac{\pi m^2}{\kappa H(\tau_*(k))}}~,
\end{equation}
which is exactly what we expect from the LZ model (\ref{fermionLZLargeChem}).

\begin{figure}[h!]
	\centering
	\includegraphics[width=17cm]{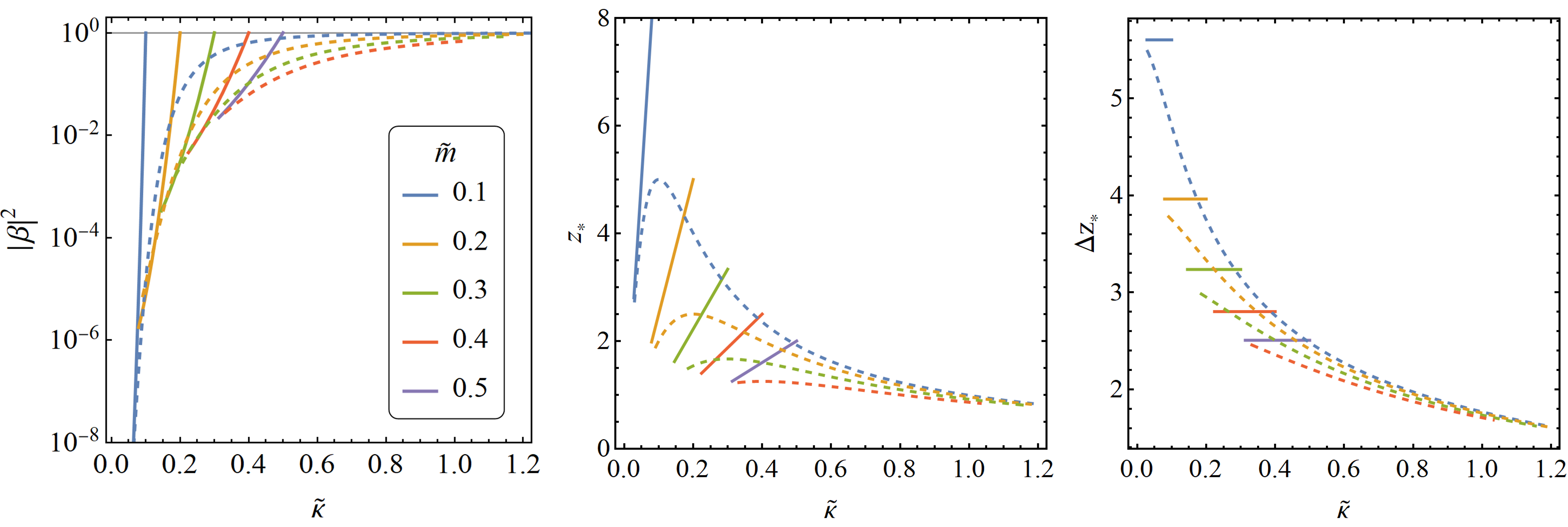}\\
	\caption{The production histories in radiation-dominated era. Left panel: Production amount as a function of the dimensionless chemical potential for different particle masses. Middle panel: The $z$-domain production time dependence on chemical potential and mass. Right panel: The $z$-domain production width. In all three plots, solid lines represent spin-1 bosons while dashed lines represent spin-1/2 fermions. Particles with different masses are distinguished by the colors of the lines according to the legend in the left panel. The parameter range is chosen according to FIG.~\ref{RadGinzburgRegion}. Due to the imposed constraint, the purple dashed line for fermions with $\tilde{m}=0.5$ is absent in all three plots, since it would correspond to artificial initial conditions, as discussed in the main text above.}\label{RadDomProdHistory}
\end{figure}

The Ginzburg criterion must also be applied to fermions. Hence, unlike the previous scenarios in dS and its deviations, the chemical potential of fermions is bounded from above (as well as below) by the applicability of our method. We plot the production histories in FIG.~\ref{RadDomProdHistory}. In the valid parameter region, we found that the production amount is monotonically decreasing with mass. Bosons are produced later with larger chemical potential, while the production time of fermions first increase and then decrease with chemical potential. The boson production width is independent of the chemical potential whereas that of fermions decreases with chemical potential, since $\tilde{\kappa}$ enters the expression for effective mass.

\subsection{Matter-dominated era}
Now we turn to the matter-dominated era with $w=0$ and a scale factor quadratically dependent on the conformal time, $a(\tau)=c_m\tau^2$. The Hubble parameter is $H(\tau)=\frac{2}{c_m\tau^3}$. Here $c_m>0$ and $\tau$ runs from $0$ to $+\infty$. The initial Big Bang singularity is understood to be removed by attaching a period of radiation domination era and some former primordial eras. A vacuum initial condition is still assumed, therefore we will still impose a Ginzburg criterion so that the Stokes-line method gives physical results. 

Defining $z=k\tau$, the EoM of a massive vector boson reads
\begin{equation}
	\frac{d^2 f(z)}{dz^2}+w^2(z,k)f(z)=0~,~w^2(z,k)=1-2\tilde\kappa(k)z^2 +\tilde{m}^2(k) z^4~,\label{MatDomSpin1}
\end{equation}
with scale-dependent chemical potential and mass
\begin{equation}
	\tilde{\kappa}(k)=\frac{\kappa c_m}{k^3}~,~~~\tilde{m}(k)=\frac{m c_m}{k^3}~.
\end{equation}
The $w(z,k)$ in the matter-dominated universe has four simple zeros on the complex $z$-plane, namely, $\pm z_c$ and $\pm z_c^*$ with
\begin{equation}
	z_c=\frac{\sqrt{\tilde{m}+\tilde{\kappa}}-i\sqrt{\tilde{m}-\tilde{\kappa}}}{\sqrt{2}\tilde{m}}~.	
\end{equation}
The $z_c$-$z_c^*$ pair lies in the right-half plane with positive real parts. Therefore, they are joined by a Stokes line that cross the real axis in the physical region. On the other hand, the Stokes line joining $-z_c$ and $-z_c^*$ crosses the real axis in the unphysical region, which can only be understood in the aforementioned continuation sense.

We are interested in the $z_c$-$z_c^*$ pair since these determines the physical particle production details. As $\kappa$ approaches $-m$, the two pairs tend to merge into one, and an analysis of the ``physical" Stokes line alone seems insufficient. In our setup, however, this is not a problem. The Ginzburg criterion keeps the production time late enough so that the turning points at $-z_c,-z_c^*$ do not have significant influence on the production history at leading order.

The singulant integral takes a relatively complicated form,
\begin{eqnarray}
	\nonumber F(z)&=&-\frac{2}{3} i z \sqrt{1-2 \tilde{\kappa }z^2+\tilde{m}^2 z^4}\\
	\nonumber&&+\frac{4}{3
		\tilde{m}}\sqrt{\frac{\tilde{m}^2-\tilde{\kappa }^2}{\tilde{\kappa
			}-i \sqrt{\tilde{m}^2-\tilde{\kappa }^2}}}\Bigg[K\left(-\frac{\tilde{m}^2-2 \tilde{\kappa }^2+2 i \tilde{\kappa
		} \sqrt{\tilde{m}^2-\tilde{\kappa }^2}}{\tilde{m}^2}\right)\\
	\nonumber&&~~~~~~~~~~~~~~~~~~~~~~~~~~~~~~~~-F\left(\arcsin\left(\frac{z
		\tilde{m}}{\sqrt{\tilde{\kappa }-i \sqrt{\tilde{m}^2-\tilde{\kappa }^2}}}\right)\Big|-\frac{\tilde{m}^2-2
		\tilde{\kappa }^2+2 i \tilde{\kappa } \sqrt{\tilde{m}^2-\tilde{\kappa }^2}}{\tilde{m}^2}\right)\Bigg]\\
	\nonumber&&+\frac{4 i \tilde{\kappa }
		\sqrt{\tilde{\kappa }+i \sqrt{\tilde{m}^2-\tilde{\kappa }^2}}}{3
		\tilde{m}^2}\Bigg[E\left(-\frac{\tilde{m}^2-2 \tilde{\kappa
		}^2+2 i \tilde{\kappa } \sqrt{\tilde{m}^2-\tilde{\kappa }^2}}{\tilde{m}^2}\right)\\
	&&~~~~~~~~~~~~~~~~~~~~~~~~~~~~~~~~-E\left(\arcsin\left(\frac{z
		\tilde{m}}{\sqrt{\tilde{\kappa }-i \sqrt{\tilde{m}^2-\tilde{\kappa }^2}}}\right)\Big|-\frac{\tilde{m}^2-2
		\tilde{\kappa }^2+2 i \tilde{\kappa } \sqrt{\tilde{m}^2-\tilde{\kappa }^2}}{\tilde{m}^2}\right)\Bigg]~,~~~
\end{eqnarray}
where $K(M)$, $E(M)$ and $E(x|M)$, $F(x|M)$ are the complete and incomplete elliptic integrals.

\begin{figure}[h!]
	\centering
	\includegraphics[width=15cm]{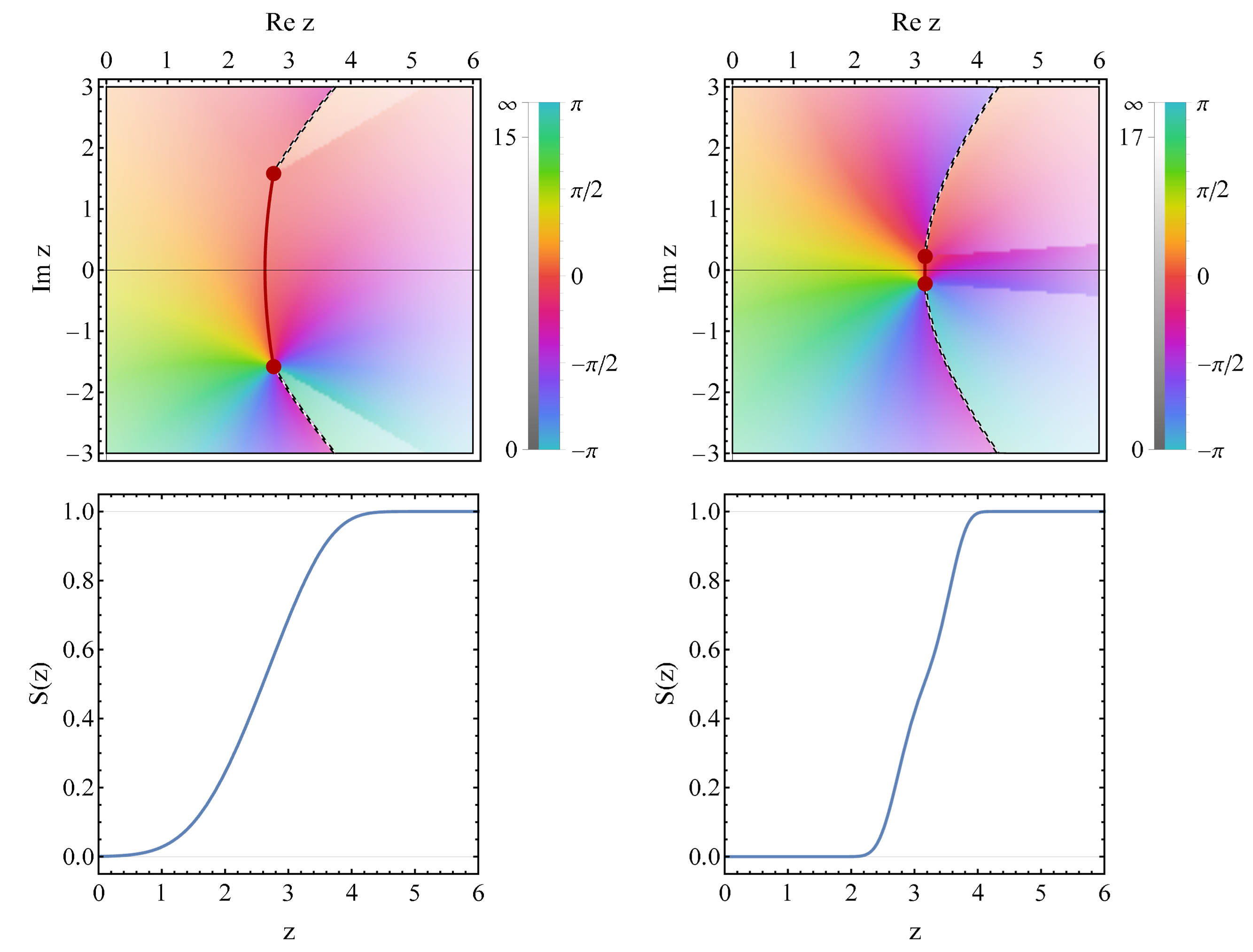}~~~~~~
	\caption{The super-adiabatic singulant $F(z)$ in matter-dominated era for $\tilde\kappa=0.05$ (upper left panel) and $\tilde{\kappa}=0.099$ (upper right panel), with the dimensionless mass $\tilde m=0.1$. The hue represents the phase $\arg F(z)$ while the brightness represents the modulus $|F(z)|$. In both panels, the dark red line is the Stokes line joining the complex turning points. The lower two panels show the Stokes multiplier $S(z)$ corresponding to the parameters chosen above.		
	}\label{MatDomSpin1Singulant}
\end{figure}
\begin{figure}[h!]
	\centering
	\includegraphics[width=16cm]{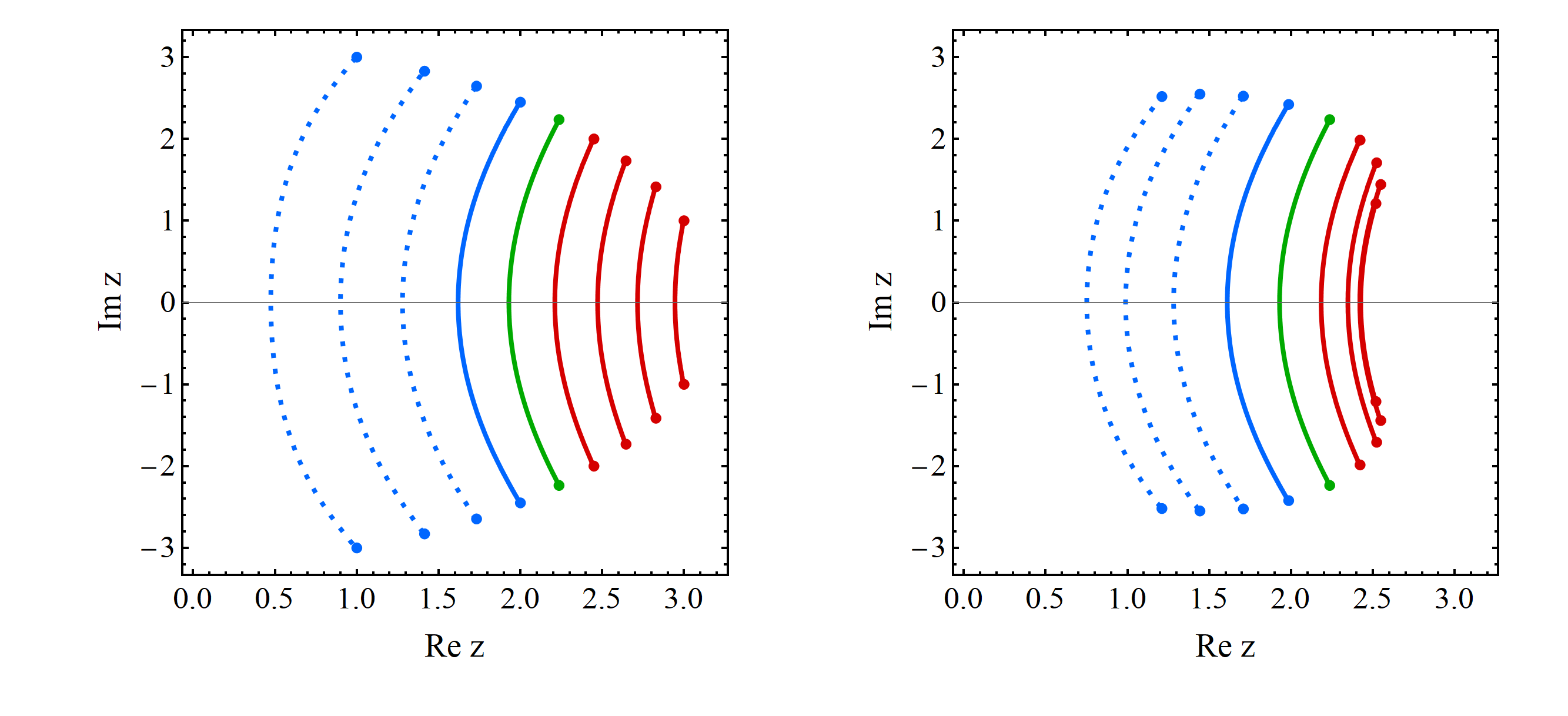}~~~~~~~~~~~~~~~~~~~~\\
	\caption{The Stokes lines for spin-1 bosons (left panel) and spin-1/2 fermions (right panel) in matter-dominated universe. In both panels, the mass is set to $\tilde{m}=0.1$, and the momentum $k$ is fixed to have a dimensionless chemical potential ranging over $\tilde{\kappa}=-0.08,-0.06,\cdots,0.06,0.08$ (from left to right, and the green lines correspond to the case without chemical potential). The gray region with $\Re z<0$ cannot be reached physically. Solid lines represent scenarios where the Ginzburg criterion (\ref{MatDomGinzburg}) is fulfilled and vacuum initial condition is satisfied, while dotted lines can only be understood in the analytically continued sense.
	}\label{MatDomStokesLines}
\end{figure}
\begin{enumerate}
	\item[$\bullet$] Production amount. Taking the real part gives $|\beta(k)|^2=e^{-2\Re F(z_i)}$ with
	\begin{eqnarray}
		\nonumber \Re F(z_i)&=&\frac{4 }{3 \tilde{m}^2}\Bigg[\sqrt{\tilde{m}^2-\tilde{\kappa }^2} \Re\left(\sqrt{\tilde{\kappa }+i
			\sqrt{\tilde{m}^2-\tilde{\kappa }^2}}
		K\left(-\frac{\tilde{m}^2-2 \tilde{\kappa }^2+2 i \tilde{\kappa } \sqrt{\tilde{m}^2-\tilde{\kappa
				}^2}}{\tilde{m}^2}\right)\right)\\
		\nonumber&&~~~~~~~~~~~~~~~~-\tilde{\kappa } \Im\left(\sqrt{\tilde{\kappa }+i
			\sqrt{\tilde{m}^2-\tilde{\kappa }^2}}
		E\left(-\frac{\tilde{m}^2-2 \tilde{\kappa }^2+2 i \tilde{\kappa } \sqrt{\tilde{m}^2-\tilde{\kappa
				}^2}}{\tilde{m}^2}\right)\right)\Bigg]\\
		&=&\frac{\Gamma \left(\frac{5}{4}\right)}{\Gamma \left(\frac{7}{4}\right)}\sqrt{\frac{\pi }{2\tilde{m}}}\left[1-\frac{2 \Gamma
			\left(\frac{3}{4}\right) \Gamma \left(\frac{7}{4}\right) \tilde{\kappa }}{\Gamma \left(\frac{1}{4}\right) \Gamma
			\left(\frac{5}{4}\right) \tilde{m}}-\frac{3 \tilde{\kappa }^2}{8 \tilde{m}^2}
		+\mathcal{O}\left(\frac{\tilde{\kappa}^3}{\tilde{m}^3}\right)\right]~.~~~
	\end{eqnarray}
	In terms of the original parameters, we have
	\begin{equation}
		|\beta(k)|^2=\exp\left\{-\frac{\Gamma \left(\frac{5}{4}\right)}{\Gamma \left(\frac{7}{4}\right) }\sqrt{\frac{2\pi k^3}{c_m m}}\left[1-\frac{2 \Gamma
			\left(\frac{3}{4}\right) \Gamma \left(\frac{7}{4}\right) \kappa}{\Gamma \left(\frac{1}{4}\right) \Gamma
			\left(\frac{5}{4}\right) m}-\frac{3 \kappa^2}{8 m^2}+\mathcal{O}\left(\frac{\kappa^3}{m^3}\right)\right]\right\}~.\label{MatDomProductionAmount}
	\end{equation}
	Thus in the matter-dominated era, the superficial degeneracy between $\kappa>0$ and $\kappa<0$ in the radiation-dominated era is explicitly broken, since there is no $Z_2$ symmetry in the EoM (\ref{MatDomSpin1}) now. This shows again that the degeneracy is unphysical. Not only the $z\to -\infty$ phase before radiation domination (inflation, etc.) can break the degeneracy, the $z\to+\infty$ stage after it (matter-domination) can do so, too. The production amount drops to zero quickly for $k\gg(c_m m)^{1/3}$.
	
	\item[$\bullet$] Production time. The production time is solved from $\Im F(z_*)=0$. The result can be expressed as a power series
	\begin{equation}
		z_*(\tilde{m},\tilde{\kappa})\simeq\frac{1}{\sqrt{\tilde{m}}}\left[0.6098+\frac{0.4659 \tilde{\kappa
		}}{\tilde{m}}-\frac{0.1011 \tilde{\kappa }^2}{\tilde{m}^2}+\frac{0.03446\tilde{\kappa}^3}{\tilde{m}^3}+\mathcal{O}\left(\frac{\tilde{\kappa}^4}{\tilde{m}^4}\right)\right]~.\label{MatDomProdTime}
	\end{equation}
	Although $z_*$ is always positive for $|\tilde\kappa|<\tilde m$, the Ginzburg criterion must still be imposed to match the vacuum initial condition.

	\item[$\bullet$] Production width. This can also be obtained as a power series,
	\begin{equation}
		\Delta z_*(\tilde{m},\tilde{\kappa})\simeq\frac{1}{\tilde{m}^{1/4}}\left[1.474-\frac{0.2971 \tilde{\kappa }}{\tilde{m}}+\frac{0.09144 \tilde{\kappa
			}^2}{\tilde{m}^2}-\frac{0.06400 \tilde{\kappa }^3}{\tilde{m}^3}+\mathcal{O}\left(\frac{\tilde{\kappa}^4}{\tilde{m}^4}\right)\right]~.\label{MatDomProdWidth}
	\end{equation}
	The Ginzburg criterion is now
	\begin{eqnarray}
		0<\frac{\Delta z_*}{2}\lesssim z_*~,~~~\text{or}~~~\left(\frac{k^3}{c_m m}\right)^{1/4}\gtrsim1.208\times\frac{1-0.2017\frac{\kappa}{m}+0.06205\frac{\kappa^2}{m^2}-0.04343\frac{\kappa^3}{m^3}+\cdots}{1+0.7641\frac{\kappa}{m}-0.1660\frac{\kappa^2}{m^2}+0.05652\frac{\kappa^3}{m^3}+\cdots}~.\label{MatDomGinzburg}
	\end{eqnarray}
	We plot this region in FIG.~\ref{MatGinzburgRegion} for both bosons and fermions.
	\begin{figure}[h!]
		\centering
		\includegraphics[width=9cm]{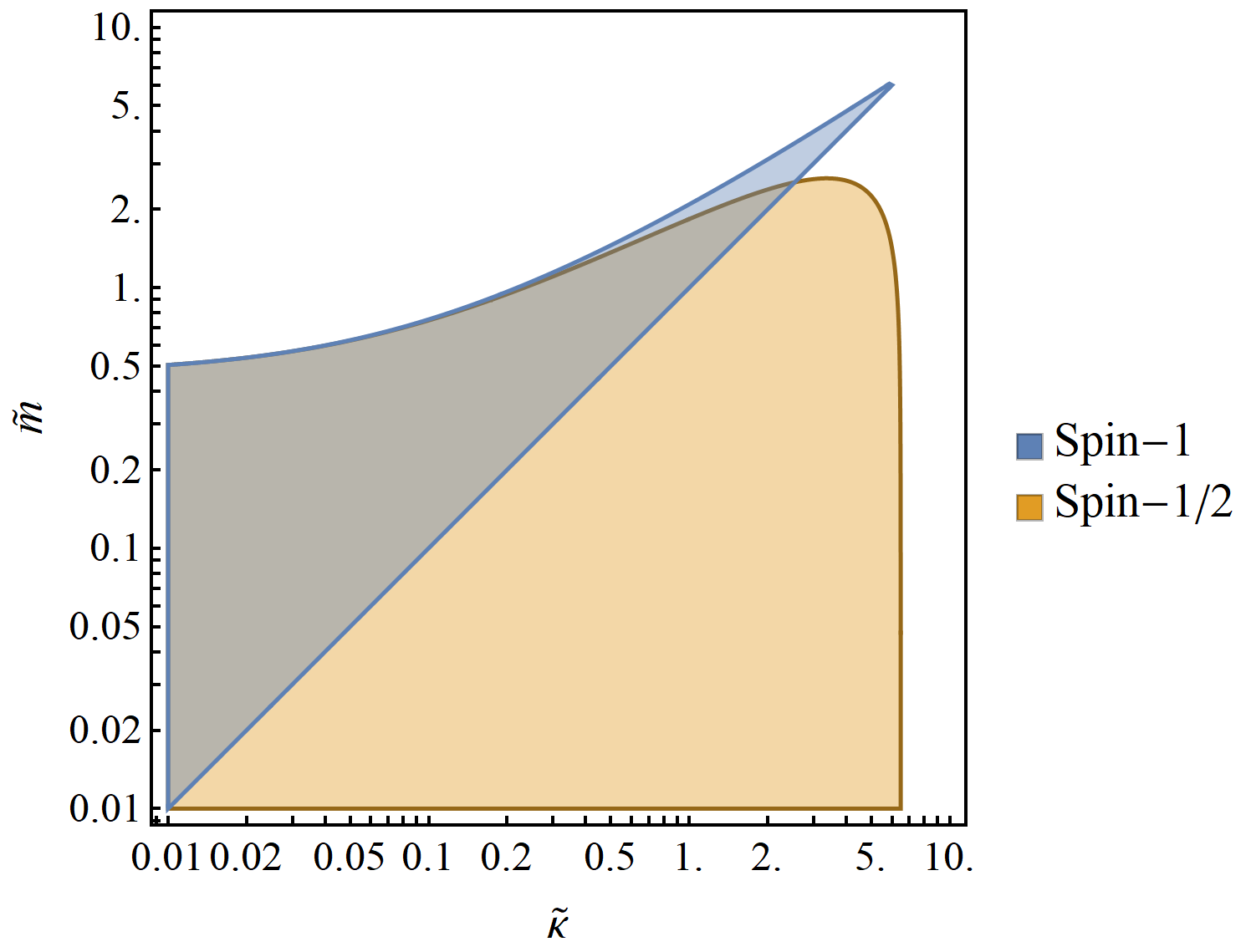}\\
		\caption{The parameter region where the Ginzburg criterion (\ref{MatDomGinzburg}) is satisfied. The blue region stands for spin-1 bosons while the yellow region stands for spin-1/2 fermions.}\label{MatGinzburgRegion}
	\end{figure}
	
\end{enumerate}

The generalization to fermions is the same as above. To check the large-chemical-potential behavior, we set $\tilde{\kappa}\gg\tilde{m}$, and obtain
\begin{equation}
	|\beta(k)|^2\xrightarrow{\tilde{\kappa}\gg\tilde{m}}e^{-\frac{\pi\tilde{m}^2}{2\tilde{\kappa}^{5/2}}}= e^{-\frac{\pi m^2}{2\kappa} \sqrt{\frac{k^3}{c_m \kappa ^3}}}=e^{-\frac{\pi m^2}{\kappa H(\tau_*(k))}}~.
\end{equation}
This again agrees with the LZ model result (\ref{fermionLZLargeChem}). In fact, we can take advantage of the LZ model and obtain some interesting sum rules for the numeric coefficients in (\ref{MatDomProdTime}) and (\ref{MatDomProdWidth}). According to the exact solution of LZ model, the $\tilde{\kappa}\gg\tilde{m}$ limit production time becomes $z_*\xrightarrow{\tilde{\kappa}\gg\tilde{m}}\tilde{\kappa}^{-1/2}$. Matching this with (\ref{MatDomProdTime}) in the fermion case, we have
\begin{equation}
	1\simeq 0.6098+0.4659-0.1011+0.03446+\cdots~.
\end{equation}
The production width predicted by the LZ model is $\Delta z_*=\sqrt{\frac{\pi}{2}}\tilde{\kappa}^{-1/4}$. Matching this with (\ref{MatDomProdWidth}), we have
\begin{equation}
	\sqrt{\frac{\pi}{2}}\simeq 1.474-0.2971+0.09144-0.06400+\cdots~.
\end{equation}
Another piece of information attainable from the LZ model is an upper bound on $\tilde{\kappa}$ due to the Ginzburg criterion:
\begin{equation}
	\frac{\Delta z_*}{2}<z_*\xrightarrow{\tilde{\kappa}\gg\tilde{m}} \tilde{\kappa}<\frac{64}{\pi^2}~,
\end{equation}
in agreement with FIG.~\ref{MatGinzburgRegion}.

\begin{figure}[h!]
	\centering
	\includegraphics[width=17cm]{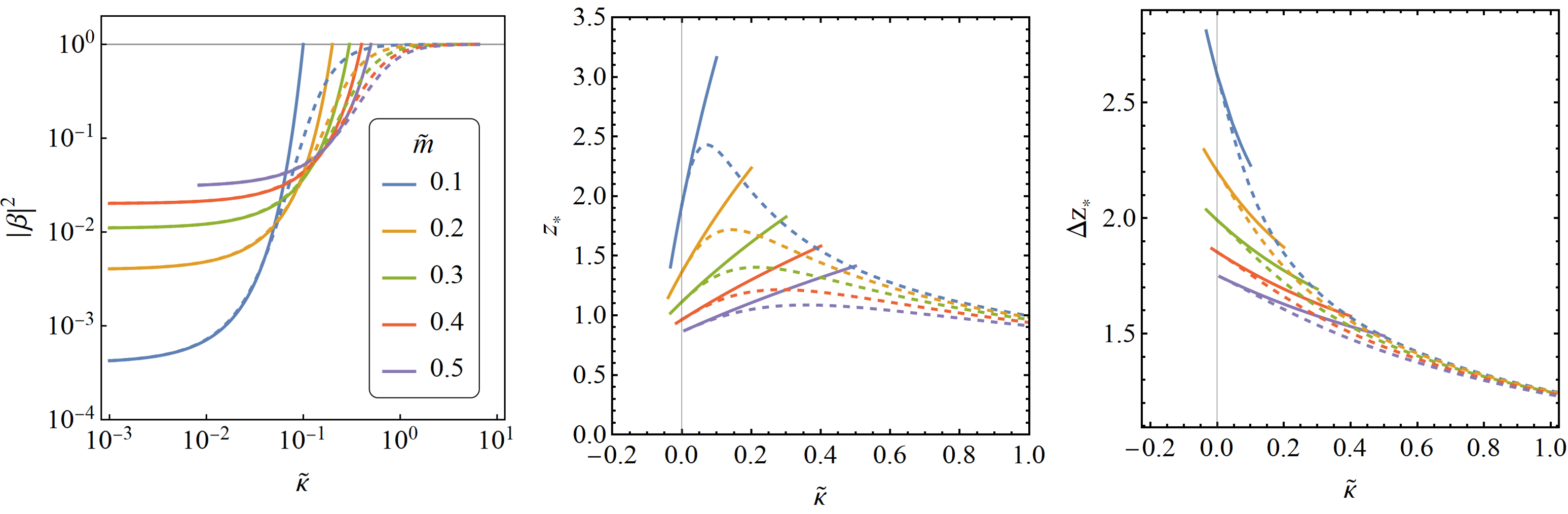}\\
	\caption{The production histories in matter-dominated era. Left panel: Production amount as a function of the dimensionless chemical potential for different particle masses. Middle panel: The $z$-domain production time dependence on chemical potential and mass. Right panel: The $z$-domain production width. In all three plots, solid lines represent spin-1 bosons while dashed lines represent spin-1/2 fermions. Particles with different masses are distinguished by the colors of the lines according to the legend in the left panel. The parameter range is chosen according to FIG.~\ref{MatGinzburgRegion}.}\label{MatDomProdHistory}
\end{figure}

As before, we plot the production histories for both bosons and fermions in FIG.~\ref{MatDomProdHistory}. The left panel clearly demonstrates the seemingly counter-intuitive mass dependence mentioned previously. Here, the valid region covers the part where the production amount increases with the mass. This shows the fact that production amount decreases with mass in the radiation-domination era is just a coincidence, and that with a time-dependent Hubble parameter, there are two opposite effects competing against each other, in which case the resulting mass dependence can be subtle. Another notable aspect is that the production time can be either earlier or later than horizon re-entry.

\section{Summary and outlook}\label{SummaryOutlook}
Ranging from cosmological collider physics in the primordial era to baryogenesis in the late universe, chemical potential plays an important role in the process of spontaneous creation of particles. In this paper, we focused on the impact of chemical potential on gravitational massive particle production. We first introduced the general form of chemical potential term and gave a necessary condition for its physical effects. After reviewing the chemical potential for particles with different spins, we extracted their essential features and likened the corresponding Bogoliubov coefficients to the coefficients of instantaneous positive/negative frequency solutions. Then the mathematical tools such as asymptotic series, Berry's smoothing techniques of Stokes-lines and Borel resummation were introduced to solve the coefficients. Having checked the applicability of this method at $|\beta|^2\lesssim 1$, we obtain the recipe of particle production histories for both spin-1 bosons and spin-1/2 fermions, which are related by a simple replacement formula. At last, applying this recipe to cosmology, we gave a fine-grained analysis of chemical-potential-assisted particle production in five common FRW spacetimes. The production amount, time and width are obtained as analytic/semi-analytical expressions, each with characteristic dependences on chemical potential and mass. 

In summary, our method demonstrates the application of uniformly smoothed Stokes-line method to fine-grained particle production. In addition, our results serve as valuable theoretical data for future studies of chemical potential as well as general particle production.

Despite the heavy mathematical machinery and the detailed analysis in this current work, there are still many questions left unanswered which we hope to address in the future. We list a few of them as outlooks below.
\begin{enumerate}
	\item[$\bullet$] Starting with vacuum initial condition, the introduction of chemical potential invites the interesting possibility of significant particle production with $|\beta|^2\sim 1$, even with large masses. This mathematically corresponds to the failure of choosing an optimal truncation order $n$ (${\rm Re}F<2$ for vector bosons and ${\rm Re}F<1$ for fermions) determined by either the stationary phase condition of the first-order perturbation or the minimum term in the asymptotic series solution, and we proposed to use the Borel summation to evaluate the whole divergent asymptotic series, extending the workable parameter regions to ${\rm Re}F\gtrsim 0.5$ for bosons and ${\rm Re}F \gtrsim 0.2$ for fermions respectively. However, large errors in evaluating the particle production amount were still found when the particle production is too large and runs outside the mentioned workable regions, and part of the reason may be attributed to the failure of approximating $w(z)$ or $E(z)$ around its complex root $z_c$. Although we are currently unable to fully resolve the problem when $|\beta|^2\sim 1$, it is interesting to note that the result with $|\beta|^2\lesssim 1$, when naively extrapolated to the $|\beta|^2\sim 1$ case, actually gives very accurate answers for the production amount ($e.g.$, in dS and radiation domination era). A systematic method of calculating particle production which can link to the limit with $|\beta|^2\sim 1$ , the tachynonic instability for bosons and the exact Landau-Zener model for fermions (\ref{eq:LZ_approximation}), may require new techniques, and we leave it for future works.
	
	\item[$\bullet$] Throughout the analysis of particle production in Sect.~\ref{TheLongTechinicalSection}, we have assumed a chemical potential constant in space and time. Although this can be justified as leading order approximations, the full understanding can only be acquired by introducing appropriate spacetime dependences according to different contexts. 
	
	\item[$\bullet$] In dS and its two types of deviations, it is natural to assume a vacuum initial condition. However, in radiation domination era and matter domination era, quantum fields do not necessarily evolve from the vacuum. In fact, it is expected to have some initial particle population produced in earlier stages of the universe such as inflation or (p)reheating. Yet a non-vacuum initial condition is highly model-dependent. In this work, we choose vacuum initial condition because we focus more on a model-independent analysis of particle production due to a later effect of chemical potential. The treatment of other initial conditions is briefly described in Sect.~\ref{RDsubsection}. However, it is worthwhile to note some interesting behaviors. If the initial particles are thermally prepared, the interference term in $|\beta(z)|^2$ is averaged out by taking the ensemble average over the phase of $\beta_i$. Thus particle number generally increases due to chemical potential, as expected. In contrast, if the initial particles are coherently prepared with a common phase of $\beta_i$, chemical potential can serve to produce or destroy particles, depending on the sign of the interference term. If the particle number decreases, one can understand it as the ``decay" of particles with energy injection into the background chemical potential sector. It would be interesting to investigate these possibilities with concrete models in the future.
	
	\item[$\bullet$] The knowledge of the production time and width can be helpful in the estimation of signal strength in cosmological collider physics. As mentioned before, the $\mathcal{O}(|\beta|)$ oscillatory signatures on the cosmological collider originate from the interference between the positive frequency part and the negative frequency part, whose presence is controlled by the Stokes multiplier $S(z)$. This fact can be useful when estimating the loop diagrams.
	At loop level, the momentum integral receives contribution from the UV region with $z\gg 1$. Usually, this UV divergent part can be regularized by a momentum cutoff at Hubble scale, $i.e.$, $\frac{k}{a}<H$. Then the signal strength follows from dimensional analysis. This is convenient if the mass of the particle running the loop is close to Hubble scale \cite{Chen:2018xck}. However, if the particle is much heavier, the dimensionless parameter $\mu> 1$ can enter in complicated ways. Adding chemical potential introduces yet another dimensionless parameter $\tilde\kappa$, thus invalidating the naive dimensional analysis \cite{Hook:2019zxa}. However, with the knowledge of particle production history, the momentum cutoff can be posed more precisely at $z_*$. This is because physically speaking, the particles that generate the signals do not get produced until their momentum drops below the production scale, $\frac{k}{a}<H (z_*\pm \Delta z_*)$. This potentially offers a better way to estimation signal strength, which deserves further explorations.
	
	\item[$\bullet$] Aside from chemical potential, there are many other sophisticated mechanisms of cosmological particle production, to which the smoothed version of Stokes-line method can be applied. For example, parametric resonance is widely used in models of preheating \cite{Traschen:1990sw,Dolgov:1989us,Kofman:1997yn}, generation of primordial black holes \cite{Cai:2018tuh,Zhou:2020kkf} and primordial gravitational waves \cite{Lin:2015nda,Cai:2020ovp}. For a periodic effective frequency $w(z)$, the turning points form periodic pairs on the complex $z$-plane, joined by periodic Stokes lines. Then the resonance condition can be viewed as the constructive interference of particle production amplitudes when crossing each Stokes line. In the literature, there are already preliminary attempts in this direction \cite{Hashiba:2021npn,Enomoto:2021hfv}, but using the traditional Stokes-line method without uniform smoothing. This will be accurate if the Stokes lines are well-separated so that one can apply the ``dilute gas" approximation, treating each crossing separately as sudden jumps in particle number. However, if the production widths are as wide as the separation between two neighboring Stokes lines, one may need to go to the fine-grained picture and perform the analysis using the smoothed version of Stokes-line method. It is interesting to compare this method with traditional ones such as the Floquet theory.
\end{enumerate}

\section*{Acknowledgment} 
We would like to thank Kaifeng Zheng and Kun-Feng Lyu for helpful discussions. This work is supported in part by GRF Grants 16301917, 16304418 and 16303819 from the Research Grants Council of Hong Kong, and the NSFC Excellent Young Scientist (EYS) Scheme (Hong Kong and Macau) Grant No. 12022516.
\appendix

\section{A checklist of results}\label{ChecklistAppendix}
In this appendix, we assemble our main results into a checklist in TABLE.~\ref{checklist}. We have explicitly spell out the schematic form of various quantities and given the full expressions/plots as references jumping into the text.

\begin{table}[h!]
	\centering
	\begin{tabular}{c|c|c|c|c|c}
		\toprule[1.2pt]
		& \textbf{Amount} & \textbf{Time}  & \textbf{Width} & \textbf{Plot} & \textbf{Valid region} \\
		\midrule[1pt]
		\multirow{4}{*}{\textbf{dS}} & $=$(\ref{dSSpin1ProdAmountAfterResum}) & $\simeq$(\ref{dSSpin1ProdTimeAfterResum}) & $\simeq$(\ref{dSSpin1ProdWidthAfterResumey}) &&\\
		& &  &  &&\\
		&$e^{-2\pi(m-\kappa)}$ &$0.6m+0.3\kappa+\cdots$& $\frac{2.1}{\sqrt{m}}-\frac{0.4\kappa}{m^{3/2}}+\cdots$&FIG.~\ref{dS01ProdHistory}&\\
		& &&&&\\
		\midrule
		\multirow{5}{*}{\textbf{$\epsilon$-dS}} & $=$(\ref{dSEpsProdAmount}) & $\approx$(\ref{dSEpsProdTime}) & $\approx$(\ref{dSEpsProdWidth}) &&\\
		& &  & &&\\
		&$e^{-2\pi(m-\kappa)[1+\epsilon(\ln k+\cdots)]}$ &$0.6m+0.3\kappa+\cdots$ & $\frac{2.1}{\sqrt{m}}-\frac{0.4\kappa}{m^{3/2}}+\cdots~~~~$  & FIG.~\ref{dSEpsProdHistory}&$\epsilon\ll 1$\\
		& &$~~~~+\epsilon(0.8 m+\cdots)$&$~~~~+\frac{\epsilon}{\sqrt{m}}(-\ln k+\cdots)$&&\\
		&& &&&\\
		\midrule
		\multirow{5}{*}{\textbf{$\eta$-dS}} & $=$(\ref{dSetaspin1ProdAmount}) & $\approx$(\ref{dSetaspin1ProdTime}) & $\approx$(\ref{dSetaspin1ProdWidth}) &&\\
		& &&&&\\
		&$e^{-2\pi(m-\kappa)[1+c(k)(\cdots)]}$&$(1+c(k))\left(0.6m+0.3\kappa+\cdots\right)$&$\left(1-\frac{c(k)}{2}\right)\left(\frac{2.1}{\sqrt{m}}-\frac{0.4\kappa}{m^{3/2}}+\cdots\right)$&FIG.~\ref{dSEtaProdHistory}& $c(k),\eta_i\ll 1$\\
		& &$+c(k)\eta_i\left(0.8m+\cdots\right)$&$+\frac{c(k)\eta_i}{\sqrt{m}}\left(\ln\frac{m}{H_i}+\cdots\right)$&&\\
		& &&&&\\
		\midrule
		\multirow{4}{*}{\textbf{RD}} & $=$(\ref{RadDomSpin1ProdAmount}) & $=$(\ref{RadDomSpin1ProdTime}) & $=$(\ref{RadDomSpin1ProdWidth}) && Ginzburg \\
		&&&&&criterion:\\
		&$e^{-\frac{\pi k^2}{c_r m^3}\left(m^2-\kappa^2\right)}$&$\frac{k^2\kappa}{c_r m^2}$&$\sqrt{\frac{\pi k^2}{c_r m}}$&FIG.~\ref{RadDomProdHistory}&(\ref{GinzburgCriterion})\\
		& &&&&\\
		\midrule
		\multirow{4}{*}{\textbf{MD}} & $=$(\ref{MatDomProductionAmount}) & $\simeq$(\ref{MatDomProdTime}) & $\simeq$(\ref{MatDomProdWidth}) && Ginzburg \\
		& &&&&criterion:\\
		&$e^{-1.0\sqrt{\frac{2\pi k^3}{c_m m}}\left(1-\frac{0.7\kappa}{m}+\cdots\right)}$ & $\sqrt{\frac{k^3}{c_m m}}\left(0.6+\frac{0.5 \kappa}{m}+\cdots\right)$& $\left(\frac{k^3}{c_m m}\right)^{1/4}\left(1.5-\frac{0.3\kappa}{m}+\cdots\right)$&FIG.~\ref{MatDomProdHistory}&(\ref{MatDomGinzburg})\\
		&&&&&\\
		\bottomrule[1.2pt]
	\end{tabular}
	\caption{A checklist of results for spin-1 vector bosons. Here RD and MD stand for radiation domination era and matter domination era, respectively.  $m$ is the mass while $\kappa$ is the chemical potential. The $=$, $\simeq$, $\approx$ symbols are used to indicate whether the result is exact, numerically exact, or empirical with 2\% error. In the schematic expressions, to display the most salient features, we have omitted the Hubble parameter and blurred the difference between $\frac{m}{H}$ and $\mu=\sqrt{\frac{m^2}{H^2}-\frac{1}{4}}$ in dS. The widths in dS, $\epsilon$-type dS and $\eta$-type dS are measured in e-folds, whereas the widths in RD and MD are measured in $z$-domain. The results for spin-1/2 fermion is obtained via a simple replacement rule $m^2\to m^2+\kappa^2$. Setting $\kappa=0$ also gives the purely gravitational production results.}\label{checklist}
\end{table}

\section{The 1/4 puzzle}\label{OneFourthPuzzle}
\begin{figure}[h!]
	\centering
	\includegraphics[width=9cm]{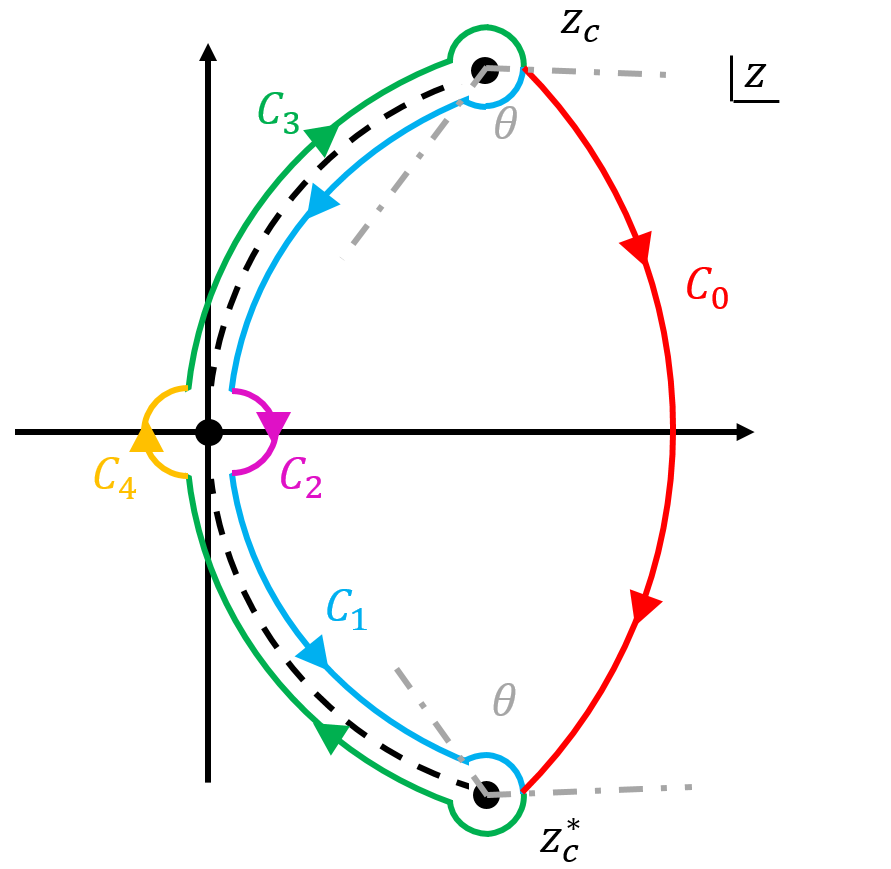}\\
	\caption{The original integration contour $\mathcal{C}_0$ and its deformations. The black dashed line is the branch cut. The angle $\theta$ at which $\mathcal{C}_1$ and $\mathcal{C}_3$ meet is chosen carefully according to a principal value prescription.}\label{contour2}
\end{figure}

As mentioned in Sect.~\ref{dSAnalysis}, the mismatch of the $1/4$ term in the dS effective mass is due to the non-vanishing correction to the IR frequency near $z=0^+$. These corrections are hidden in the asymptotic series. In order to take them into account, we must use $W$ instead of $W^{(0)}=w$ and compute Dingle's singulant $F$ order-by-order. We first deform the integration contour from $\mathcal{C}_0:~\Im F(z)=0$ to lie along the branch cut joining $z_c$, $0$ and $z_c^*$, which is defined by
\begin{equation}
	\arg w^2(z)=\pm\pi~.
\end{equation}
This is illustrated by the path $\mathcal{C}_1\cup\mathcal{C}_2$ in FIG.~\ref{contour2}. All the super-adiabatic corrections $\delta W^{(n)}$ are proportional to odd powers of $w$ and therefore possess the same branch cut. It is straightforward to check that $0,z_c,z_c^*$ and the branch cut are the only singularities for $\delta W^{(n)}$ with $n\geqslant 1$.

Now consider a closed integration path $\mathcal{C}_1\cup\mathcal{C}_2\cup\mathcal{C}_3\cup\mathcal{C}_4$ that tours around the branch cut. The integral can be converted to a residue at $z\to\infty$,
\begin{equation}
	\oint_{\mathcal{C}_{1,2,3,4}} W(z)dz=\oint_{\mathcal{C}'_{1,2,3,4}} W(1/z')\frac{dz'}{z'^2}=-2\pi i \Res_{z'\to 0}\frac{W(1/z')}{z'^2}~.
\end{equation}
Separating the full frequency into the zeroth order and higher orders, we have
\begin{equation}
	\Res_{z'\to 0}\frac{W(1/z')}{z'^2}=\Res_{z'\to 0}\left[\frac{w(1/z')}{z'^2}+\sum_{n=1}^{\infty}\frac{\delta W^{(n)}(1/z')}{z'^2}\right]=-\tilde\kappa~,
\end{equation}
where we have used the fact that $\delta W^{(n)}(z)$ drops as
\begin{equation}
	\delta W^{(n)}(z)\xrightarrow{z\to \infty}\mathcal{O}\left(\frac{1}{z^{2n+1}}\right)
\end{equation}
at infinity and therefore does not contribute to the residue. 

Then we separate the closed contour into two parts,
\begin{equation}
	2\pi i\tilde\kappa=\oint_{\mathcal{C}_{1,2,3,4}} W(z)dz=\int_{\mathcal{C}_1\cup\mathcal{C}_3}W(z)dz+\int_{\mathcal{C}_2\cup\mathcal{C}_4}W(z)dz~.\label{ClosedContourInt}
\end{equation}
Across the branch cut, the phase of $W(z)$ jumps by $\pi$ and its modulus remains continuous. Therefore, along the branch cut, the integral on $\mathcal{C}_1$ and $\mathcal{C}_3$ gives the same result. The arcs around $z_c$ and $z_c^*$ on $\mathcal{C}_1$ and $\mathcal{C}_3$, however, must be taken with a grain of salt. As these two singularities are of high orders, the integral there is ill-defined when the radius of the arc goes to zero. As a result, we need to manually impose a principal value prescription. By symmetry, the function
\begin{equation}
	H(\theta)\equiv i\int_{\mathcal{C}_1(\theta)} W(z)dz
\end{equation}
is always real. Hence when we adjust the meeting points of $\mathcal{C}_1$ and $\mathcal{C}_3$ by changing $\theta$ (see FIG.~\ref{contour2}), the integral varies from $H(0^+)$ to $H(2\pi^-)$. Then by the intermediate value theorem of a continuous real function, there always exists an angle $0<\theta_m<2\pi$ such that
\begin{equation}
	H(\theta_m)=\frac{H(0^+)+H(2\pi^-)}{2}~.
\end{equation}
If we choose $\theta_m$ to be the angle of the meeting point of $\mathcal{C}_1$ and $\mathcal{C}_3$, then the integrals along them give the same result:
\begin{equation}
	\int_{\mathcal{C}_1(\theta_m)}W(z)dz=\int_{\mathcal{C}_3(\theta_m)}W(z)dz~.
\end{equation}
In addition, due to the sign flip across the branch cut, the integral along $\mathcal{C}_2$ and $\mathcal{C}_4$ are related by
\begin{equation}
	\int_{\mathcal{C}_2}W(z)dz=-\int_{\mathcal{C}_4}W(z)dz=-\pi i \Res_{z\to 0^+} W(z)=-i\pi\mu~.
\end{equation}
Thus (\ref{ClosedContourInt}) simplifies into
\begin{equation}
	\int_{\mathcal{C}_1(\theta_m)}W(z)dz=i\pi\tilde\kappa~.
\end{equation}
Finally, we obtain the original integral as
\begin{equation}
	\int_{\mathcal{C}_0} W(z)dz=\int_{\mathcal{C}_1(\theta_m)\cup\mathcal{C}_2} W(z)dz=-i\pi (\mu-\kappa)~,
\end{equation}
which essentially gives the resummed production amount (\ref{dSBetaResummed}) in dS.

\bibliographystyle{utphys}
\bibliography{reference}

\end{document}